\documentclass[twoside]{article}

\usepackage{url,intmacros,graphicx}
\usepackage{amssymb,amsmath}
\usepackage{capt-of}
\usepackage{comment}

\def\BibTeX{{\rm B\kern-.05em{\sc i\kern-.025em b}\kern-.08em
    T\kern-.1667em\lower.7ex\hbox{E}\kern-.125emX}}

\numberwithin{equation}{section}

\newcommand{\email}[1]{\\ \small{\url{#1}} \\}
\newcommand{\institution}[1]{\\ \parbox{3.0in}{\small{#1}}}
\newcommand{\keywords}[1]{\small\textbf{Keywords: }#1}
\newcommand{\AMSsubj}[1]{\noindent\textbf{AMS subject classifications: }#1}
\newcommand\whenaccepted{Submitted: May 20, 2006;
                         Revised: November 10, 2010; July 18, 2011; September 1, 2012; Match 27, 2014; }
\newcommand{\tilt}{\!\!\sim\!\!}
\newcommand{\eqspace}{\;\;\;}

\title{Precision Arithmetic\footnote{\whenaccepted}}

\author{Chengpu Wang
\institution{40 Grossman Street, Melville, NY 11747, USA}
\email{Chengpu@gmail.com}}

\date{}

\begin{document}
\maketitle
\begin{abstract}
A new deterministic floating-point arithmetic called \emph{precision arithmetic} is developed to track precision for arithmetic calculations.  It uses a novel rounding scheme to avoid the excessive rounding error propagation of conventional floating-point arithmetic.  Unlike interval arithmetic, its uncertainty tracking is based on statistics and the central limit theorem, with a much tighter bounding range.  Its stable rounding error distribution is approximated by a truncated Gaussian distribution.  Generic standards and systematic methods for comparing uncertainty-bearing arithmetics are discussed.  The precision arithmetic is found to be superior to interval arithmetic in both uncertainty-tracking and uncertainty-bounding for normal usages.
  
The arithmetic code is published at \url{http://precisionarithm.sourceforge.net}.
\end{abstract}

\AMSsubj{65-00}

\clearpage
\section{Introduction}
\label{sec: introduction}

\subsection{Measurement Precision}

Except for the simplest counting, scientific and engineering measurements never give completely precise results \cite{Statistical_Methods}\cite{Precisions_Physical_Measurements}.  The precision of measured values ranges from an order-of-magnitude estimation of astronomical measurements to $10^{-2}$ to $10^{-4}$ of common measurements to $10^{-14}$ of state-of-art measurements of basic physics constants \cite{Basic_Constants_Measurements}.  Such value which has uncertainty is called an \emph{imprecise value}.

In scientific and engineering measurements, the uncertainty of a measurement $x$ usually is characterized by the sample deviation $\delta x$ \cite{Statistical_Methods}\cite{Precisions_Physical_Measurements}\cite{Probability_Statistics}. In certain cases, such as raw reading from an ideal analog-to-digital converter, the uncertainty of a measurement $x$ is given as a bounding range $\Delta x$\footnote{$x$ is normally an integer as the output of an ADC (Analog-to-Digital Converter).  Ideally, $\Delta x$ equals a half bit of ADC.  $\Delta x$ can be larger if the settle time is not long enough, or if the ADC is not ideal.} \cite{Electronics}.  If [$x-\Delta$x, $x+\Delta x$] crosses 0, $x$ is neither positive nor negative for certainty due to the following two possibilities: 

\begin{enumerate}
\item  Either $\Delta x$ is too large to give a precise measurement of $x$;
\item  Or $x$ itself is a measurement of zero.
\end{enumerate}

\noindent To distinguish which case it is, additional information is required so that the measurement $x \pm$ $\Delta x$ itself is \emph{insignificant} if $[x - \Delta x, x + \Delta x]$ crosses 0.  An insignificant value also has conceptual difficulty in participating in many mathematical operations, such as calculating the square root or acting as a divisor.

$P \equiv \delta x /| x | $ is defined here as the (relative) \emph{precision} of the measurement, whose inverse is commonly known as the \emph{significance} \cite{Statistical_Methods}\cite{Precisions_Physical_Measurements}.  Precision represents the reliable information content of a measurement.  Finer precision means higher reliability and thus better reproducibility of the measurement \cite{Statistical_Methods}\cite{Precisions_Physical_Measurements}.  Taking the traditional definition in measurement, precision in this paper does not mean the maximal bit count of significand as in the term ``arbitrary precision arithmetic"\footnote{Arbitrary precision integer means a digital integer which has arbitrary number of bits, while arbitrary precision arithmetic usually means fixed-point arithmetic \cite{Fixed_Point_Arithmetic} which has arbitrary fractional bits.} \cite{Arbitrary_Precision_Arithmetic}.

\subsection{Problem of Conventional Floating-Point Arithmetic}

The \emph{conventional floating-point arithmetic} \cite{Computer_Architecture}\cite{Floating_Point_Arithmetic}\cite{Floating_Point_Standard} assumes a constant and best-possible precision for each value all the time, and constantly generates artificial information during the calculation \cite{Arithmetic_Digital_Computers}.  For example, the following calculation is carried out precisely in integer format:
\begin{equation}
\begin{split}
\label{eqn: int num calc}
64919121 \times 205117922 &- 159018721 \times 83739041=\\
13316075197586562 &- 13316075197586561 = 1;
\end{split}
\end{equation}

If Formula \eqref{eqn: int num calc} is carried out using conventional floating-point arithmetic: 
\begin{equation}
\begin{split}
\label{eqn: float num calc}
64919121 \times 205117922 &- 159018721 \times 83739041 =\\ 
64919121.000000000 &\times 205117922.000000000 \\- 159018721.000000000 &\times 83739041.000000000 =\\
13316075197586562. &- 13316075197586560. = 2. = 2.0000000000000000;
\end{split}
\end{equation}
\begin{enumerate}
\item  The multiplication results exceed the maximal significance of the 64-bit IEEE floating-point representation; so they are rounded off, generating rounding errors;
\item  The normalization of the subtraction result amplifies the rounding error to most significant bit (MSB) by padding zeros.
\end{enumerate}

\noindent Formula \eqref{eqn: float num calc} is a showcase for the problem of conventional floating-point arithmetic.  Because normalization happens after each arithmetic operation \cite{Computer_Architecture}\cite{Floating_Point_Arithmetic}\cite{Floating_Point_Standard}, such generation of rounding errors happens very frequently for addition and multiplication, and such amplification of rounding errors happens very frequently for subtraction and division.  The accumulation of rounding errors is an intrinsic problem of conventional floating-point arithmetic \cite{Numerical_Recipes}, and in the majority of cases such accumulation is almost uncontrollable \cite{Arithmetic_Digital_Computers}.  For example, because a rounding error from lower digits quickly propagates to higher digits, the $10^{-7}$ precision resolution of the 32-bit IEEE floating-point format \cite{Computer_Architecture}\cite{Floating_Point_Arithmetic}\cite{Floating_Point_Standard} is usually not fine enough for calculations involving input data of $10^{-2}$ to $10^{-4}$ precision.

Self-censored rules are developed to avoid such rounding error propagation \cite{Numerical_Recipes}\cite{Precise_Numerical_Methods}, such as avoiding subtracting results of large multiplication, as in Formula \eqref{eqn: float num calc}.  However, these rules are not enforceable, and in many cases are difficult to follow, e.g., even a most carefully crafted algorithm can result in numerical instability after extensive usage.  Because the propagation speed of a rounding error depends on the nature of a calculation itself, e.g., generally faster in nonlinear algorithms than linear algorithms\footnote{A classic example is the contrast of the uncertainty propagation in the solutions for the 2nd-order linear differential equation vs. in those of Duffing equation (which has a $x^3$ term in addition to the $x$ term in a corresponding 2nd-order linear differential equation).} \cite{Chaotic_Dynamics}, propagation of rounding error in conventional floating-point arithmetic is very difficult to quantify generically \cite{Stochastic_Arithmetic}.  Thus, it is difficult to tell if a calculation is improper or becomes excessive for a required result precision.  In common practice, reasoning on an individual theoretical base is used to estimate the error and validity of calculation results, such as from the estimated transfer functions of the algorithms used in the calculation \cite{Numerical_Recipes}\cite{Error_Analysi_Digital_Filters}\cite{Floating-point_Digital_Filters}.  However, such analysis is both rare and generally very difficult to carry out in practice.  

Today most experimental data are collected by an ADC (Analog-to-Digital Converter) \cite{Electronics}.  The result obtained from an ADC is an integer with fixed uncertainty; thus, a smaller signal value has a coarser precision.  When a waveform containing raw digitalized signals from ADC is converted into conventional floating-point representation, the information content of the digitalized waveform is distorted to favour small signals since all converted data now have the same and best possible precision.  However, the effects of such distortion in signal processing are generally not clear.

What is needed is a floating-point arithmetic that tracks precision automatically.  When the calculation is improper or becomes excessive, the results become insignificant.  All existing uncertainty-bearing arithmetics are reviewed below.

\subsection{Interval Arithmetic}

\emph{Interval arithmetic} \cite{Precise_Numerical_Methods}\cite{Interval_Analysis}\cite{Worst_Case_Error_Bounds}\cite{Interval_Analysis_Theory_Applications}\cite{Interval_Arithmetic}\cite{Interval_Analysis_Notations} is currently a standard method to track calculation uncertainty.  It ensures that the value x is absolutely bounded within its \emph{bounding range} $[x] \equiv [\underbar{x}, \bar{x}]$, in which $\underbar{x}$ and $\bar{x}$ are lower and upper bounds for $x$, respectively. In this paper, interval arithmetic is simplified and tested as the following arithmetic formulas\footnote{For the mathematical definition of interval arithmetic, please see \cite{Interval_Analysis_Notations}.}  \cite{Interval_Analysis_Theory_Applications}:
\begin{align}
\label{eqn: interval +} & 
\left[x_1\right] + \left[x_2\right] = \left[\underbar{x}_{1} + \underbar{x}_{2}, \bar{x}_{1} + \bar{x}_{2}\right]; \\
\label{eqn: interval -} & 
\left[x_1\right] - \left[x_2\right] = \left[\underbar{x}_{1} - \bar{x}_{2}, \bar{x}_{1} - \underbar{x}_{2}\right]; \\
\label{eqn: interval *} &
\left[x_1\right] \times \left[x_2\right] = \left[min(\underbar{x}_{1}  \underbar{x}_{2}, \underbar{x}_{1}  \bar{x}_{2}, \bar{x}_{1}  \underbar{x}_{2}, \bar{x}_{1}  \bar{x}_{2}), max(\underbar{x}_{1}  \underbar{x}_{2}, \underbar{x}_{1}  \bar{x}_{2}, \bar{x}_{1}  \underbar{x}_{2}, \bar{x}_{1}  \bar{x}_{2})\right]; \\
\label{eqn: interval /} & 0 \notin \left[x_2\right]: \;
\left[x_1\right] / \left[x_2\right] = \left[x_1\right] \times \left[1/\bar{x}_{2}, 1/\underbar{x}_{2}\right];
\end{align}

If interval arithmetic is implemented using a floating-point representation with limited resolution, its resulting bounding range is widened further \cite{Worst_Case_Error_Bounds}.

A basic problem is that the bounding range used by interval arithmetic is not compatible with usual scientific and engineering measurements, which instead use the statistical mean and deviations to characterize uncertainty \cite{Statistical_Methods}\cite{Precisions_Physical_Measurements}.  Most measured values are well approximated by a Gaussian distribution \cite{Statistical_Methods}\cite{Precisions_Physical_Measurements}\cite{Probability_Statistics}, which has no limited bounding range.  Let \emph{bounding leakage} be defined as the possibility of the true value to be outside a bounding range.  If a bounding range is defined using a statistical rule on bounding leakage, such as the $6\sigma-10^{-9}$ rule for Gaussian distribution \cite{Probability_Statistics} (which says that the bounding leakage is about $10^{-9}$ for a bounding range of mean $\pm$ 6-fold of standard deviations), there is no guarantee that the calculation result will also obey the $6\sigma-10^{-9}$ rule using interval arithmetic, since interval arithmetic has no statistical foundation\footnote{There is some attempt \cite{Statistics_For_Interval_Arithmetic} to connect intervals in interval arithmetic to confidence interval or the equivalent so called p-box in statistics. Because this attempt seems to rely heavily on 1) specific properties of the uncertainty distribution within the interval and/or 2) specific properties of the functions upon which the interval arithmetic is used, this attempt does not seem to be generic.  Anyway, this attempt seems to be outside the main course of interval arithmetic, which has no statistics in mind.}.  

Another problem is that interval arithmetic only provides the worst case of uncertainty propagation, so that it tends to over-estimate uncertainty in reality.  For instance, in addition and subtraction, it gives the result when the two operands are +1 and -1 correlated respectively \cite{Affine_Arithmetic}.  However, if the two operands are -1 and +1 correlated respectively instead, the actual bounding range after addition and subtraction reduces, which is called the best case in random interval arithmetic \cite{Random_Interval_Arithmetic}.  The vast overestimation of bounding ranges in these two worst cases prompts the development of affine arithmetic \cite{Affine_Arithmetic}\cite{Affine_Arithmetic_book}, which traces error sources using a first-order model.  Being expensive in execution and depending on approximate modeling even for such basic operations as multiplication and division, affine arithmetic has not been widely used.  In another approach, random interval arithmetic \cite{Random_Interval_Arithmetic} reduces the uncertainty over-estimation of standard interval arithmetic by randomly choosing between the best-case and the worst-case intervals.  

A third problem is that the results of interval arithmetic may depend strongly on the actual expression of an analytic function $f(x)$.  For example, Formula \eqref{eqn: interval depend 1}, Formula \eqref{eqn: interval depend 2} and Formula \eqref{eqn: interval depend 3} are different expressions of the same $f(x)$; however, the correct result is obtained only through Formula \eqref{eqn: interval depend 1}, and uncertainty may be exaggerated in the other two forms, e.g., by 67-fold and 33-fold at input range [0.49, 0.51] using Formula \eqref{eqn: interval depend 2} and Formula \eqref{eqn: interval depend 3}, respectively.  This is called the dependence problem of interval arithmetic \cite{Interval_Arithmetic}.  
\begin{align}
\label{eqn: interval depend 1} & 
f(x) = (x - 1/2)^{2} - 1/4; \\
\label{eqn: interval depend 2} & 
f(x) = x^{2} - x; \\
\label{eqn: interval depend 3} & 
f(x) = (x - 1) x;
\end{align}

Interval arithmetic has very coarse and algorithm-specific precision but constant zero bounding leakage.  It represents the other extreme from conventional floating-point arithmetic.  To meet practical needs, a better uncertainty-bearing arithmetic should be based on statistical propagation of the rounding error, while also allowing reasonable bounding leakage for normal usages.

\subsection{Statistical Propagation of Uncertainty}

If each operand is regarded as a random variable, and the statistical correlation between the two operands is known, the resulting uncertainty is given by the \emph{statistical propagation of uncertainty} \cite{Statistical_Arithmetic}\cite{Statistical_Analysis}, with the following arithmetic equations, in which $\sigma$ is the deviation of a measured value $x$, $P$ is its precision, and $\gamma$ is the correlation between the two operands $x_{1}$ and $x_{2}$:

\begin{align}
\label{eqn: stat +} 
(x_1 \pm \sigma _1) + (x_2 \pm \sigma _2) & = (x_1 + x_2) & \pm \sqrt{\sigma _1^{2} + \sigma _2^{2} + 2 \sigma _1 \sigma _2 \gamma}; \\
\label{eqn: stat -} 
(x_1 \pm \sigma _1) - (x_2 \pm \sigma _2) & = (x_1 - x_2) & \pm \sqrt{\sigma _1^{2} + \sigma _2^{2} - 2 \sigma _1 \sigma _2 \gamma}; \\
\label{eqn: stat *} 
(x_1 \pm \sigma _1) \times (x_2 \pm \sigma _2) & = (x_1 \times x_2) & \pm |x_1 \times x_2|\sqrt{P_1^{2} + P_2^{2} + 2 P_1 P_2 \gamma}; \\
\label{eqn: stat /} 
(x_1 \pm \sigma _1) / (x_2 \pm \sigma _2) & = (x_1/x_2) & \pm |x_1 / x_2|\sqrt{P_1^{2} + P_2^{2} - 2 P_1 P_2 \gamma};
\end{align}

Tracking uncertainty propagation statistically seems an ideal solution.  However, in practice, the correlation between two operands is generally not precisely known, so the direct use of statistical propagation of uncertainty is very limited.  In this paper, as a proxy for statistical propagation of uncertainty, an \emph{independence arithmetic} always assumes that no correlation exists between any two operands, whose arithmetic equations are Formula \eqref{eqn: stat +}, Formula \eqref{eqn: stat -}, Formula \eqref{eqn: stat *} and Formula \eqref{eqn: stat /}, where $\gamma=0$.  Independence arithmetic is actually de facto arithmetic in engineering data processing, such as in the common belief that uncertainty after averaging reduces by the square root of number of measurements \cite{Statistical_Methods}\cite{Precisions_Physical_Measurements}, or the ubiquitous Monte Carlo method\footnote{Most but not all applications of Monte Carlo methods assume independence between any two random variables.  In a minority of applications, a Monte Carlo method can be used to construct specified correlation between two random variables \cite{Monte_Carlo_Statistics}.} \cite{Monte_Carlo_Method}\cite{Monte_Carlo_Statistics}, or calculating the mean and variance of a Taylor expansion \cite{Taylor_Expansion_Uncertainty}.

\subsection{Significance Arithmetic}

\emph{Significance arithmetic} \cite{Significance_Arithmetic} tries to track reliable bits in an imprecise value during the calculation.  In the two early attempts \cite{Digital_Significance_Arithmetic}\cite{Unnormalized_Arithmetic}, the implementations of significance arithmetic are based on simple operating rules upon reliable bit counts, rather than on formal statistical approaches.  They both treat the reliable bit counts as integers when applying their rules, while in reality a reliable bit count could be a fractional number \cite{Mathematica_Significance_Arithmetic}, so they both can cause artificial quantum reduction of significance.  The significance arithmetic marketed by Mathematica \cite{Mathematica_Significance_Arithmetic} uses a linear error model that is consistent with a first-order approximation of interval arithmetic \cite{Precise_Numerical_Methods}\cite{Interval_Analysis_Theory_Applications}\cite{Interval_Arithmetic}, and further provides an arbitrary precision representation which is in the framework of  conventional floating-point arithmetic. It is definitely not a statistical approach. 

Stochastic arithmetic \cite{Stochastic_Arithmetic}\cite{CADNA_library}, which can also be categorized as significance arithmetic, randomizes the least significant bits (LSB) of each of input floating-point values, repeats the same calculation multiple times, and then uses statistics to seek invariant digits among the calculation results as significant digits.  This approach may require too much calculation since the number of necessary repeats for each input is specific to each algorithm, especially when the algorithm contains branches.  Its sampling approach may be more time-consuming and less accurate than direct statistical characterization \cite{Probability_Statistics}, such as directly calculating the mean and deviation of the underlying distribution.  It is based on modeling rounding errors in conventional floating-point arithmetic, which is quite complicated.  A better approach may be to define arithmetic rules that make error tracking by probability easier.

As the mathematical foundation to significance arithmetic, when a uncertainty-bearing value is multiplied by a constant, the significance or relative precision still holds, while the absolute precision \cite{Statistical_Methods}\cite{Precisions_Physical_Measurements} scales with the constant.  In this respect, fixed-point arithmetic \cite{Fixed_Point_Arithmetic}, which assumes a fixed absolute precision, does not have a sounding mathematical foundation.

\subsection{ An Overview of This Paper}

In this paper, a new floating-point arithmetic called \emph{precision arithmetic} \cite{Prev_Precision_Arithmetic} is developed to track uncertainty during floating-point calculations, as described in Section \ref{sec: precision arithmetic}.  Generic standards and systematic methods for validating uncertainty-bearing arithmetics are discussed in Section \ref{sec: validation}.  Precision arithmetic is compared with other uncertainty-bearing arithmetics in Section \ref{sec: FFT} to Section \ref{sec: integration}.  A brief discussion is provided in Section \ref{sec: conclusion and discussion}.

\clearpage
\section{Precision Arithmetic}
\label{sec: precision arithmetic}

\subsection{Assumptions for Precision Arithmetic}

As stated previously, the precision $P$ is defined as the (relative) precision of a measurement in this paper. Precision arithmetic tracks uncertainty distribution during calculations using specially designed arithmetic rules.  It has the \emph{uncorrelated uncertainty assumption} as its basic assumption, presuming that the uncertainties of any two different values can be regarded as uncorrelated of each other.  This assumption can be turned into a realistic statistical requirement for input data for precision arithmetic.

Because it is not realistic to track the actual uncertainty distributions, which may vary according to each specific algorithm, the objectives of precision arithmetic are to enclose the actual uncertainty distribution with a \emph{bounding distribution}:
\begin{enumerate}
\item  The bounding distribution is symmetric around an expected value which is the value given by mathematics when there is no uncertainty.  
\item  The bounding distribution is Gaussian, with deviations calculated by precision arithmetic.  
\end{enumerate}
As shown later in this paper, the objectives of precision arithmetic are extended from the central limit theorem \cite{Probability_Statistics}, so that the bounding distribution is a truncated Gaussian distribution, which approximates the actual uncertainty closely when there is decent amount of arithmetic calculations.

In addition, precision arithmetic uses heavily the \emph{scaling principle} which says that the result precision should not change when an imprecise value is either multiplied or divided by a non-zero constant.  The scaling principle can be concluded from Formula \eqref{eqn: stat *} and Formula \eqref{eqn: stat /} for statistical propagation of uncertainty.  It is also the foundation for significance arithmetic. 

Related to the scaling principle, the \emph{recovering principle} says that the imprecise result should restore the original imprecise value if mathematically the original value is restored conceptually, such as when an imprecise value is inverted twice. In precision arithmetic, the value of the imprecise value obeys the recovering principle, while it is questionable if the uncertainty of the imprecise value can be recovered.

\subsection{The Uncorrelated Uncertainty Assumption}

When there is a good estimation of the sources of uncertainty, the uncorrelated uncertainty assumption can be judged directly, e.g., if noise \cite{Statistical_Methods}\cite{Precisions_Physical_Measurements} is the major source of uncertainty, the uncorrelated uncertainty assumption is probably true.  This criterion is necessary to ascertain repeated measurements of the same signal.   Otherwise, the uncorrelated uncertainty assumption can be judged by the correlation and the respectively precisions of two measurements.  

Let $X$, $Y$, and $Z$ denote three mutually independent random variables \cite{Probability_Statistics} with variance $\sigma^{2}(X)$, $\sigma^{2}(Y)$ and $\sigma^{2}(Z)$, respectively.  Let $\alpha$ denote a constant.  Let $Cov()$ denote the covariance function.  Let $\gamma$ denote the correlation between $(X + Y)$ and $(\alpha X + Z)$. And let:
\begin{align}
& \eta _{1} ^{2} \equiv \frac{\sigma ^{2} (Y)}{\sigma ^{2} (X)}; \eqspace
 \eta _{2} ^{2} \equiv \frac{\sigma ^{2} (Z)}{\sigma ^{2} (\alpha X)} =\frac{\sigma ^{2} (Z)}{\alpha ^{2} \sigma ^{2} (X)}; \\
\label{eqn: correlation}
& \gamma =\frac{Cov(X+Y,\alpha X+Z)}{\sqrt{\sigma ^{2} (X+Y)} \sqrt{\sigma ^{2} (\alpha X+Z)}}  =\frac{\alpha /|\alpha |}{\sqrt{1+\eta _{1} ^{2} } \sqrt{1+\eta _{2} ^{2}}} \equiv \frac{\alpha /|\alpha |}{1+\eta ^{2}};
\end{align}

Formula \eqref{eqn: correlation} gives the correlation $\gamma$ between two random variables, each of which contains a completely uncorrelated part and a completely correlated part, with $\eta$ being the average ratio between these two parts.  Formula \eqref{eqn: correlation} can also be interpreted reversely: if two random variables are correlated by $\gamma$, each of them can be viewed as containing a completely uncorrelated part and a completely correlated part, with $\eta$ being the average ratio between these two parts.

\begin{figure}[p]
\centering
\includegraphics[height=2.5in]{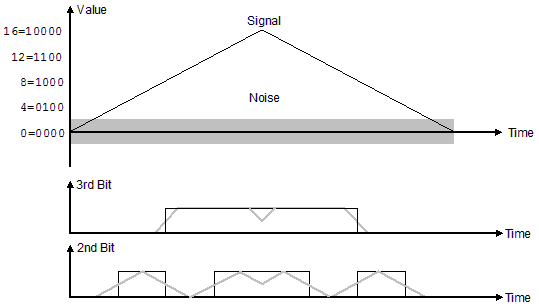}
\captionof{figure}{Effect of noise on bit values of a measured value.  The triangular wave signal and the added white noise are shown at top using the thin black line and the grey area, respectively.  The values are measured by a theoretical 4-bit Digital-to-Analog Converter in ideal condition, assuming LSB is the 0th bit.  The measured 3rd and 2nd bits without the added noise are shown using thin black lines, while the mean values of the measured 3rd and 2nd bits with the added noise are shown using thin grey lines.}
\label{fig: Signal_and_Noise}
\end{figure}

One special application of Formula \eqref{eqn: correlation} is the correlation between a measured signal and its true signal, in which noise is the uncorrelated part between the two.  Figure \ref{fig: Signal_and_Noise} shows the effect of noise on the most significant two bits of a 4-bit measured signal when $\eta=1/4$.  Its top chart shows a triangular waveform between 0 and 16 as a black line, and a white noise between -2 and +2, using the grey area.  The measured signal is the sum of the triangle waveform and the noise.  The middle chart of Figure \ref{fig: Signal_and_Noise} shows the values of the 3rd digit of the true signal as a black line, and the mean values of the 3rd bit of the measurement as a grey line.  The 3rd bit is affected by the noise during its transition between 0 and 1.  For example, when the signal is slightly below 8, only a small positive noise can turn the 3rd digit from 0 to 1.  The bottom chart of Figure \ref{fig: Signal_and_Noise} shows the values of the 2nd digit of the signal and the measurement as a black line and a grey line, respectively.  Figure \ref{fig: Signal_and_Noise} clearly shows that the correlation between the measurement and the true signal is less at the 2nd digit than at the 3rd digit.  Quantitatively, according to Formula \eqref{eqn: correlation}:

\begin{enumerate}
\item  The overall measurement is 99.2\% correlated to the signal with $\eta=1/8$;
\item  The 3rd digit of the measurement is 97.0\% correlated to the signal with $\eta=1/4$;
\item  The 2nd digit of the measurement is 89.4\% correlated to the signal with $\eta=1/2$;
\item  The 1st digit of the measurement is 70.7\% correlated to the signal with $\eta=1$;
\item  The 0th digit of the measurement is 44.7\% correlated to the signal with $\eta=2$.
\end{enumerate}
The above conclusion agrees with the common experiences that, below the noise level of measured signals, noises rather than true signals dominate each digit.  

Similarly, while the correlated portion between two values has exactly the same value at each bit of the two values, the ratio of the uncorrelated portion to the correlated portion increases by 2-fold for each bit down from MSB of the two values, regardless of the nature of the uncorrelated portion.  Quantitatively, let $P$ denote the larger precision of the two values, and let $\eta_{P}$ denote the ratio of the uncorrelated portion to the correlated portion at level of uncertainty; then $\eta_{P}$ increases with decreased $P$ according to Formula \eqref{eqn: uncertainty level}. According to Formula \eqref{eqn: correlation}, if two significant values are overall correlated with $\gamma$, at the level of uncertainty the correlation between the two values decreases to $\gamma_P$ according to Formula \eqref{eqn: uncertainty correlation}.
\begin{align}
\label{eqn: uncertainty level}
& \eta_{P} = \frac{\eta}{P}, \eqspace P < 1; \\
\label{eqn: uncertainty correlation}
& \frac{1}{\gamma_{P}} - 1 = \left(\frac{1}{\gamma} -1\right) \frac{1}{P^2}, \eqspace P < 1;
\end{align}

\begin{figure}[p]
\centering
\includegraphics[height=2.5in]{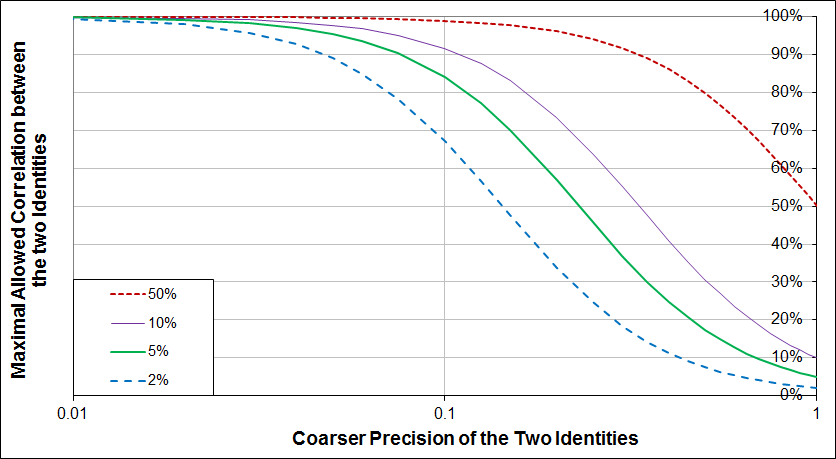} 
\captionof{figure}{Allowed maximal correlation between two values vs. input precisions and independence standard (as shown in legend) for the independence uncertainty assumption of precision arithmetic to be true.}
\label{fig: Independent_Uncertainty_Assumption}
\end{figure}

Figure \ref{fig: Independent_Uncertainty_Assumption} plots the relation of $\gamma$ vs. $P$ for each given $\gamma_{P}$ in Formula \eqref{eqn: uncertainty correlation}.  When $\gamma_{P}$ is less than a predefined maximal threshold (e.g., 2\%, 5\% or 10\%), the two values can be deemed virtually uncorrelated of each other at the level of uncertainty.  If the two values are independent of each other at their uncertainty levels, their uncertainties are uncorrelated of each other.  Thus for each independence standard $\gamma_{P}$, there is a maximal allowed correlation between two values below which the uncorrelated uncertainty assumption of precision arithmetic holds.  The maximal allowed correlation is a function of the larger precision of the two values according to Formula \eqref{eqn: uncertainty correlation}.  Figure \ref{fig: Independent_Uncertainty_Assumption} shows that for two precisely measured values, their correlation $\gamma$ is allowed to be quite high.  To be acceptable in precision arithmetic, each of the low-resolution values should contain enough noise in its uncertainty, so that they do not have much correction through the systematic error \cite{Statistical_Methods}\cite{Precisions_Physical_Measurements}.  Thus, the uncertainty assumption uncertainty assumption has much weaker statistical requirement than the assumption for independence arithmetic, which requires the two values to be independent of each other.

It is tempting to add noise to otherwise unqualified values to make their uncertainties uncertainty assumption of each other.  As an extreme case of this approach, if two values are constructed by adding noise to the same signal, they are 50\% correlated at the uncertainty level so that they will not satisfy the uncorrelated uncertainty assumption\footnote{The 50\% curve in Figure \ref{fig: Independent_Uncertainty_Assumption} thus defines the maximal possible correlations between any two measured signals. This other conclusion of Formula \eqref{eqn: uncertainty correlation} makes sense because the measurable correlation between two measurements should be limited by the precisions of their measurements.}.

\subsection{Precision Representation and Precision Round Up Rule}

Let the content of a floating-point number be denoted as $S 2^E$, in which $S$ is the significand\footnote{While ``significand'' is the official word \cite{Floating_Point_Standard} to describe ``The component of a binary floating-point number that consists of an explicit or implicit leading bit to the left of its implied binary point and a fraction field to the right'', ``mantissa'' is often unofficially used instead.} and $E$ is the exponent of 2 of the floating-point number.  In addition, the precision representation $S\tilt 2^E$ contains a carry $\;\tilt\;$  to indicate its rounding error, which can be:
\begin{itemize}
\item  +: The rounding error is positive;
\item  -: The rounding error is negative;
\item  ?: The sign of the rounding error is unknown;
\item  \#: The precision value contains an error code.  Each error code is generated due to a specific illegal arithmetic operation such as dividing by zero.  An operand error code is directly transferred to the operation result.  In this way, illegal operations can be traced back to the source.
\end{itemize}

Because there is only limited bits to hold $S$, a \emph{round up} is needed, which proceeds according to the following round up rule:
\begin{itemize}
\item  A value of $(2S)\tilt 2^E$ is rounded up to $S\tilt 2^{E+1}$.  
\item  A value of $(2S+1)\!+\!2^E$ is rounded up to $(S+1)\!-\!2^{E+1}$.  
\item  A value of $(2S+1)\!-\!2^E$ is rounded up to $S\!+\!2^{E+1}$.
\item  A value of $(2S+1)?2^E$ is rounded up to $(S+1)\!-\!2^{E+1}$.  
\end{itemize}
Let the value before any rounding up be the original value, the round-up rule ensures that $S 2^E$ is always the closest value with exponent E to the original value.  After each round up, the original rounding error is reduced by half for the new significand.  If the original significand is odd, the round up generates a new rounding error of 1/2, which is added to the existing rounding error.  Since the newly generated rounding error always cancels the existing rounding error, the rounding error range is limited to half bit of the significand, or the bounding range for the rounding error is [-1/2,+1/2].  The precision arithmetic also tracks the rounding error bounding range $R$ so that the precision representation becomes $S\tilt R\; 2^E$.  

If the initial $\sim$ is wrong, it will be corrected by the first round up when $S$ is odd, or R will be reduced to half after each round up when S is even.  Hence, the precision round up process is stable and self-correcting.

\subsection{Probability Distribution of Rounding Errors}

\begin{figure}[p]
\centering
\includegraphics[height=2.5in]{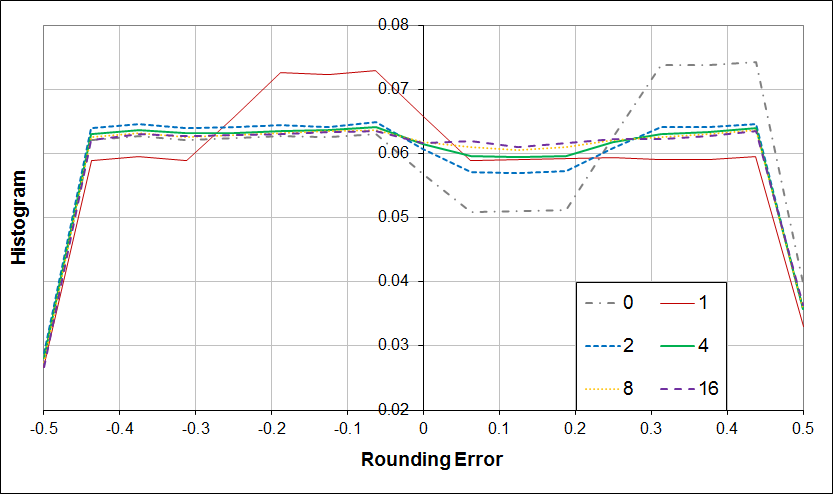} 
\captionof{figure}{Measured probability distribution of rounding errors of precision round-up rule for the minimal significand thresholds 0, 1, 2, 4, and 8 respectively.  Mathematically the probability is usually defined either in range (-1/2, +1/2] or in range [-1/2, +1/2), but not in range [-1/2, +1/2].  Because -1/2 and +1/2 in bounding range have different meaning in precision representation, the probability range is defined as [-1/2, +1/2], which introduces the artificially smaller count of histogram in sections containing either -1/2 or +1/2.  }
\label{fig: Prec_Rnd_Err_Dist}
\end{figure}

An ideal floating-point calculation is carried out conceptually to infinitesimal precision before it is rounded up to representation precision \cite{Floating_Point_Standard}\cite{Precise_Numerical_Methods}\cite{Stochastic_Arithmetic}.  Thus, rounding up should be a process independent of any calculation, and it should be evaluated separately.  To estimate the rounding error distribution within its bounding range [-1/2, +1/2], a large number\footnote{For each minimal significand threshold, 64K random integers are used.  The actual number of random integers is not important as far as 1) it gives a stable empirical histogram, and 2) the random integers are uniformly distributed without repeat in values.} of positive random integers are converted into precision values and then rounded up once at a step time until each of them has a significand smaller than a predefined minimal significand threshold.  The precision value at each step is compared with the original value for the rounding error.  Figure \ref{fig: Prec_Rnd_Err_Dist} shows the result histogram of rounding errors for the minimal significand thresholds 0, 1, 4 and 16, respectively.  When each significand bit has an equal chance to be either 0 or 1, the result distribution of the rounding errors is expected to be uniformly distributed within the range [-1/2, +1/2] \cite{Digit_Value_Distribution}.  However, the precision round up rule changes this equal chance for a few lowest digits of a significand.  So when the minimal significand threshold is smaller, the bias in rounding error distribution is larger, as shown in Figure \ref{fig: Prec_Rnd_Err_Dist}, and the result distribution is close to uniform only when the minimal significand threshold is 4 and above.

\subsection{Result Uncertainty For Addition and Subtraction}

In floating-point arithmetic, rounding errors are uncertainties \cite{Floating_Point_Standard}\cite{Precise_Numerical_Methods}\cite{Stochastic_Arithmetic}.  The precision round-up rule incorporates all randomness of an imprecise value into its carry and bounding range so that it preserves the uncorrelated uncertainty assumption between any two values.  The uncorrelated uncertainty assumption suggests that the result rounding error distribution of addition is the convolution of the two operand rounding error distributions, while the result rounding error distribution of subtraction is the convolution of the first operand rounding error distribution and the mirror image of the second operand rounding error distribution \cite{Probability_Statistics}.  Thus, when the exponents of two operands are equal, the results of addition and subtraction are:
\begin{equation}
\label{eqn: rounding error +-}
S_{1}\tilt _{1}R_{1}2^E \pm S_{2}\tilt _{2}R_{2}2^E = (S_{1} \pm S_{2})\tilt(R_{1}+R_{2})2^E;
\end{equation}
Table \ref{tab: uncertainty for addition} shows the result $\;\tilt\;$ for addition, while Table \ref{tab: uncertainty for subtraction} shows the result $\;\tilt\;$ for subtraction.  It will be shown that the $\;\tilt\;$ immediately after a calculation is actually not important because the precision round up rule frequently is applied after each calculation in precision arithmetic as its normalization process.  

\begin{table}[h]
\centering
\begin{tabular}{|c|c|p{0.5in}|p{0.5in}|c|c|c|}
\hline 
$\tilt_{1}$ vs. $\;\tilt_{2}$ & $\tilt_{1} = \;\; \tilt_{2}$ & \multicolumn{5}{c|}{$\tilt_{1} \neq \;\; \tilt_{2}$} \\ 
\hline 
 &  & $\tilt_{1}$ = ? & $\tilt_{2}$ = ? & $R_{1} > R_{2}$ & $R_{1} < R_{2}$ & $R_{1} = R_{2}$ \\ 
\hline 
$\tilt$ & $\tilt_{1}$ & $\tilt_{2}$ & $\tilt_{1}$  & $\tilt_{1}$ & $\tilt_{2}$ & ? \\ \hline 
\end{tabular}
\captionof{table}{Result $\:\tilt\:$ in $S_{1}\tilt_{1}R_{1}2^E + S_{2}\tilt_{2}R_{2}2^E = (S_{1} + S_{2})\tilt(R_{1} + R_{2})2^E$}
\label{tab: uncertainty for addition}
\end{table}

\begin{table}[h]
\centering
\begin{tabular}{|c|p{0.5in}|c|c|c|p{0.5in}|p{0.5in}|} 
\hline 
$\tilt_{1}$ vs. \;$\tilt_{2}$ & \multicolumn{4}{|c|}{$\tilt_{1} = \;\; \tilt_{2}$} & \multicolumn{2}{|c|}{$\tilt_{1} \neq \;\; \tilt_{2}$} \\ 
\hline 
 & $\tilt_{1} = \; ?$ & $R_{1} > R_{2}$ & $R_{1} < R_{2}$ & $R_{1} = R_{2}$ & $\tilt_{1} = \; ? $ & $\tilt_{1} \neq \; ?$ \\
\hline 
$\tilt$ & ? & $\tilt_{1}$ & $-\tilt_{2}$ & ? & $-\tilt_{2}$ & $\tilt_{1}$ \\
 \hline 
\end{tabular}
\captionof{table}{Result $\:\tilt\:$ in $S_{1}\tilt_{1}R_{1}2^E - S_{2}\tilt_{2}R_{2}2^E = (S_{1} - S_{2})\tilt(R_{1} + R_{2})2^E$.}
\label{tab: uncertainty for subtraction}
\end{table}

Let $P_{\frac{1}{2}}(x)$ be the rounding error distribution after rounding up, which is uniformly distributed between [-1/2, +1/2] according to Formula \eqref{eqn: rounding error 1/2 distribution}.  Let $P_{\frac{n}{2}}(x)$ be the convolution of $P_{\frac{1}{2}}(x)$ according to Formula \eqref{eqn:rounding error n/2 distribution}: 
\begin{align}
\label{eqn: rounding error 1/2 distribution}
& P_{\frac{1}{2}}(x) \equiv 1, \eqspace  -1/2 \leq x \leq +1/2;  \\
\label{eqn:rounding error n/2 distribution}
& P_{\frac{n}{2}}(x) \equiv \int _{-\infty}^{+\infty}P_{\frac{1}{2}}(y)P_{\frac{n-1}{2}}(x-y)dy=\int _{-1/2}^{+1/2}P_{\frac{n-1}{2}}(x-y) dy,\eqspace n=2,3,4\dots;
\end{align}
Formula \eqref{eqn:rounding error n/2 distribution} shows that $P_{\frac{n}{2}}(x)$ has a bounding range of $R \equiv \frac{n}{2}$, in which case it is easy to prove that the deviation $\sigma$ of $P_{\frac{n}{2}}(x)$ is determined by its bounding range $R$:
\begin{equation}
\label{eqn: rounding error range vs deviation}
\sigma^{2} = R/6;
\end{equation}
Also, the same bounding range can be reached in any combination:
\begin{equation}
\label{eqn: rounding error range convergence}
P_{\frac{m+n}{2}}(x)=\int _{-\infty }^{+\infty }P_{\frac{m}{2}}(y)P_{\frac{n}{2}}(x-y) dy;
\end{equation}

\begin{figure}[p]
\centering
\includegraphics[height=2.5in]{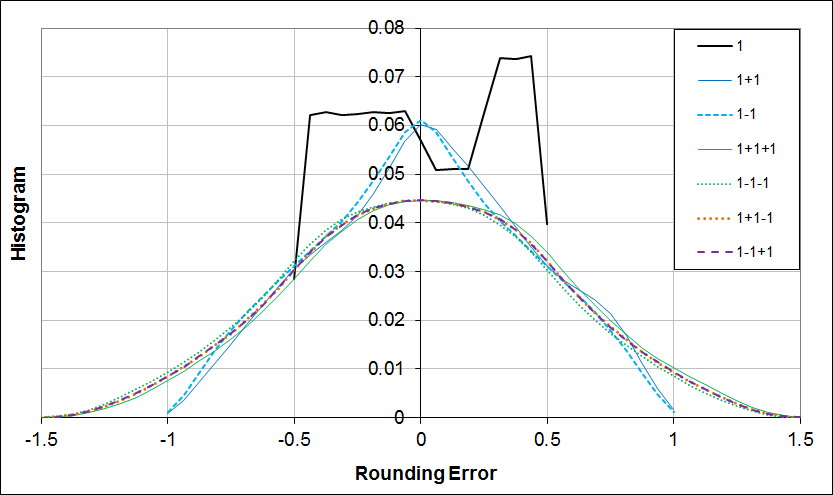} 
\captionof{figure}{Measured probability distribution of the rounding error after addition and subtraction.  In the legend, ``1'' for measured rounding error distribution for the minimal significand thresholds 0, ``1+1'' for addition once and ``1-1'' for subtraction once using the rounding error distribution of ``1'', while ``1+1+1'' for addition twice, ``1-1-1'' for subtraction twice, ``1+1-1'' for addition once then subtraction once, and ``1-1+1'' for subtraction once then addition once.}
\label{fig: Prec_Add_Err_Dist}
\end{figure}

In reality, $P_{\frac{1}{2}}(x)$ is not strictly uniformly distributed in its bounding range [-1/2, +1/2].  As the worst case, let $P_{1/2}(x)$ be the rounding error distribution with the minimal significand threshold of 0 in Figure \ref{fig: Prec_Rnd_Err_Dist}. Figure \ref{fig: Prec_Add_Err_Dist} shows the rounding error distribution after addition and subtraction, in which:
\begin{itemize}
\item  R=1/2:  ``1'' for no addition or subtraction.
\item  R=2/2:  ``1+1'' for addition once, and ``1-1'' for subtraction once.
\item  R=3/2:  ``1+1+1'' for addition twice, ``1-1-1'' for subtraction twice, ``1+1-1'' for addition once then subtraction once, and ``1-1+1'' for subtraction once then addition once.
\end{itemize}
Figure \ref{fig: Prec_Add_Err_Dist} shows that the rounding error distributions for the same bounding range largely repeat each other, confirming Formula \eqref{eqn: rounding error range convergence}.  Addition and subtraction have a slightly different result distribution due to uneven $P_{\frac{1}{2}}(x)$.  In all cases, the distributions quickly approach Gaussian with the increase of bounding ranges.  

Even for the worst-case $P_{\frac{1}{2}}(x)$, the deviation $\sigma$ relates to the bounding range $R$ empirically as $\sigma = 0.423005 R^{0.50000}$ with a reliable factor of $0.9999999$, confirming Formula \eqref{eqn: rounding error range vs deviation} empirically.  

The Lyapunov form of the central limit theorem \cite{Probability_Statistics} states that if $X_i$ is a random variable with mean $\mu_i$ and variance $\sigma^2_i$ for each $i$ among a series of $n$ mutually independent random variables, then with increased $n$, the sum $\sum\limits_{i}^{n} X_i$ converges in distribution to the Gaussian distribution with mean $\sum\limits_{i}^{n} \mu_i$ and variance $\sum\limits_{i}^{n} \sigma^2_i$. Applying the central limit theorem to precision arithmetic when a calculation ends with summation:
\begin{itemize}
\item $P_{n/2}(x)$ converges in distribution to a Gaussian distribution of mean $0$ and deviation $\sigma$ with an increased $n$
\item Figure \ref{fig: Prec_Add_Err_Dist} shows that such convergence to Gaussian distribution is very quick. 
\item The stable rounding error distribution is \emph{independent} of any initial rounding error distribution, so that we can extend the rounding error distribution to uncertainty distribution in general. 
\end{itemize} 

Because of rounding, $P_{\frac{n}{2}}(x)$ is extended to $P_{R}(x)$ for 2's fractional $R$, which is further extended to characterize uncertainty distribution in general.

\subsection{Uncertainty Distribution}

The probability density function $D_{y}(y)$ after linear transformation ($y = \alpha x + \beta)$  of a generic probability density function $D_{x}(x)$ is \cite{Probability_Statistics}:
\begin{equation}
\label{eqn: distribution of linear transformation}
D_{y}(y) = D_{x}((y- \beta)/\alpha )/\alpha;
\end{equation}
Letting $N(x)$ be the density function of a normal distribution, the density function of uncertainty distribution in precision arithmetic is thus:
\begin{equation}
\label{eqn: central limit theorem}
\rho(y) \equiv N(y/\sigma)/\sigma, \eqspace y\in[-R,+R];
\end{equation}

The deviation $\delta x$ and bounding range $\Delta x$ of $S\tilt R\; 2^E$ is:
\begin{align}
\label{eqn: uncertainty deviation}
& \delta x = \sigma \times 2^{E}, \eqspace \sigma < \sqrt{\hat{R}/6}; \\
\label{eqn: uncertainty range}
& \Delta x = R \times 2^{E}, \eqspace R < \hat{R}; \\
\label{eqn: uncertainty range vs deviation}
& \Delta x / \delta x = R/ \sigma  = \sqrt{6R};
\end{align}
Using $\delta x$ and $\Delta x$, Formula \eqref{eqn: central limit theorem} becomes the probability density function $\rho (\widetilde{y})$ in Formula \eqref{eqn: uncertainty distribution}:  
\begin{equation}
\label{eqn: uncertainty distribution}
\rho (\widetilde{y}) = N(\widetilde{y} / \delta x)/ \delta x, 
  \eqspace \widetilde{y} \in [-\Delta x, + \Delta x];
\end{equation}
And the precision representation $S\tilt R\; 2^E$ is interpreted as:. 
\begin{equation}
\label{eqn: uncertainty interpretation}
x \pm \delta x = x + \widetilde{y}, \eqspace \widetilde{y} \in \rho(\widetilde{y});
\end{equation}

\subsection{Uncertainty Rounding and Normalization}

\begin{figure}[p]
\centering
\includegraphics[height=2.5in]{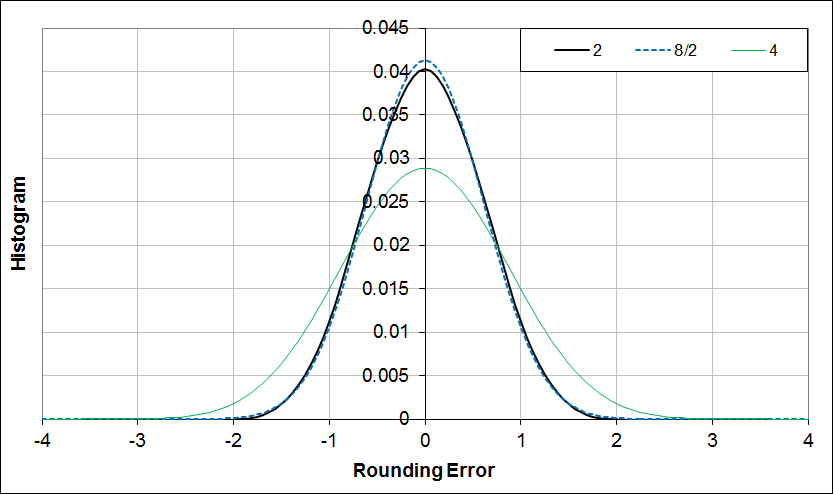} 
\captionof{figure}{The result rounding error distribution $R=8/2$ after the original error distribution $R=8$ is rounded up once.  The $R=8/2$ distribution is compared with the $R=4$ distribution and the $R=2$ distribution, which have the same bounding range and deviation, respectively.}
\label{fig: Prec_RndByDev_Dist}
\end{figure}

According to Formula \eqref{eqn: distribution of linear transformation}, when an imprecise value is rounded up once, its density function becomes $N(y/\frac{\sigma}{2})/\frac{\sigma}{2} ,y\in [-\frac{R}{2},+\frac{R}{2}]$, however the two ways to carry out rounding up can not reach this new distribution ideally:
\begin{itemize}
\item  By range:  When $R$ is reduced to 1/2-fold, $\sigma$ is reduced to $1/\sqrt{2}$-fold.  The new $\rho(y)$ becomes $N(y/\frac{\sigma}{\sqrt{2}} )/\frac{\sigma}{\sqrt{2}} ,y\in [-\frac{R}{2} ,+\frac{R}{2}]$.  This approach distorts the probability but retains strict bounding, which aligns with interval arithmetic.
\item  By deviation:  When $\sigma$ is reduced to 1/2-fold, $R$ is reduced to 1/4-fold.  The new $\rho(y)$ becomes $N(y/\frac{\sigma}{2} )/\frac{\sigma}{2} ,y\in [-\frac{R}{4} ,+\frac{R}{4}]$. This approach ignores the probability distribution on the Gaussian tails outside $[-\frac{R}{4} ,+\frac{R}{4}]$, but preserves the overall characteristics of the distribution.  
\end{itemize}
Figure \ref{fig: Prec_RndByDev_Dist} compares these two ways of rounding up when the original rounding error range is R=8, in which R=4 is rounded up by range, while R=2 is rounded up by deviation.  It clearly shows that rounding up by deviation results in a more similar rounding error distribution.  Rounding up by deviation is required by the scaling principle, so it is used universally in precision arithmetic.  Rounding up by deviation also introduces bounding leakage called \emph{round-up leakage}.  In Figure \ref{fig: Prec_RndByDev_Dist}, the 8/2 distribution of the rounding error outside the range [-2, +2] contributes to a round-up leakage of 0.05\%. 

Because the tail of the Gaussian distribution deceases by $e^{-6 R y^2}$, the round-up leakage decreases exponentially with the increased bounding range $R$.  Smaller round-up leakage also means that the actual rounding error distribution becomes more similar to the rounding error distribution with increased bounding range $R$. When $R$ is above a threshold $\hat{R}$, round-up leakage is small enough so that rounding up by deviation can be applied repeatedly.  This is the \emph{normalization} process in precision arithmetic. When $\hat{R}=16$, the maximal normalization leakage is $10^{-6}$, which is small enough for most applications, and which has a comparable bounding range as the de facto $6 \sigma -10^{-9}$ rule for negligible bounding leakages in statistics.  In addition to limit the bit count for both $R$ and $S$, normalization also enforce the correctness of carry sign $\;\tilt\;$ in precision representation $S\tilt R\; 2^E$.

Because the precision round up rule only looks at the sign of the current rounding error, rounding up by either deviation or range will not change the rounding error distribution. 

As an inverse operation to rounding up by deviation, a precision round-down rule is defined using the scaling principle.  After rounding down once, $S\tilt R\; 2^E$ becomes $(2S)\tilt (4R)2{E-1}$.  Round down reduces bounding leakage.  To add or subtract two operands with different exponents, the operand with a larger exponent is first rounded down to the other exponent, and the result of addition or subtraction using Formula \eqref{eqn: rounding error +-} is normalized afterwards.

\subsection{Uncertainty Initiation}

An integer $S$ is initialized as $S 2^0$.

A conventional 64-bit floating-point value $S 2^E$ is usually initialized as $S?\frac{1}{2}2^E$ because the IEEE floating-point standard \cite{Floating_Point_Standard} guarantees accuracy to half bit of a significand.  

A mean-deviation pair $(x \pm \delta x)$ of 64-bit conventional floating-point values is initialized as $S\tilt R\; 2^E$ by:
\begin{enumerate}
\item  rounding up $\delta x$ until Formula \eqref{eqn: uncertainty deviation} is satisfied;
\item  obtaining $R$ and $E$ from final $\delta x$;
\item  rounding up $x$ to $E$; and
\item  obtaining $S$ and $\;\tilt\;$ from final $x$.
\end{enumerate}

\subsection{Result Uncertainty For Multiplication}

After $S\tilt R\; 2^E$ is multiplied by $2$, both its range and deviation increase by 2-fold.  If the scaling principle is applied, the result is $2S\tilt 2^2R\; 2^E$.  When $S\tilt R\; 2^E$ is normalized, the result is then normalized as $S\tilt R\; 2^{E+1}$, and there is neither bound widening nor bounding leakage.  Generally, the direct result of multiplying $S_1\tilt_1 R_1 2^{E_1}$ by $S_2 2^{E_2}$ is:
\begin{equation}
\label{eqn: multiplication by a constant}
S_1\tilt_1 R_1 2^{E_1} \times S_2 2^{E_2} = (S_1 S_2) \tilt_1 ({S_2}^2 R_1) 2^{E_1 + E_2};
\end{equation}
Because Formula \eqref{eqn: multiplication by a constant} obeys scaling principle, the result uncertainty is still $\rho$-distributed.

According to uncorrelated uncertainty assumption, the product bounding range of multiplying $0\tilt_1 R_1 2^{E_1}$ by $0\tilt_2 R_2 2^{E_2}$ is $R_1 R_2 2^{E_1+E_2}$. Thus:
\begin{multline}
\label{eqn: multiplication uncertainty}
S_1\tilt_1 R_1 2^{E_1} \times S_2\tilt_2 R_2 2^{E_2} = (S_1 S_2) 2^{E_1+E_2} + 0(\tilt_1\tilt_2)(R_1 R_2) 2^{E_1+E_2}\\
 + 0\tilt_1 ({S_2}^2 R_1) 2^{E_1+E_2} + 0\tilt_2 ({S_1}^2 R_2) 2^{E_1+E_2};
\end{multline}
In Formula \eqref{eqn: multiplication uncertainty}, both $0\tilt_1 ({S_2}^2 R_1) 2^{E_1+E_2}$ and $0\tilt_2 ({S_1}^2 R_2)2^{E_1+E_2}$ are $\rho$-distributed.  The probability density function for $0(\tilt_1\tilt_2)(R_1 R_2)2^{E_1+E_2}$ is calculated as:
\begin{equation}
\label{eqn: multiplication distribution 1}
\rho_{mul}(x) = \frac{d}{dx} \int\limits _{x_1 \times x_2<x} \frac{1}{\sqrt{2 \pi}\delta x_1} e^{-\frac{x_1^2}{2 (\delta x_1)^2}}
     \frac{1}{\sqrt{2 \pi}\delta x_2} e^{-\frac{x_2^2}{2 (\delta x_2)^2}} d x_1 d x_2;
\end{equation}
Letting $\delta x \equiv (\delta x_1)(\delta x_2)$, $y_1 \equiv x_1 / (\delta x_1)$, $y_2 \equiv x_2 / (\delta x_2)$ and $y \equiv x / (\delta x)$, Formula \eqref{eqn: multiplication distribution 1} is simplified as:
\begin{equation}
\label{eqn: multiplication distribution 2}
\rho_{mul}(x) = \frac{1}{2 \pi \delta x} \frac{d}{dy} \int\limits _{y_1 \times y_2 < y} e^{-\frac{y_1^2 + y_2^2}{2}} d y_1 d y_2;
\end{equation}
Using polar coordinate $(r, \theta)$ instead Cartesian coordinate $(y_1, y_2)$, Formula \eqref{eqn: multiplication distribution 2} is simplified as:
\begin{multline}
\label{eqn: multiplication distribution 3}
\rho_{mul}(x) = \frac{1}{2 \pi \delta x} \frac{d}{dy} \int _{0}^{2\pi} d \theta
    \int _{0}^{\frac{y}{2\sin(\theta) \cos(\theta)}} e^{-\frac{r^2}{2}} d \frac{r^2}{2} \\
= \frac{1}{2 \pi \delta x} \frac{d}{dy} \int _{0}^{2\pi} d \theta
    \left( 1 - e^{-\frac{y}{\sin(2\theta)}} \right) 
= \frac{1}{2 \pi \delta x} \int _{0}^{\pi} \frac{e^{-\frac{|y|}{\sin(\theta)}}}{\sin(\theta)} d \theta;
\end{multline}
Similarly, letting $\delta x \equiv \sqrt{(\delta x_1)^2 + (\delta x_2)^2}$, 
$y_1 \equiv x_1 / (\delta x_1)$, $y_2 \equiv x_2 / (\delta x_2)$ and 
$y \equiv x / \delta x$, the distribution for $0\tilt R_1 2^E + 0\tilt R_2 2^E = 0\tilt (R_1+R_2) 2^E$ is calculated as:
\begin{multline}
\label{eqn: addition distribution 3}
\rho_{add}(x) = \frac{1}{2 \pi \delta x} \frac{d}{dy} \int _{0}^{2\pi} d \theta
    \int _{0}^{\frac{y}{\sin(\theta) + \cos(\theta)}} e^{-\frac{r^2}{2}} r d r \\
= \frac{1}{2 \pi \delta x} \frac{d}{dy} \int _{0}^{2\pi} d \theta
    \left( 1 - e^{-\frac{y^2}{2}\frac{1}{1 + \sin(2\theta)}} \right) 
= \frac{1}{2 \pi \delta x} \int _{0}^{2\pi} d \theta |y| \frac{e^{-\frac{y^2}{2}\frac{1}{1 + \sin(2\theta)}}}{1 + \sin(2\theta)};
\end{multline}
The result of Formula \eqref{eqn: addition distribution 3} is known as Formula \eqref{eqn: addition distribution 4}.  Because 
$2\sin(\theta)|_{\theta\in[0,\pi]}$ almost repeats $1+\sin(\theta)|_{\theta\in[-\frac{\pi}{2},\frac{3\pi}{2}]}$, the result of Formula \eqref{eqn: multiplication distribution 3} is estimated as Formula \eqref{eqn: multiplication distribution 4} and Formula \eqref{eqn: multiplication distribution 5}:
\begin{align}
\label{eqn: addition distribution 4}
y = \frac{x}{\sqrt{{\delta x_1}^2+{\delta x_2}^2}}: \eqspace & \rho_{add}(x) = \frac{1}{\sqrt{2 \pi}\sqrt{{\delta x_1}^2+{\delta x_2}^2}} e^{-\frac{y^2}{2}}; \\
\label{eqn: multiplication distribution 4}
y = \frac{x}{\delta x_1\delta x_2}: \eqspace & \rho_{mul}(x) = \frac{1}{\sqrt{2 \pi}(\delta x_1\delta x_2)} \frac{e^{-2|y|}}{\sqrt{|y|}}; \\
\label{eqn: multiplication distribution 5}
z \equiv 2 \sqrt{y} = 2 \sqrt{\frac{x}{\delta x_1\delta x_2}}: \eqspace & \rho_{mul}(x) = \frac{1}{\sqrt{2 \pi} \sqrt{\delta x_1\delta x_2}} e^{-\frac{z^2}{2}};
\end{align}
Formula \eqref{eqn: multiplication distribution 5} shows that $0(\tilt_1\tilt_2)R_1 R_2@(E_1\!+\!E_2)$ is $\rho$-distributed with deviation $\sqrt{(\delta x_1)(\delta x_2)}$ or range $R_1 R_2$, which adds to the rest two $\rho$-distributed terms of Formula \eqref{eqn: multiplication uncertainty}.  Thus, the result of multiplication is also $\rho$-distributed.

The result precision $P$ of multiplication is:
\begin{equation}
\label{eqn: multiplication precision}
P^2 = \frac{R_1 R_2 + {S_2}^2 R_1 + {S_1}^2 R_2}{(S_1 S_2)^2} 
= {P_1}^2 + {P_2}^2 + \frac{1}{6} {P_1}^2 {P_2}^2;
\end{equation}
Formula \eqref{eqn: multiplication precision} shows that precision cannot be improved during multiplication.  It is identical to Formula \eqref{eqn: stat *} except their cross term, representing difference in their statistical requirements.

\subsection{Result Uncertainty For Division}

The reverse of Formula \eqref{eqn: multiplication by a constant} defines:
\begin{equation}
\label{eqn: division by a constant}
\frac{S_1\tilt_1 R_1 2^{E_1}}{S_2 2^{E_2}} = \frac{S_1}{S_2} \tilt_1 \frac{R_1}{{S_2}^2} 2^{E_1-E_2}
\end{equation}
In Formula \eqref{eqn: division by a constant}, the rounding error decreases by $S_2$-fold, but the bounding range decreases by ${S_2}^2$-fold, so there is bounding leakage.  To limit the bounding leakage to acceptable level, the result is rounded down until normalized.  In this case, rounding down the direct result is equivalent to round down the dividend $S_1\tilt R_1 2^{E_1}$ by the same amount before the division.

\begin{figure}[p]
\centering
\includegraphics[height=2.5in]{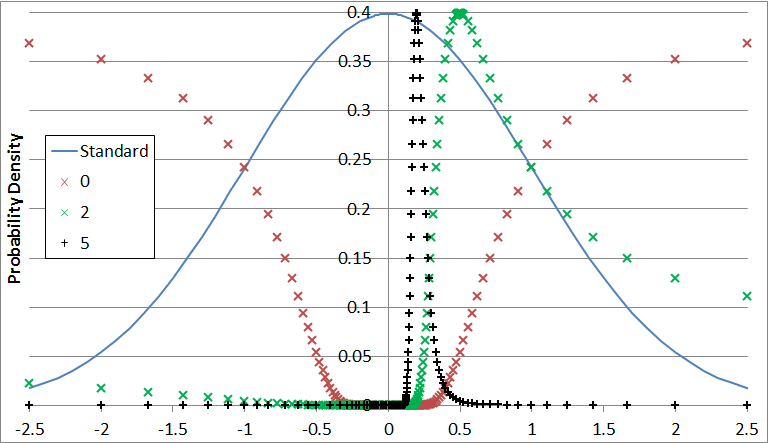} 
\captionof{figure}{The probability density function for $1/(x + \delta x)$ in which $\delta x = 1$ and $x=0, 2, 5$ respectively, as shown in the inlet. For comparison, the density function of a standard distribution is shown. }
\label{fig: Inversion_Distribution}
\end{figure}

Using the methodology defined in \cite{Probability_Statistics}, the probability density function for $1/(x+\delta x)$ is calculated as:
\begin{equation}
\label{eqn: inversion uncertainty distribution}
|\frac{1}{y} - x| < \Delta x: \eqspace \rho_{inv}(y) = \frac{1}{\sqrt{2 \pi}\delta x} \frac{e^{-\frac{(\frac{1}{y} - x)^2}{2(\delta x)^2}}}{y^2};
\end{equation}
Letting $z \equiv (1/y - x)/\delta x$, whose range is $\Delta x /\delta x\!=\!\sqrt{6 R}$ according to Formula \eqref{eqn: uncertainty range vs deviation}, the probability density function for $1/(x+\delta x)$ becomes:
\begin{equation}
\label{eqn: inversion uncertainty distribution 2}
|z| < \sqrt{6 R}: \eqspace \rho_{inv}(\frac{1}{z\delta x + x}) = \frac{1}{\sqrt{2 \pi}\delta x} e^\frac{-z^2}{2}
\end{equation}
Figure \ref{fig: Inversion_Distribution} shows the probability density function $p(x)$ for $1/(x + \delta x)$ in which $\delta x = 1$:
\begin{itemize}
\item when $x=0$: $p(\infty)=N(0)$, in which $N(x)$ is normal distribution, which is also displayed in the figure.  This distribution has no moment defined, thus it has no mean, deviation and etc.
\item when $x=2$: $p(\infty)=N(2)$.  This distribution has mode at $x=1/2$.
\item when $x=5$: $p(\infty)=N(5)$.  This distribution has mode at $x=1/5$. It has finer precision and it looks closer to Gaussian than the previous case.
\end{itemize}

The mean for $1/(x+\delta x)$ is calculated as: 
\begin{equation}
\label{eqn: inversion mean}
\mu(\frac{1}{x}) = \frac{1}{\sqrt{2 \pi}\delta x} \int_{|\frac{1}{y} - x| < \Delta x} y \rho_{inv}(y) dy 
= \frac{1}{x \sqrt{2 \pi}} \int_{-\frac{\Delta x}{\delta x}}^{+\frac{\Delta x}{\delta x}}
    \frac{e^{-\frac{z^2}{2}}}{z P(x) + 1} dz; \\
\end{equation}
Formula \eqref{eqn: inversion mean} diverges at $z = -1/P(x)$, which is excluded from the integration range when $\Delta x \ll |x|$.  In such a case, replacing $\Delta x /\delta x$ with $\bar{R} \equiv \sqrt{6 \hat{R}}$ for the integration range, the mean is calculated as:
\begin{align}
\label{eqn: uncertainty moment}
& M(j) \equiv \frac{1}{\sqrt{2 \pi}}  \int_{-\bar{R}}^{+\bar{R}} z^j e^{-\frac{z^2}{2}} dz; \\
\label{eqn: inversion mean 1}
& \mu(\frac{1}{x}) = \frac{1}{x} \sum _{j=0}^{+\infty} {P(x)}^{j} M(j);
\end{align}
$\hat{R}$ has to be small enough so that Formula \eqref{eqn: inversion mean 1} can converge for non-zero $P(x)$.  

\begin{figure}[p]
\centering
\includegraphics[height=2.5in]{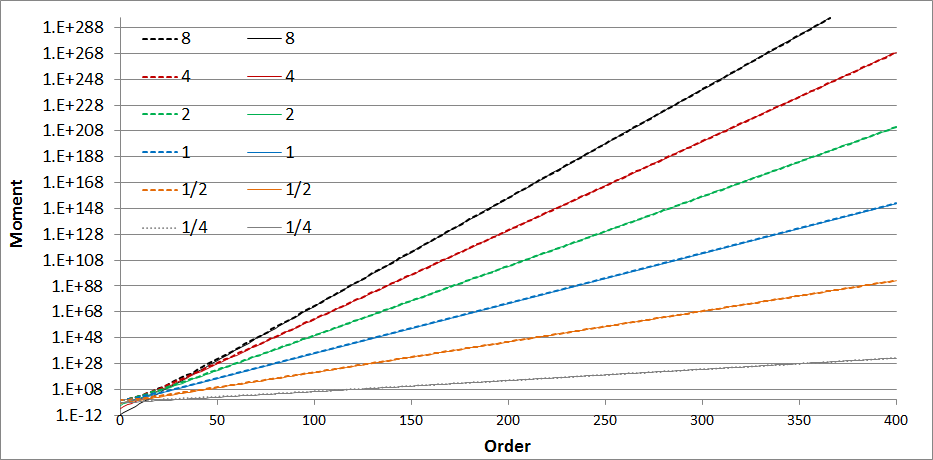} 
\captionof{figure}{The moments up to 200-order for rounding error distribution for different bounding range $\hat{R}$ of the legend. The moments for each bounding range $\hat{R}$ are drawn in one color using dashed line, and exponentially fitted, whose fitting lines are displayed in the same color using thin solid lines. }
\label{fig: Prec_Moments}
\end{figure}

The moment for uncertainty distribution is generally defined in Formula \eqref{eqn: uncertainty moment}, which is 0 when $j$ is odd. $M(2j)$ is calculated as: 
\begin{equation}
\begin{split}
\label{eqn: uncertainty moment 1}
M(2j) & = (2j-1)M(2j-2) - \frac{2}{\sqrt{2\pi}} \bar{R}^{2j - 1} e^{-\frac{\bar{R}^2}{2}} \\
& = (2j - 1)!! (M(0) - \frac{2}{\sqrt{2\pi}} e^{-\frac{\bar{R}^2}{2}} \sum_{k=1}^{j} \frac{\bar{R}^{2k - 1}}{(2k - 1)!!});
\end{split}
\end{equation}
Formula \eqref{eqn: uncertainty moment 2} \cite{Double_Factorial} shows that $M(2j)$ eventually becomes 0 for large enough $j$:
\begin{equation}
\begin{split}
\label{eqn: uncertainty moment 2}
& \lim_{j \to \infty} \sum_{k=1}^{j} \frac{\bar{R}^{2k - 1}}{(2k - 1)!!} = e^{\frac{\bar{R}^2}{2}} \int_{0}^{+\bar{R}} e^{-\frac{z^2}{2}} dz; \\
& \lim_{j \to \infty} M(2j) = (2j - 1)!!(M(0) - M(0)) = 0;
\end{split}
\end{equation}
Unfortunately, such convergence to 0 is very slow, and lower orders of $M(2j)$ actually increases exponentially according to Figure \ref{fig: Prec_Moments}. Empirically:
\begin{equation}
\label{eqn: uncertainty moment empirical}
j<400: \eqspace M(2j) = (2.4429 \hat{R}^{0.4998})^{2j};
\end{equation}
Specifically, $\hat{R}=4: M(2j) \sim 4.8844^{2j}$.  When $P(x) < 1/4.8844 = 0.20473 $, $1/(x+\delta x)$ converges so that inversion can be defined statistically: 
\begin{equation}
\label{eqn: inversion mean 2}
\mu(\frac{1}{x}) = \frac{1}{x} (1 + {P(x)}^2 + 3 {P(x)}^4 + 15 {P(x)}^6 + ...); 
\end{equation}

However, Formula \eqref{eqn: inversion mean 2} does not obey the recovering principle, so that a better approach is to let the value of inversion to be $1/x$ and the inversion deviation defined around $1/x$, as:
\begin{equation}
\begin{split}
\label{eqn: inversion deviation}
(\delta \frac{1}{x})^2 & \equiv \frac{1}{\sqrt{2 \pi}\delta x} \int_{|\frac{1}{y} - x| < \Delta x} (y - \frac{1}{x})^2 \rho_{inv}(y) dy \\
& = \frac{1}{x^2 \sqrt{2 \pi}} \int_{-\bar{R}}^{+\bar{R}} e^{-\frac{z^2}{2}} (\frac{1}{z P(x) + 1} - 1)^2 dz; \\
P(\frac{1}{x})^2 & = \sum _{j=1}^{+\infty} {P(x)}^{2j} M(2j) (2j - 1)
= {P(x)}^2 + 9 {P(x)}^4 + 75 {P(x)}^6 + ...;
\end{split}
\end{equation}

An equivalent way to calculate Formula \eqref{eqn: inversion deviation} is to use Taylor expansion around $1/x$ in $\rho(y)$ space instead of in $\rho_{inv}(y)$ space: 
\begin{multline}
\label{eqn: inversion deviation 1}
(\delta\frac{1}{x})^2 \equiv \frac{1}{\sqrt{2 \pi}\delta x} \int_{|\frac{1}{y} - x| < \Delta x} (y - \frac{1}{x})^2 \rho(y)_{inv} dy \\
= \frac{1}{\sqrt{2 \pi}\delta x} \int_{|\frac{1}{y} - x| < \Delta x} (\frac{1}{y} - \frac{1}{x})^2 \rho(y) dy 
= \frac{1}{x^2 \sqrt{2 \pi}} \int_{-\bar{R}}^{+\bar{R}} e^{-\frac{z^2}{2}} (\frac{1}{z P(x) + 1} - 1)^2 dz;
\end{multline}

Formula \eqref{eqn: inversion deviation} obeys the recovering principle for uncertainty only approximately when $P(x)^2 \ll 1$.  

The sign $~$ of the representation $S\tilt R\;2^E$ becomes negation of itself after inversion.

Dividing an operand by itself results in precise 1.

\subsection{Result Uncertainty For Square and Square Root}

A special case of multiplication is between an operand and itself.  If $x$ is $N(x)$ distributed, $x^2$ is $\chi^2$-distributed with freedom 1 \cite{Probability_Statistics}, which has mean 1 and variance of 2.  $N(x)$ and $\chi^2(x)$ have quite different characteristics, e.g., $\chi^2(x)$ only roughly resembles half of $N(x)$.  The bounding goal of precision arithmetic extends $\chi^2$ distribution to the other side of the mathematically expected value, and absorb the square of mean of the $\chi^2$ distribution into final variance:
\begin{equation}
\label{eqn: square uncertainty}
(S_1\tilt R_1 2^{E_1})^2 \equiv {S_1}^2 2^{2 E_1} + 0\tilt 4 R_1 {S_1}^2 2^{2 E_1} + 0\tilt 3 {R_1}^2 2^{2 E_1};
\end{equation}
The result precision of square is:
\begin{equation}
\label{eqn: square precision}
P^2 = \frac{4 R_1 {S_1}^2 + 3 {R_1}^2}{{S_1}^4} = 4 {P_1}^2 + 3 {P_1}^4;
\end{equation}

Similar to Formula \eqref{eqn: inversion deviation 1}, with $z=(y - x)/\delta x$ the deviation for $x^2$ is calculated as Formula \eqref{eqn: square deviation 1}, which confirms Formula \eqref{eqn: square precision}:
\begin{equation}
\begin{split}
\label{eqn: square deviation 1}
(\delta x^2)^2
& = \frac{1}{\sqrt{2 \pi}\delta x} \int_{|y^2 - x| < \Delta x} (y^2 - x^2)^2 \rho(y) dy \\
& = \frac{x^4}{\sqrt{2 \pi}} \int_{-\bar{R}}^{+\bar{R}} e^{-\frac{z^2}{2}} ((1 + z P(x))^2 - 1)^2 dz; \\
P(x^2)^2 & = 4 M(2) P(x)^2 + 4 M(3) P(x)^3 + M(4) P(x)^4 = 4 P(x)^2 + 3 P(x)^4;
\end{split}
\end{equation}
And the deviation for $\sqrt{x}$ is calculated as Formula \eqref{eqn: sqrt deviation 1}.
\begin{equation}
\begin{split}
\label{eqn: sqrt deviation 1}
(\delta \sqrt{x})^2
& = \frac{1}{\sqrt{2 \pi}\delta x} \int_{|\sqrt{y} - x| < \Delta x} (\sqrt{y} - \sqrt{x})^2 \rho(y) dy \\
& = \frac{|x|}{\sqrt{2 \pi}} \int_{-\bar{R}}^{+\bar{R}} e^{-\frac{z^2}{2}} (\sqrt{1 + z P(x)} - 1)^2 dz \\
& = \frac{|x|}{\sqrt{2 \pi}} \int_{-\bar{R}}^{+\bar{R}} e^{-\frac{z^2}{2}} (\frac{z P(x)}{2} - \frac{z^2 P(x)^2}{8} + \frac{z^3 P(x)^3}{16} + ...)^2 dz; \\
P(\sqrt{x})^2 & = \frac{1}{4} P(x)^2 + \frac{15}{64} P(x)^4 + ...;
\end{split}
\end{equation}
The above Taylor-expansion method can be extended to calculate power $x^n$ and root $\sqrt[n]{x}$ for any integer $n$.

The combination of Formula \eqref{eqn: square deviation 1} and Formula \eqref{eqn: sqrt deviation 1} as Formula \eqref{eqn: sqrt deviation 2} shows again that the uncertainty deviation obeys the recovery principle only approximately when $P(x)^2 \ll 1$.  Furthermore, the result of $\sqrt{x^2}$ and $(\sqrt{x})^2$ are different in Formula \eqref{eqn: sqrt deviation 2}, showing dependency problem.  The reason why the uncertainty deviation can not obey the recovery principle strictly is not clear at this moment. 
\begin{equation}
\begin{split}
\label{eqn: sqrt deviation 2}
P(\sqrt{x^2})^2 & = \frac{1}{4} (4 P(x)^2 + 3 P(x)^4) + \frac{15}{64} (4 P(x)^2 + 3 P(x)^4) + ... \\
& = P(x)^2 + \frac{9}{2} P(x)^4 + ...; \\
P((\sqrt{x})^2)^2 & = 4(\frac{1}{4} P(x)^2 + \frac{15}{64} P(x)^4 + ...)
    + 3 (\frac{1}{4} P(x)^2 + \frac{15}{64} P(x)^4 + ...)^2; \\
& = P(x)^2 + \frac{9}{8} P(x)^4 + ...;
\end{split}
\end{equation}

\subsection{Result Uncertainty For Function Evaluation}

Extending Formula \eqref{eqn: inversion deviation 1}, the uncertainty of the function $f(x)$ at $(x \pm \delta x)$ is evaluated by the set $\{f(x+\widetilde{y})-f(x)\}$, as Formula \eqref{eqn: uncertainty variance}, in which $\widetilde{y}$ and $\rho(\widetilde{y})$ are defined by Formula \eqref{eqn: uncertainty distribution}.  
\begin{equation}
\label{eqn: uncertainty variance}
(\delta f)^{2} \equiv \int (f(x+ \widetilde{y})-f(x))^{2} \rho ( \widetilde{y}) \: d \widetilde{y};
\end{equation}
For example, Formula \eqref{eqn: polynomial uncertainty} provides uncertainty for generic polynomial to $J$-th order.  
\begin{equation}
\begin{split}
\label{eqn: polynomial uncertainty}
& C_{j}^{k} \equiv \frac{j!}{k!(j - k)!}; \\
(\delta \sum_{j=0}^{J} a_j x^j)^2 &= \int (\sum_{j=0}^{J} a_j (x + \widetilde{y})^j - \sum_{j=0}^{J} a_j x^j)^2 \rho(\widetilde{y}) d \widetilde{y} \\
&= \int (\sum_{j=1}^{J} a_j \sum_{k=1}^{j} C_{j}^{k} \widetilde{y}^k x^{j-k})^2 \rho(\widetilde{y}) d \widetilde{y} \\
&= \sum_{j_1=1}^{J} \sum_{j_2=1}^{J} a_{j_1} a_{j_2} \sum_{k_1=1}^{j_1} \sum_{k_2=1}^{j_2} 
    C_{j_1}^{k_1} C_{j_2}^{k_2} M(k_1 + k_2) (\delta x)^{k_1 + k_2} x^{j_1 + j_2 - k_1 - k_2};
\end{split}
\end{equation}

If the function $f(x)$ is Taylor expandable at $x$, $f(x+\widetilde{y})-f(x)$ is calculated according to Formula \eqref{eqn: Taylor 1d}, in which $f^{(n)}$ denotes the $n$th derivatives of $f(x)$ at $x$.  $(\delta f)^{2}$ is given by Formula \eqref{eqn: uncertainty 1d}, in which $M(j)$ is defined by Formula \eqref{eqn: uncertainty moment} and \eqref{eqn: uncertainty moment}.
\begin{align}
\label{eqn: Taylor 1d}
& f(x + \widetilde{y})-f(x) = \sum _{n=1}^{\infty}\frac{f^{(n)}(x)}{n!}  \widetilde{y}^{n}; \\
\label{eqn: uncertainty 1d}
& (\delta f)^{2} =\sum _{n=0}^{\infty}\sum _{j=0}^{\infty}\frac{f^{(n)}(x)}{n!} \frac{f^{(j)} (x)}{j!} (\delta x)^{n+j} M(n+j) - {f(x)}^2;
\end{align}

Let $P(f(x)) \equiv \delta f(x)/|f(x)|$ be defined as the precision for $f(x)$; and let $\alpha$ be a constant.  According to Formula \eqref{eqn: uncertainty 1d}:
\begin{align}
\label{eqn: scaling}
\delta(\alpha x) = \alpha (\delta x); \eqspace & P(\alpha x) = P(x); \\
\label{eqn: inverse}
(\delta\frac{1}{x})^2 = \frac{1}{x^2} \sum _{j=1}^{\infty} (2j-1) M(2j) {P(x)}^{2j};
   \eqspace & P(\frac{1}{x})^2 = \sum _{j=1}^{\infty} (2j-1) M(2j) {P(x)}^{2j}; \\
\label{eqn: square}
(\delta x^2)^2 = 4 x^2(\delta x)^2 + 3 (\delta x)^4; \eqspace & {P(x^2)}^2 = 4 P(x)^{2} + 3 P(x)^{4};
\end{align}
Formula \eqref{eqn: scaling} confirm the scaling principle, while Formula \eqref{eqn: square} and \eqref{eqn: inverse} confirm Formula \eqref{eqn: square precision} and \eqref{eqn: inversion deviation}, respectively.

Due to uncorrelated uncertainty assumption, Taylor expansion can also be used to find the result $(\delta f)^{2}$ of the function $f(x_{1}, x_{2})$, in which $f^{(m,n)}$ denotes the $m$th and $n$th partial derivatives of $x_{1}$ and $x_{2}$, respectively; and the uncorrelated uncertainty assumption between $x_{1}$ and $x_{2}$ leads to independence between the random variables $\widetilde{y}_{1}$ and $\widetilde{y}_{2}$ in Formula \eqref{eqn: uncertainty 2d}:
\begin{align}
\label{eqn: Taylor 2d}
& f(x_{1} +  \widetilde{y}_{1} ,x_{2} +  \widetilde{y}_{2}) - f(x_{1} ,x_{2}) = 
 \sum _{m=0}^{\infty} \sum _{n=0}^{\infty} \frac{f^{(m,n)}}{m!n!}  \widetilde{y}_{1}^{m}  \widetilde{y}_{2}^{n} - f(x_{1} ,x_{2}); \\
\label{eqn: uncertainty 2d}
& (\delta f)^{2} = \sum _{m=0}^{\infty} \sum _{n=0}^{\infty} \sum _{i=0}^{\infty} \sum _{j=0}^{\infty} \frac{f^{(m,n)}}{m!n!}   \frac{f^{(i,j)}}{i! j!} (\delta x_{1})^{m+i} (\delta x_{2})^{n+j} M(m+i) M(n+j) - {f(x_{1} ,x_{2})}^2;
\end{align}
Such an approach can be extended to a function of an arbitrary number of input variables.  Formula \eqref{eqn: uncertainty 2d} shows that an input contributes to the result uncertainty in more than one way, in the same way as $\Delta x$ appears in more than one term in the Taylor expansion of $\Delta f(x,y,z)$.  

According to Formula \eqref{eqn: uncertainty 2d}:
\begin{align}
\label{eqn: uncertainty +-}
& \delta (x_1 \pm x_2)^2 = (\delta x_1)^2 + (\delta x_2)^2; \\
\label{eqn: uncertainty *}
& \delta (x_1 \times x_2)^2 = (\delta x_1)^2 {x_2}^2 + {x_1}^2 (\delta x_2)^2 + (\delta x_1)^2 (\delta x_2)^2; \\
\label{eqn: uncertainty * precision}
& P(x_1 \times x_2)^2 = P(x_1)^2 + P(x_2)^2 + \frac{1}{6} P(x_1)^2 P(x_2)^2; \\
\label{eqn: uncertainty /}
& \delta (\frac{x_1}{x_2})^2 = \frac{(\delta x_1)^2}{{x_2}^2}  + {x_1}^2 (\delta \frac{1}{x_2})^2 + (\delta x_1)^2 (\delta \frac{1}{x_2})^2
\simeq \frac{\delta (x_1 \times x_2)^2}{{x_2}^4}; 
\end{align}
Formula \eqref{eqn: uncertainty +-} and \eqref{eqn: uncertainty * precision} confirm Formula \eqref{eqn: rounding error +-} and \eqref{eqn: multiplication precision}, respectively.

If $f(x)$ is a black-box function, because the normal distribution $N(x)$ is well known, standard methods exist to divide the range $[x - \Delta x, x + \Delta x]$ into equal-probability quantiles \cite{Probability_Statistics}, and $\delta f$ can be found numerically by sampling.  For example, a $\kappa$-point monotonic sampling requires that 1) each numerically monotonic region contains at least $\kappa$ consecutive sampling points; and 2) the whole range $[x - \Delta x, x + \Delta x]$ has been divided into monotonic regions only.  When $\kappa$ is sufficiently large, the chance of missing a peak or a valley is small enough so that the sampling is fair enough.  The range $[x - \Delta x, x + \Delta x]$ is first divided into $\kappa$ equal-probability quantiles, and at each additional step, each quantile is further divided into an equal number of sub quantiles until both sampling requirements are met.  Then $\delta f^{2}$ is calculated as the sum of 1) the sample variance and 2) the square of the sample mean.

\subsection{Dependence Problem}

Formula \eqref{eqn: uncertainty 1d} and its multi-dimension extensions such as Formula \eqref{eqn: uncertainty 2d} accurately account for all contribution to the result uncertainty within $f(x)$, providing a clean and deterministic solution for $(\delta f)^{2}$, e.g., it gives the same result for Formula \eqref{eqn: interval depend 1}, Formula \eqref{eqn: interval depend 2} and Formula \eqref{eqn: interval depend 3}.  Therefore, precision arithmetic has no expression-based dependence problem.

It is tempting to define basic arithmetic operations as Formula \eqref{eqn: uncertainty +-}, Formula \eqref{eqn: uncertainty *} and Formula \eqref{eqn: uncertainty /}, and apply them progressively to calculate $f(x)$, similar to how basic arithmetic operations are used in conventional floating arithmetic.  However, such an approach may apply the uncorrelated uncertainty assumption wrongly between a value and its mathematical expression, such as between $x$ and $x^{2}$, so that it may result in the dependence problem similar to that of interval arithmetic \cite{Interval_Arithmetic}.  For example, for such a use of precision arithmetic, only Formula \eqref{eqn: interval depend 1} gives the correct result $4 (x-\frac{1}{2})^2 (\delta x)^2 + 3 (\delta x)^4$, while Formula \eqref{eqn: interval depend 2} over-estimates the result uncertainty by $4 x (\delta x)^2$, which has the largest fold of over-estimation at $x=\frac{1}{2}$ when $\delta x < 1$.  Let $x$, $y$ and $z$ be three values satisfying the uncorrelated uncertainty assumption. Functions $f(x,y)$ and $g(x,z)$ are correlated through $x$, and they need to be tested for the uncorrelated uncertainty assumption before they can be used to calculate $h(f,g)$ using precision arithmetic.  For example, using precision arithmetic, the correlation $\gamma$ between $(x \pm \delta x)$ and $(x \pm \delta x)^2$ is calculated by Formula \eqref{eqn: correlation between x and x^2}, which shows that $\gamma$ increases with decreased precision $P(x) \equiv \delta x/|x|$ of $x$, in contrast with how the uncorrelated uncertainty assumption favors finer precision $P$ in Formula \eqref{eqn: uncertainty correlation}.  After applying Formula \eqref{eqn: uncertainty correlation}, the correlation on the uncertainty level $\gamma_P$ is no less than $\frac{16}{19}$, so that precision arithmetic rejects calculating Formula \eqref{eqn: interval depend 2} progressively using the basic arithmetic operations.

\begin{equation}
\label{eqn: correlation between x and x^2}
\gamma = \frac{\int \left((x + \widetilde{y})^2 - x^2 \right) \left((x + \widetilde{y}) - x \right) \rho (\widetilde{y}) \: d \widetilde{y}}{\sqrt{\int \left((x + \widetilde{y})^2 - x^2\right)^2 \rho (\widetilde{y}) \: d \widetilde{y}} \sqrt{\int \left((x + \widetilde{y}) - x)^2 \rho (\widetilde{y}\right) \: d \widetilde{y}}} = 
\frac{1}{\sqrt{1 + \frac{3}{4} P(x)^2}}
\end{equation}

In other words, converting a numerical algorithm from using conventional floating-point arithmetic to using precision arithmetic may be more complicated than directly replacing the variable types and the arithmetic being used.  To avoid the dependence problem, the safest approach is to obtain an analytic form of the final expression of an algorithm before applying Formula \eqref{eqn: uncertainty 1d} and its multi-dimension extensions, similar to how symbolic calculations are currently used in affine arithmetic \cite{Symbolic_Affine_Arithmetic}.

\subsection{Conditional Execution}

Conditional execution based on the comparison relation between two values is frequently used in practical algorithms \cite{Numerical_Recipes}.  When each value has associated uncertainty, the comparison relation between two values becomes quite different.  This is particularly true for interval arithmetic, in which a value can be anywhere inside its bounding range \cite{Interval_Analysis}.  In precision arithmetic, each value has a mathematically expected value plus a well-defined bounding distribution for uncertainty.  The comparison relation between two imprecise values in precision representation can be defined either by their mathematically expected values, or by their statistical comparison relations based on confidence \cite{Probability_Statistics}.  

However, the usage of condition execution in a traditional algorithm needs to be re-evaluated conceptually with uncertainty statistics in mind when upgrading an algorithm to use precision arithmetic, because most conditional executions are created to optimize implementation.  For example, LU decomposition \cite{Numerical_Recipes} carefully chooses the sequence of execution to minimize rounding errors, so that it introduces additional dependence problem due to conditional execution, e.g., small value change of a matrix item can result in different conditional execution path and large result difference.  In other words, to solve the linear equation $A x = b$, in which $A$ is a matrix, $x$ and $b$ are two vectors, with the uncertainty of $A^{-1}$ analytically solvable using Taylor expansion (as demonstrated in Section \ref{sec: matrix}), precision arithmetic prefers to solve it as $x = A^{-1} b$ than to use the LU decomposition method.

\subsection{Calculation Inside Uncertainty}

Formula \eqref{eqn: addition with 0-bit inside uncertainty} shows that the current choice of $\hat{R}$ calculates 0-bit inside uncertainty, e.g., 1 with precision ${10}^{-3}$ is represented as $1024$ in $S$.  In contrast, all other arithmetic represents the value as $2^{53}/2^{53}$.  While calculating many bits inside uncertainty does not seem meaningful according to significance arithmetic \cite{Significance_Arithmetic}, not calculating at all inside uncertainty may not be an optimal approach either.  Thus, Formula \eqref{eqn: uncertainty deviation} is modified as Formula \eqref{eqn: uncertainty bits inside}, in which $\chi$ is a small constant positive integer, to introduce the $\chi$-bit calculation inside uncertainty by providing an altered interpretation of the precision for $S\tilt R\; 2^E$. 
\begin{equation}
\begin{split}
\label{eqn: addition with 0-bit inside uncertainty}
1/3\pm 0.001 + 2/3\pm 0.001 & = 341+6.29\; 2^{-10} + 683-6.29\; 2^{-10} \\ & = 1024?12.6\; 2^{-10} = 1\pm 0.001\sqrt{2};
\end{split}
\end{equation}

\begin{equation}
\label{eqn: uncertainty bits inside}
\delta x = \sqrt{R / 6 \cdot 2^{-\chi}} 2^{E + \chi};
\end{equation}
Table \ref{tab: precision arithmetic with different bit calculated inside} shows examples of precision arithmetic with different $\chi$ for $\hat{R}=16$, e.g., with $\chi = 2$, precision arithmetic represents the expected value of $1.000 \pm 0.001$ as $2^{12}/2^{12}$.  $\chi$ will be set to 4 empirically later in this paper.  The value of $\delta x/2^\chi$ is defined as \emph{resolution} for the corresponding precision arithmetic.
\begin{table}[h]
\centering
\begin{tabular}{|p{1.25in}|c|c|c|} 
\hline 
$\chi$ & 0 & 1 & 2 \\ 
\hline 
$0.5\pm0.001$ = & $512?6.29\; 2^{-10}$ & $1024?12.58\; 2^{-11}$ & $2048?25.16\; 2^{-12}$ \\ 
\hline 
$1\pm0.001$ = & $1024?6.29\; 2^{-10}$ & $2048?12.58\; 2^{-11}$ & $4096?25.16\; 2^{-12}$ \\ 
\hline 
$1\pm0.002$ = & $512?6.29\; 2^{-9}$ & $1024?12.58\; 2^{-10}$ & $2048?25.16\; 2^{-10}$ \\ 
\hline 
\end{tabular}
\captionof{table}{Examples of precision arithmetic with different $\chi$ for $\hat{R}=16$, in which $\chi$ stands for bits calculated inside uncertainty.}
\label{tab: precision arithmetic with different bit calculated inside}
\end{table}

The limited calculation inside uncertainty does not necessarily mean that precision arithmetic has a larger calculation error.  In Formula {eqn: addition with 0-bit inside uncertainty}, the mathematically expected value for the result is precisely 1, even though the mathematically expected values of the two operands for addition are not precisely 1/3 and 2/3 after the uncertainty initiation, respectively.

\subsection{Implementation}

The conventional 64-bit floating-point standard IEEE-754 \cite{Floating_Point_Arithmetic}\cite{Floating_Point_Standard} has:
\begin{itemize}
\item  11 bits for storing exponent $E$;
\item  53 bits for storing significand $S$ (with a hidden MSB).
\item  1 bit for storing sign;
\end{itemize}

To be a super set of the conventional 64-bit floating-point standard, an 80-bit implementation of precision arithmetic has:
\begin{itemize}
\item  11 bits for storing exponent $E$;
\item  53 bits for storing significand $S$ (without using the hidden MSB);
\item  1 bit for storing sign;
\item  2 bits for storing carry $\sim$;
\item  13 bits to store the bounding range $R$ as a fixed-point value.
\end{itemize}

Precision arithmetic is implemented in C++.  With heavy additional codes to count for statistics and to detect implementation errors, it runs about seven times slower than the implementation of interval arithmetic using Formula \eqref{eqn: interval +}, \eqref{eqn: interval -}, \eqref{eqn: interval *} and \eqref{eqn: interval /}.  It is probably faster in speed than the implementation of interval arithmetic without the dependence problem \cite{Interval_Arithmetic}.  With code weight trimming and optimization, its speed is expected to be improved at least 3-fold.  Unlike conventional floating-point arithmetic, it only calculates a limited number of significand bits, e.g., 12 bits instead of all the 53 bits when the precision is $10^{-3}$ and $\chi$ is 2.  Its slowest but very frequent operation is to find the position of the highest non-zero significand digit, which can be found instantly with a decoder \cite{Electronics}.  Thus, future hardware optimization can also improve the speed of precision arithmetic by another estimated 10-fold.

\subsection{Alternative Form of Precision Arithmetic}

Because the uncorrelated uncertainty assumption can lead directly to 1) the Gaussian distribution as the underlying distribution for rounding errors, and 2) Formula \eqref{eqn: uncertainty 1d} and its multi-dimension extensions such as Formula \eqref{eqn: uncertainty 2d} for generic Taylor expansion, an \emph{alternative form of precision arithmetic} is to represent each uncertainty-bearing value as $x \pm \delta x$ in Formula \eqref{eqn: uncertainty interpretation}.  The bounding range is then calculated from $\delta x$ as the confidence interval \cite{Probability_Statistics} for any required upper limit on bounding leakage, e.g., if the required bounding leakage is $10^{-9}$ or less, the bounding interval is $[x - 6\delta x, x + 6\delta x]$.  This alternative form of precision arithmetic is \emph{not} adopted in this paper for the following reasons:
\begin{itemize}
\item For the actual numerical calculation, if conventional floating-point arithmetic is used separately for $x$ and $\delta x$, then $x$ and $\delta x$ will be contaminated by unspecified amount of rounding errors.  Because the calculation for $\delta x$ is more complex than that for $x$,  $\delta x$ probably contains more rounding error than $x$.  Thus, the current form of precision arithmetic defines its own floating-point representation for $x \pm \delta x$ as $S\tilt R\; 2^E$.
\item Another effect of using conventional floating-point arithmetic for $x$ and $\delta x$ is to calculate many bits inside uncertainty, whose validity is not clear at this moment.  In contrast, as demonstrated by Table \ref{tab: precision arithmetic with different bit calculated inside}, the current form of precision arithmetic controls the number of bits calculated inside uncertainty.
\end{itemize}
However, the alternative form could be valuable in theoretical discussions of precision arithmetic.

\subsection{Types of Uncertainties Included in Precision Arithmetic}

There are four sources of result uncertainty after a calculation \cite{Statistical_Methods}\cite{Numerical_Recipes}:
\begin{itemize}
\item input uncertainties
\item rounding errors
\item truncation errors
\item modelling errors
\end{itemize}

As described previously, both input uncertainties and rounding errors are included in the uncertainty specification of precision arithmetic.

In many cases, because a numerical algorithm approaches its analytic counterpart only after infinitive execution, a good numerical algorithm should have an estimator of the \emph{truncation error} toward its analytic counterpart, such as the Cauchy reminder estimator for Taylor expansion \cite{Numerical_Recipes}, or the residual error for numerical integration \cite{Numerical_Recipes}.  Using conventional floating-point arithmetic, a subjective upper limit is chosen for the truncation error, to stop the numerical algorithm at limited execution \cite{Numerical_Recipes}. However, such arbitrary upper limit may not be achievable with the amount of rounding errors accumulated during calculation, so that such upper limit may actually give a falsely small result precision. Because precision arithmetic tracks rounding errors of a calculation efficiently, it can be used to search for the optimal execution termination point for the numerical algorithm when the truncation error is no longer significant, which is named as the \emph{truncation rule} in this paper. In other words, using precision arithmetic, the result precision of a calculation is determined by the inputs and the calculation itself.  Section \ref{sec: taylor expansion} and \ref{sec: integration} will provide such cases of applying truncation rule to Taylor expansion and numerical integration, respectively.

Modelling errors arise when an approximate analytic solution is used, or when a real problem is simplified to obtain the solution.  For example, Section \ref{sec: FFT} demonstrates that the discrete Fourier transformation is only an approximation for the mathematically defined Fourier transformation.  Conceptually, modelling errors originate from mathematics, so they are outside the domain for precision arithmetic.

\clearpage
\section{Standards and Methods for Comparing Uncertainty-Bearing Arithmetic}
\label{sec: validation}

\subsection{Comparing Standards and Methods}

Algorithms each with a known analytic result are used to characterize uncertainty-bearing arithmetic.  The difference between the arithmetic result and the analytic result is defined as the \emph{value error}.  The question is whether the uncertainty bounding range or the uncertainty deviation is enough to cover the value error with an increased amount of calculation for any input.  Corresponding to two different goals for uncertainty-bearing, there are actually two different sets of measurements to characterize an uncertainty-bearing arithmetic:

\begin{itemize}
\item  The ratio of the absolute value error to the uncertainty deviation is defined as the \emph{tracking ratio} for each output value.  An ideal uncertainty-tracking arithmetic should have an average tracking ratio close to 1.  
\item  The ratio of the absolute value error to the uncertainty bounding range is defined as the \emph{bounding ratio} for each output value.  An ideal uncertainty-bounding arithmetic should have a maximal bounding ratio either 1 or less than but close to 1.  If the maximal bounding ratio is larger than 1, \emph{bounding leakage} measures the probability for errors to be outside uncertainty bounding range.  
\end{itemize}

In both cases, all measurements should be stable for an algorithm so that they should not change significantly for different input deviation, input data, or the amount of calculation.  For example, if different branches of conditional executions contain very different amounts of calculations, such stability is crucial for obtaining a valid estimation of result precision.

Without dependency problem, the tracking ratios for precision arithmetic is expected to be normal distributed, and the average tracking ratio is a constant $\frac{2}{\sqrt{2 \pi}} \simeq 0.7979 $ when the bounding leakage is ignored.

\subsection{Comparing Uncertainty-Bearing Arithmetics}

Precision arithmetic tracks both the uncertainty bounding range and the uncertainty deviation, so it can be evaluated for both goals.  Independence arithmetic has no uncertainty bounding range, while interval arithmetic has no uncertainty deviation.  To be able to compare all the three arithmetics, $[x - 6\delta x, x + 6\delta x]$ is used artificially as the bounding range for an average value $x$ with deviation $\delta x$ for independence arithmetic, and vice versa for interval arithmetic.

As stated previously:
\begin{itemize}
\item Independence arithmetic assumes that any two operands are independent of each other, which may not be true in most cases.
\item Precision arithmetic assumes that the uncertainties of any two operands are independent of each other, but allows the two operands themselves to be correlated.
\item Interval arithmetic has the worst-case assumption because it needs to have zero bounding leakage unconditionally.    
\end{itemize} 
The statistical assumption of precision arithmetic is weaker than that of independence arithmetic but stronger than that of interval arithmetic, so after executing the same algorithm on the same input data, the output deviation and the bounding range of precision arithmetic are expected to be larger than those of independence arithmetic but smaller than those of interval arithmetic.
\begin{itemize}

\item According to Formula \eqref{eqn: rounding error +-} and Formula \eqref{eqn: rounding error range vs deviation}, the result deviation of addition and subtraction by precision arithmetic propagates in the same way as that of independence arithmetic, while the result bounding range propagates in the same way as that of interval arithmetic. Hence addition and subtraction cannot differentiate the three arithmetics.

\item According to Formula \eqref{eqn: uncertainty *} and Formula \eqref{eqn: uncertainty /}, the result precision of multiplication and division by precision arithmetic is always larger than that by independence arithmetic.  However, if both operands have precisions much less than 1, the result precision of multiplication and division is very close to that of independence arithmetic.  Thus, the result of precision arithmetic should be much closer to that of independence arithmetic.

\item The uncertainty distribution of precision arithmetic is a truncated Gaussian distribution according to Formula \eqref{eqn: uncertainty distribution}.  When an imprecise value is multiplied by a constant, because its uncertainty bounding range and its uncertainty distribution deviation cannot be scaled linearly simultaneously according to Formula \eqref{eqn: uncertainty deviation} and Formula \eqref{eqn: uncertainty range}, precision arithmetic chooses to preserve the distribution deviation rather than the bounding range, thus introducing bounding leakages.  Figure \ref{fig: Prec_RndByDev_Dist} suggests that the bounding range of precision arithmetic should be much narrower than that of interval arithmetic, while the shape of Gaussian distribution suggests that such introduced bounding leakage should be small when the truncation range is much larger than the distribution deviation, e.g., less than $10^{-6}$ for the chosen normalization method whose truncation range is about $\pm 5$ deviations.

\item  Formula \eqref{eqn: uncertainty 1d} and its multi-dimensional expansions such as Formula \eqref{eqn: uncertainty 2d} are mathematically strict so that precision arithmetic has no dependence problem on expression differences.  In contrast, there seems no similar solution for generic Taylor expansion using interval arithmetic, because there seems no general analytic solution to find maxima and minima for generic polynomial at any range \cite{Numerical_Recipes}.  In this respect, precision arithmetic is mathematically simpler than interval arithmetic.

\end{itemize}

\subsection{Comparing Algorithms for Tests}

Algorithms of completely different nature with each representative for its category are needed to test the generic applicability of uncertainty-bearing arithmetic.  

An algorithm can be categorized by comparing the amount of its input and output data:
\begin{itemize}
\item \emph{Transforming}: A transforming algorithm has about equal amounts of input and output data.  The information contained in the data remains about the same after transforming.  The Discrete Fourier Transformation is a typical transforming algorithm, which contains exactly the same amount of input and output data, and its output data can be transformed back to the input data using essentially the same algorithm.  Matrix inversion is another such reversible algorithm.  For reversible transformations, a unique requirement for uncertainty-bearing arithmetic is to introduce the least amount of additional uncertainty after forward and reverse transformation, which provides an objective testing standard for a uncertainty-bearing arithmetic.  A test of uncertainty-bearing arithmetic using FFT algorithms is provided in Section \ref{sec: FFT}, and a test of matrix inversion is provided in Section \ref{sec: matrix}.

\item \emph{Generating}:  A generating algorithm has much more output data than input data.  Solving differential equations numerically and generating a numerical table of a specific function are two typical generating algorithms.  The generating algorithm codes mathematical knowledge into data, so there is an increase of information in the output data.  From the perspective of encoding information into data, Taylor expansion is also a generating algorithm. In generating algorithms, input uncertainty should also be considered when deciding if the result is good enough so that the calculation can stop.  Some generating algorithms are theoretical calculations which involve no imprecise input so that all result uncertainty is due to rounding errors.  Section \ref{sec: recursion} demonstrates such an algorithm, which calculates a table of the sine function using trigonometric relations and two precise input data, $sin(0)=0$ and $sin(\pi/2)=1$.  In some other generating algorithms, the accumulation of rounding errors and input uncertainty should stop the algorithm at an optimal termination point using the truncation rule, which is demonstrated in Section \ref{sec: taylor expansion}.

\item \emph{Reducing}:  A reducing algorithm has much less output data than input data such as numerical integration and statistical characterization of a data set.  Some information of the data is lost while other information is extracted during reducing.  Conventional wisdom is that a reducing algorithm generally benefits from a larger input data set \cite{Probability_Statistics}.  Such a notion needs to be re-evaluated when uncertainty accumulates during calculation.  A test of uncertainty-bearing arithmetic using numerical integration is provided in Section \ref{sec: integration}.
\end{itemize}

Other relations between the input and output can also be used to categorize an algorithm.  
\begin{itemize}
\item  In an \emph{expressive} algorithm, each output is implemented as an analytic mathematical expression of inputs.  Formula \eqref{eqn: uncertainty 1d} and Formula \eqref{eqn: uncertainty 2d} of precision arithmetic are powerful tools to solve expressive algorithms using precision arithmetic.  

\item  In a \emph{progressive} algorithm, each output is based on partial inputs and previously generated outputs.  If an output depends on the state which is defined by previous inputs and outputs, the algorithm is also progressive.  Most practical algorithms are progressive.  Even if there may be an expected analytic mathematical expression between its input and output, an algorithm may not be expressive due to its progressive implementation.  The dependence problem usually exists in a progressive algorithm.  Section \ref{sec: Comparison Using Progressive Moving-Window Linear Regression} discusses the behaviours of all the three arithmetic for a progressive algorithm.
\end{itemize}

\subsection{Input Data to Use}

To test input data of any precision, a precise input value can be cast to any specific input deviation using precision representation.  There exist two ways of implementing such a casting:
\begin{itemize}
\item  A \emph{clean} signal is obtained by directly casting a perfect signal to a specific precision.  Such casting may contain systematic rounding errors.  For instance, if a perfect sine signal repeats $2^{n}$ times in $2^{n+2}$ samples, the signal contains only values 0, $\pm 1$ and $\pm1/\sqrt{2}$, with each value repeated multiple times in the signal.  The symmetry of the arithmetic may be tested by the output symmetry of clean input signals, e.g., in the discrete Fourier transformation, the frequency space should be conjugately symmetrical \cite{Numerical_Recipes} for a clean signal in signal space.

\item  A \emph{noisy} signal is obtained by adding Gaussian noise of the same deviation as the input deviation to a perfect signal before casting.  It represents a realistic signal, and it should be used in validating arithmetic on uncertainty propagation.  
\end{itemize}

\clearpage
\section{ Comparison Using FFT}
\label{sec: FFT}

\subsection{Frequency Response of DFT (Discrete Fourier Transformation)}

Each testing algorithm needs to come under careful scrutiny.  One important issue here is whether the digital implementation of the algorithm is faithful for the original analytic algorithm.  For example, the DFT is only faithful for continuous Fourier transformation at certain frequencies, and it has a different degree of faithfulness for other frequencies.  This is called the frequency response of the DFT in this paper.

For each signal sequence $h[k], k = 0, 1 \dots  N-1$, in which $N$ is a positive integer, the DFT $H[n], n = 0, 1 \dots  N-1$ and its reverse transformation is given by Formula \eqref{eqn: Fourier} \cite{Numerical_Recipes}, in which $k$ is the \emph{index frequency} for the DFT:
\begin{align}
\label{eqn: Fourier}
& H[n]=\sum _{k=0}^{N-1}h[k] \; e^{i 2\pi \frac{k}{N} n};
& h[k]=\frac{1}{N} \sum _{n=0}^{N-1}H[n] \; e^{-i 2\pi \frac{n}{N} k};
\end{align}

The $H[n]$ of a pure sine signal $h[k] = \sin \left(2\pi f k/N \right)$ is calculated by Formula \eqref{eqn: sin Fourier}, in which $f$ is the frequency of the sine wave.  When $f$ is an index frequency for $H[n]$, Formula \eqref{eqn: sin Fourier} becomes Formula \eqref{eqn: sin Fourier integer frequency}. Otherwise, the general solution for Formula \eqref{eqn: sin Fourier} is Formula \eqref{eqn: sin Fourier fractional frequency}, which approaches \eqref{eqn: sin Fourier integer frequency} when $f$ approaches its closest integer $F$, or Formula \eqref{eqn: sin Fourier 1/2 frequency} when $f$ approaches $F \pm 1/2$.
\begin{align}
\centering
\label{eqn: sin Fourier}
& H[n] = \frac{\sum _{k=0}^{N-1}e^{i 2\pi (n+f)\frac{k}{N}}  - \sum _{k=0}^{N-1} e^{i 2\pi (n-f)\frac{k}{N}}}{2 i}; \\
\label{eqn: sin Fourier integer frequency}
& H[n]=i \delta _{n,F} N/2; \\
\label{eqn: sin Fourier fractional frequency}
& H[n] = \frac{1}{2} \frac{\sin(2\pi f - 2\pi \frac{f}{N}) + \sin(2\pi \frac{f}{N})-\sin(2\pi f) e^{-i 2\pi \frac{n}{N}}}{\cos(2\pi \frac{n}{N})-\cos(2\pi \frac{f}{N})}; \\
\label{eqn: sin Fourier 1/2 frequency}
& H[n] = N/ \pi;
\end{align}

\begin{figure}[p]
\includegraphics[width=4.5in,height=2.75in]{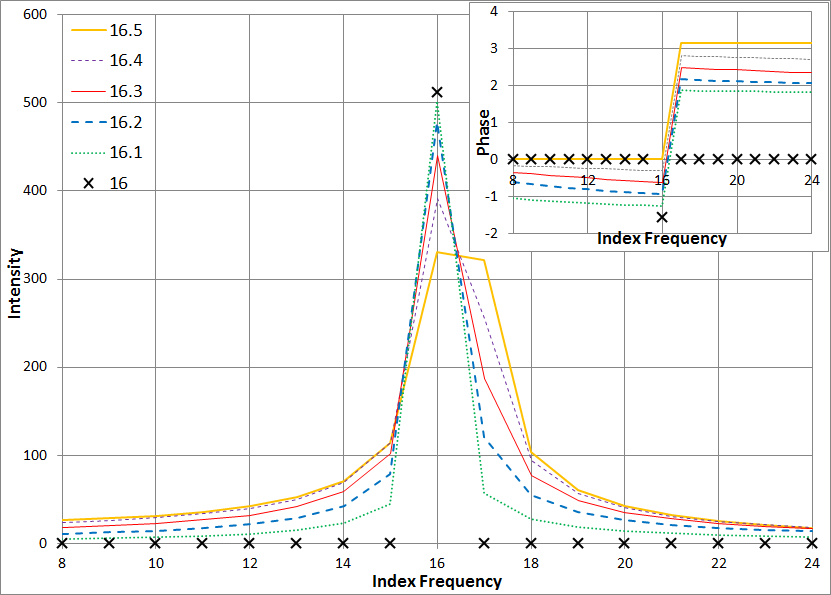} 
\captionof{figure}{Unfaithful representations of perfect sine signals in the Discrete Fourier Transformation.  The calculation is done on 1024 samples using FFT on a series of perfect sine signals having amplitude of 1 and slightly different frequencies as shown in legends.  In the drawing, x axis shows frequency, y axis shows either intensity or phase (inlet).  A faithful representation is also included for comparison, whose phase is $\pi /2$ at the index frequency, and undetermined at other frequencies.}
\label{fig: FFT_Unfaithful}
\end{figure}

The DFT $H[n]$ of the signal $h[k]$ is the digital implementation of the continuous Fourier transformation $H(s)$ of the signal $h(t)$ \cite{Numerical_Recipes}, in which $H(s)=i \delta (s-f)$ for $h[k]=\sin(2 \pi f)$.  From Formula \eqref{eqn: sin Fourier fractional frequency}, when the signal frequency of the original signal falls between two index frequencies of the transformation, the peak is lower and wider with a wrong phase, depending on the fractional frequency $|f - F|$.  Thus, the DFT is only faithful for signal components with exactly one of the index frequencies of the transform, and it suppresses and widens unfaithful signal components, each of which has a phase different from its closest faithful representation, with the phase of a sine wave distorted toward that of a cosine wave, and vise visa.  Examples of unfaithful representations of fractional frequency by the DFT are shown in Figure \ref{fig: FFT_Unfaithful}.  

Due to its width, a frequency component in an unfaithful transformation may interact with other frequency components of the Discrete Fourier spectrum, thus sabotaging the whole idea of using the Fourier Transformation to decompose a signal into independent frequency components.  Because the reverse DFT mathematically restores the original $\{h[k]\}$ for any $\{H[n]\}$, it exaggerates and narrows all unfaithful signal components correspondingly.  This means that the common method of signal processing in the Fourier space \cite{Numerical_Recipes}\cite{Stochastic_Arithmetic}\cite{Floating-point_Digital_Filters} may generate artefacts due to its uniform treatment of faithful and unfaithful signal components, which probably coexist in reality.  Unlike aliasing \cite{Electronics}\cite{Numerical_Recipes}\cite{Floating-point_Digital_Filters}, unfaithful representation of the DFT has an equal presence in the whole frequency range so that it cannot be avoided by sampling the original signal differently.

An unfaithful representation arises from the implied assumption of the DFT.  The continuous Fourier transformation has an infinitive signal range so that:
\begin{equation}
\label{eqn: Fourier continuous shift}
h(t) \Leftrightarrow H(s): \eqspace h(t - \tau) \Leftrightarrow H(s) e^{i 2\pi s \tau};
\end{equation}
As an analog, the DFT $G[n]$ of the signal $h[k], k = 1 \dots N$ can be calculated mathematically from the DFT $H[n]$ of $h[k], k = 0\dots N-1$:
\begin{equation}
\label{eqn: Fourier discrete shift}
G[n] = (H[n] + h[N] - h[0]) e^{i 2\pi n/N};
\end{equation}
Applying Formula \eqref{eqn: Fourier continuous shift} to Formula \eqref{eqn: Fourier discrete shift} results in Formula \eqref{eqn: Fourier discrete assumption}.
\begin{equation}
\label{eqn: Fourier discrete assumption}
h[N] = h[0];
\end{equation}
Thus, the DFT has an implied assumption that the signal $h[k]$ repeats itself outside the region of $[0, N-1]$ \cite{Numerical_DFT}.  For an unfaithful frequency, $h[N-1]$ and $h[N]$ are discontinuous in regard to signal periodicity, resulting in larger peak width, lower peak height, and the wrong phase.  

The most convenient signals to test uncertainty-bearing arithmetic are perfect sine or cosine signals with index frequencies.  A linear signal with the slope $\lambda, h[k] = \lambda k$, provides a generic test for input frequencies other than index frequencies, whose Fourier spectrum is:
\begin{equation}
H[n] = -\lambda \frac{N}{2} \left(1 + \frac{i}{\tan(\pi n / N)}\right);
\end{equation}

\subsection{FFT (Fast Fourier Transformation)}

When $N = 2^{L}$, in which $L$ is a positive integer, the generalized Danielson-Lanczos lemma \cite{Numerical_Recipes} can be applied to the DFT as FFT \cite{Numerical_Recipes}, in which $m = L, L-1,\dots 1,0$ indicates progress of the transformation, and $j$ is the bit-reverse of $n$:
\begin{align}
m=L: \eqspace & H[n,\frac{k}{2^{m}}] = h[j],  k,n = 0,1\dots N-1; \\
\label{eqn: Danielson-Lanczos}
m=L-1 \dots 0: \eqspace & H[n,\frac{k}{2^{m}}] = H[n,\frac{k}{2^{m+1}}] + H[n,\frac{k}{2^{m+1}}]\; \exp{(+i 2\pi \frac{n}{2^{L-m}})}; \\
m=0: \eqspace & H[n] = H[n,\frac{k}{2^{m}}];
\end{align}

\noindent Thus, each output value is obtained after applying Formula \eqref{eqn: Danielson-Lanczos} $L$ times.  $L$ is called FFT order in this paper.  

The calculation of the term $\exp{(i 2\pi \frac{n}{2^{L-m}})}$ in Formula \eqref{eqn: Danielson-Lanczos} can be simplified.  Let $<<$ denote a bit left-shift operation and let \& denote a bitwise AND operation:
\begin{align}
\label{eqn: FFT phase array}
\varphi [n] \equiv \exp{(i 2\pi \frac{n}{2^{L}})}: \eqspace & \exp{(i 2\pi \frac{n}{2^{L-m}})} = \varphi [(n<<m)\& ((1<<L)-1)];
\end{align}
It is important to have an accurate phase factor array $\varphi [n]$ when tracking the FFT calculation error.  The accuracy of $\varphi [n]$ can be checked rigidly within itself by trigonometric relations so that no significant error is introduced from trigonometric functions.  

Formula \eqref{eqn: Danielson-Lanczos} always sums up two mutually independent operands, so the error propagation in a FFT algorithm is precisely tracked by independence arithmetic, and the dependency problem should not be a concern for interval arithmetic and precision arithmetic.  

FFT is one of the most widely used algorithms \cite{Numerical_Recipes}.  By providing a balanced usage of addition, subtraction and multiplication involving trigonometric functions, it services as one of the most important benchmarks in testing processors for overall mathematical performance \cite{DSP}.  Since all three uncertainty-bearing arithmetics are generic in nature without special optimization for FFT algorithms, the testing result using FFT algorithms should be generic for expressive algorithms.  FFT algorithms provide a good linear platform to test any uncertainty-bearing arithmetic, with 1) a clearly defined value measuring the amount of calculation, 2) a known error propagation mechanism, 3) no conditional execution in the algorithm, and 4) using only basic arithmetic operations without the dependence problem.

\subsection{Evaluating Calculation Inside Uncertainty}

\begin{figure}[p]
\centering
\includegraphics[width=4.5in,height=2.4in]{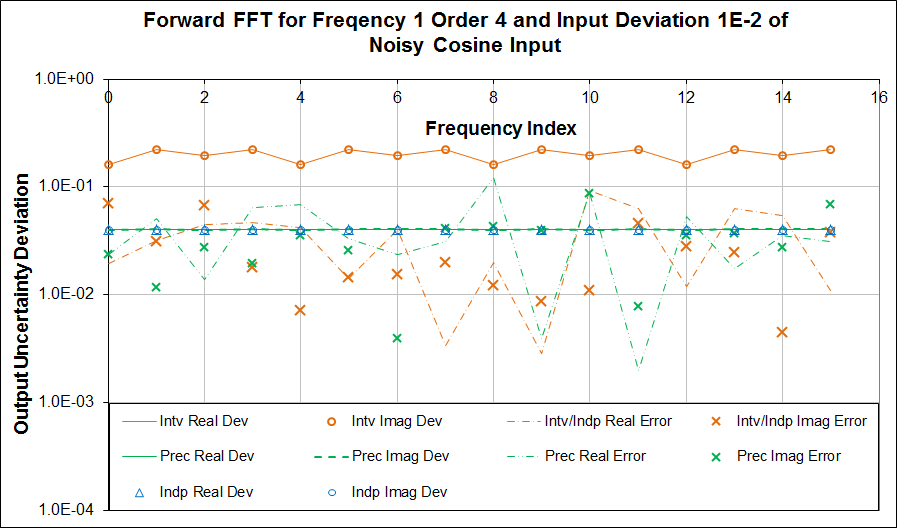} 
\captionof{figure}{The output deviations and value errors of the forward FFT on a noisy sine signal of FFT order 4, index frequency 1 and input deviation $10^{-2}$.  In the legend, "Intv" means interval arithmetic, "Indp" means independence arithmetic, "Prec'' means precision arithmetic, ``Dev'' means output uncertainty deviations, ``Error'' means output value errors, ``Real'' means real part, and ``Imag'' means imaginary part.  Because both interval arithmetic and independence arithmetic using conventional floating arithmetic for underlying calculations, they have the same value errors.}
\label{fig: Prec4_FFT_Sin_Profile_Freq1}
\end{figure}

Figure \ref{fig: Prec4_FFT_Sin_Profile_Freq1} shows the output deviations and value errors for a noisy sine signal after forward FFT.  It shows that the output deviations using precision arithmetic are slightly larger than the output deviations using independence arithmetic, but much less than those using interval arithmetic.  For a fixed input deviation, the output deviation using independence arithmetic is a constant for each FFT.  Because the value and uncertainty interact with each other through normalization in precision arithmetic, output deviations of Formula \eqref{eqn: Danielson-Lanczos} are no longer a constant.  One interesting consequence is that only in precision arithmetic the output deviations for a noisy input signal are larger than those for a corresponding clean input signal.

Figure \ref{fig: Prec4_FFT_Sin_Profile_Freq1} shows that the value errors calculated using precision arithmetic are comparable to those using conventional floating-point arithmetic, and they are both comparable to the output deviations using either precision arithmetic or independence arithmetic.  In other words, the result of calculating 2-bit or 53-bit into uncertainty are quite comparable so that the limited calculation inside uncertainty is reasonable.  

\begin{figure}[p]
\centering
\includegraphics[width=4.5in,height=2.4in]{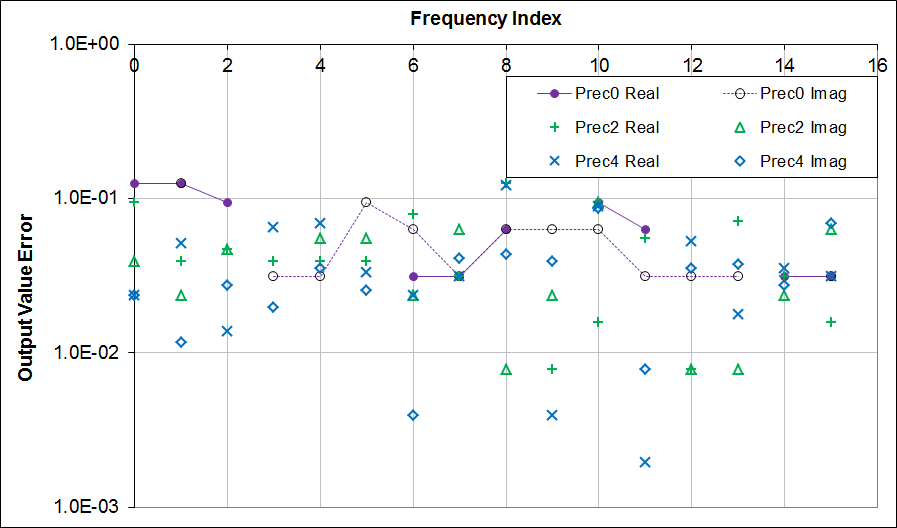} 
\captionof{figure}{The output value errors of the forward FFT on a noisy sine signal of index frequency 1 and input deviation $10^{-2}$ using precision arithmetic with different bit inside uncertainty.  In the legend, ``Prec0'' means precision arithmetic with 0-bit calculated inside uncertainty, ``Prec2'' means precision arithmetic with 2-bit calculated inside uncertainty, and "Prec4'' means precision arithmetic with 4-bit calculated inside uncertainty.}
\label{fig: Precx_FFT_Sin_Profile_Freq1}
\end{figure}

Figure \ref{fig: Precx_FFT_Sin_Profile_Freq1} compares the output value errors of precision arithmetic calculating different bits inside uncertainty.  With no calculation inside uncertainty, the output value errors exist only on four levels.  Such quantum distribution is reduced noticeably by the 2-bit calculation inside uncertainty, and is further reduced by the 4-bit calculation inside uncertainty.  Compared with Figure \ref{fig: Prec4_FFT_Sin_Profile_Freq1}, Figure \ref{fig: Precx_FFT_Sin_Profile_Freq1} shows that the result using precision arithmetic with the 4-bit calculation inside uncertainty approaches that using independence arithmetic so that the 4-bit calculation inside uncertainty seems sufficient.  Precision arithmetic with the 4-bit calculation inside uncertainty is used for further tests.

\subsection{Evaluating Uncertainty Distribution}

\begin{figure}[p]
\centering
\includegraphics[width=4.5in,height=2.75in]{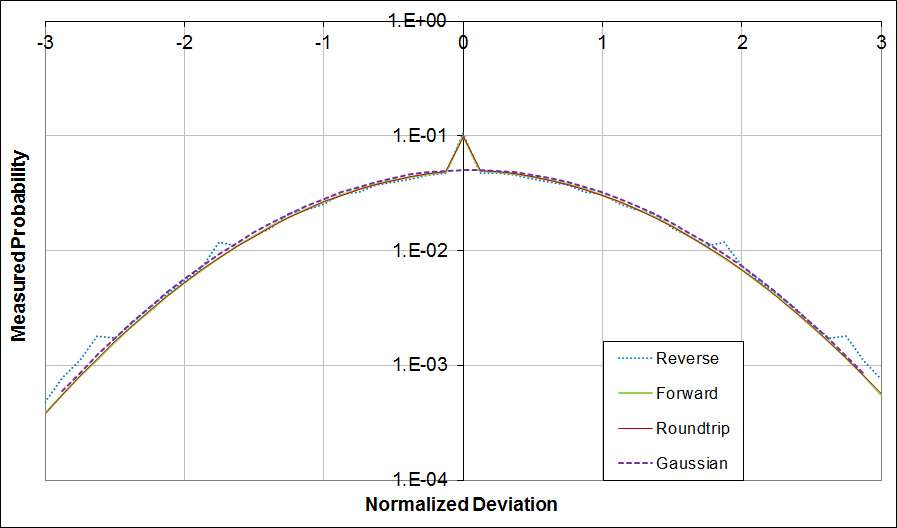} 
\captionof{figure}{ The measured tracking ratio distributions using independence arithmetic for FFT algorithms (as shown in legend).  They are best fitted by a Gaussian distribution with the mean of 0.06 and deviation of 0.98.}
\label{fig: Indp_Sin_NormDist}
\end{figure}

\begin{figure}[p]
\centering
\includegraphics[width=4.5in,height=2.75in]{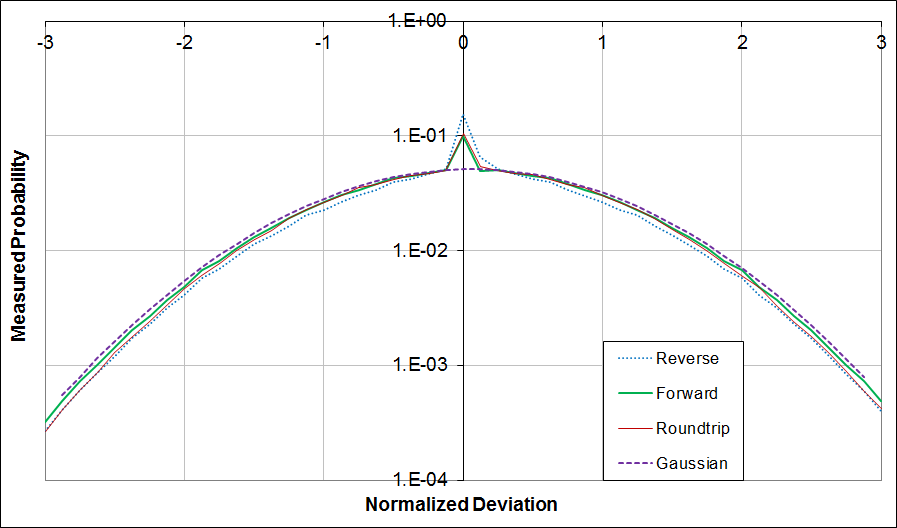} 
\captionof{figure}{ The measured tracking ratio distributions using precision arithmetic for FFT algorithms (as shown in legend).  They are best fitted by a Gaussian distribution with the mean of 0.06 and deviation of 0.97. }
\label{fig: Prec4_Sin_NormDist}
\end{figure}

Each output value error is normalized with the corresponding output uncertainty deviation before it is counted for histogram. If output value errors are Gaussian-distributed with the deviation given precisely by the corresponding output uncertainty deviation, then the normalized histogram should be normal-distributed.  Figure \ref{fig: Indp_Sin_NormDist} and Figure \ref{fig: Prec4_Sin_NormDist} show that such histograms using wither independence arithmetic or precision arithmetic with 4-bit calculated inside uncertainty are both best fit by Gaussian distribution with the deviation of 0.98 and the mean of 0.06.  Due to limited bits calculated inside uncertainty and normalization, the population at where the value errors are zero is expected to be larger for the precision arithmetic, e.g., during normalization, all values which is less than 4-fold of resolution becomes 0. This phenomenon is confirmed by Figure \ref{fig: Prec4_Sin_NormDist}.

Precision arithmetic tracks all increases of rounding errors, but it cannot track decreases of the rounding error due to mutual cancellations during arithmetic operations.  Hence the uncertainty distribution provided by precision arithmetic serves as the bounding distribution for value errors, and the actual distribution could be narrower than the bounding distribution.  FFT provides a good test for such probability bounding.  Its forward and reverse algorithms are identical except for a constant so that they result in exactly the same bounding probability distributions.  On the other hand, the forward FFT condenses a sine signal into only two non-zero imaginary values by mutual cancellation of signal components, while the reverse FFT spreads only two non-zero imaginary values to construct a sine signal.  Thus, the forward FFT is more sensitive to calculation errors than the reverse FFT, and should have a broader actual uncertainty distribution.  Indeed in Figure \ref{fig: Prec4_Sin_NormDist}, the histogram for the reverse algorithm for a sine signal with added noise is narrow than that of the forward algorithm, with that for the round-trip algorithm in the middle of the two.

\subsection{Evaluating Uncertainty-Tracking}

\begin{figure}[p]
\centering
\includegraphics[height=2.5in]{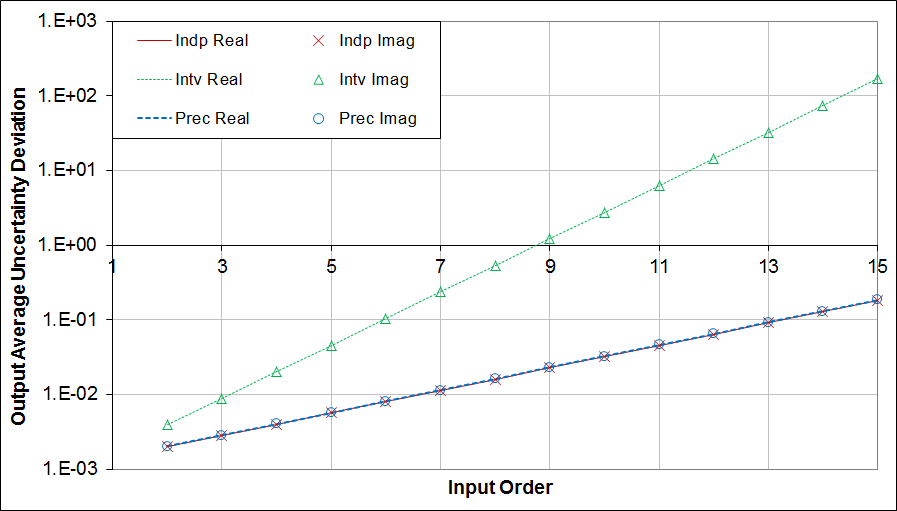} 
\captionof{figure}{ For the same input deviation of $10^{-3}$, the empirical average output deviations of the forward FFT increase exponentially with the FFT order for all uncertainty-bearing arithmetics.  In the legend, "Intv" means interval arithmetic, "Indp" means independence arithmetic, "Prec'' means precision arithmetic, ``Real'' means real part, and ``Imag'' means imaginary part.}
\label{fig: All_For_AvgDev_vs_FFTOrder}
\end{figure}

\begin{figure}[p]
\centering
\includegraphics[height=2.5in]{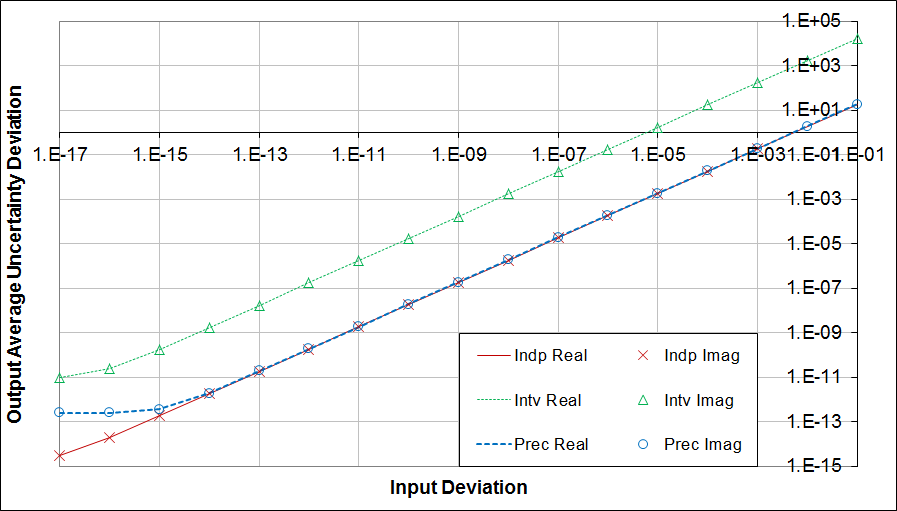} 
\captionof{figure}{ For the same order of the FFT calculation of 15, the empirical average output deviations of the forward FFT increases linearly with the input deviation for all uncertainty-bearing arithmetics.  In the legend, "Intv" means interval arithmetic, "Indp" means independence arithmetic, "Prec'' means precision arithmetic, ``Real'' means real part, and ``Imag'' means imaginary part.}
\label{fig: All_For_AvgDev_vs_InDev}
\end{figure}

Figure \ref{fig: All_For_AvgDev_vs_FFTOrder} shows that for the same input deviation, the output deviations of the forward FFT increase exponentially with the FFT order using all three arithmetics.  Figure \ref{fig: All_For_AvgDev_vs_InDev} shows that for the same FFT order, the output deviations of the forward FFT increase linearly with the input deviation using all three arithmetics.  The output deviation does not change with input frequency so that all data of the same input deviation and the same FFT order but with different input frequencies can be pooled together during analysis.  The trends in Figure \ref{fig: All_For_AvgDev_vs_FFTOrder} and Figure \ref{fig: All_For_AvgDev_vs_InDev} are modeled by Formula \eqref{eqn: fitting error}, in which $L$ is the FFT order, $\delta x$ is the input deviation, $\delta$y is the average output deviation, and $\alpha$ and $\beta$ are empirical fitting constants: 
\begin{equation}
\label{eqn: fitting error}
\delta y = \alpha \beta ^{L} \delta x;
\end{equation}
$\beta$ measures the propagation speed of the deviation with an increased amount of calculation in Formula \eqref{eqn: fitting error}.  It is called \emph{propagation base rate}.  Unless $\beta$ is close to 1, $\beta$ dominates $\alpha$ in fitting, thus determining characteristics of Formula \eqref{eqn: fitting error}.  

\begin{figure}[p]
\centering
\includegraphics[height=2.5in]{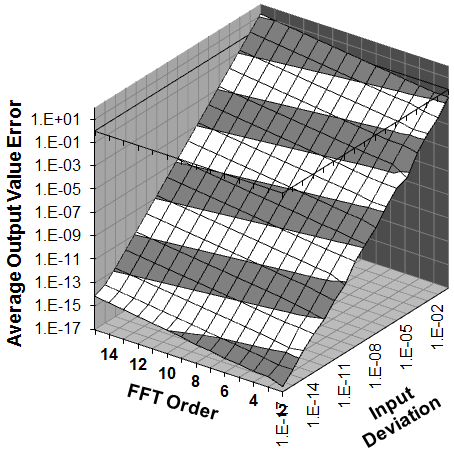} 
\captionof{figure}{ The empirical average output value errors using precision arithmetic increase exponentially with the FFT order and linearly with the input deviation, respectively.}
\label{fig: Prec4_For_AvgErr_vs_FFTOrder_InDev}
\end{figure}

It turns out that Formula \eqref{eqn: fitting error} is a very good fit for both average output deviations and value errors for all three arithmetics, such as demonstrated in Figure \ref{fig: Prec4_For_AvgErr_vs_FFTOrder_InDev}.  Because uncertainty-tracking is a competition between error propagation and uncertainty propagation, the average output tracking ratio for the forward FFT is expected to fit Formula \eqref{eqn: fitting significand} and Formula \eqref{eqn: fitting significand coefficient}, in which $z$ is the average output tracking ratio, $L$ is the FFT order, $(\alpha _{dev}, \beta _{dev})$ and $(\alpha _{err}, \beta _{err})$ are fitting parameters of Formula \eqref{eqn: fitting error} for average output deviations and value errors, respectively:
\begin{equation}
\label{eqn: fitting significand} 
z = \alpha \beta ^{L};
\end{equation}
\begin{equation}
\label{eqn: fitting significand coefficient} 
\alpha = \alpha _{err} / \alpha _{dev}; \eqspace \beta = \beta _{err} / \beta_{dev};
\end{equation}

\begin{figure}[p]
\centering
\includegraphics[height=2.5in]{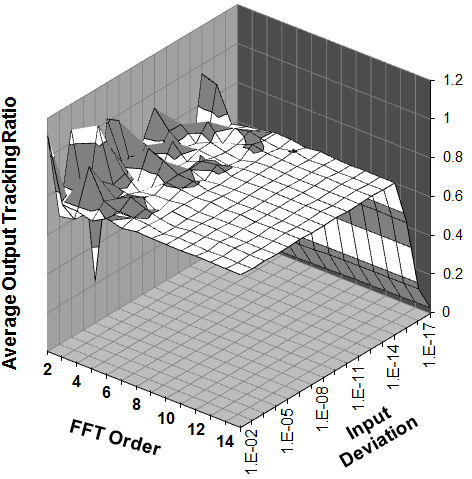} 
\captionof{figure}{ The empirical average output tracking ratios using precision arithmetic is a constant when the input deviation is larger than $10^{-14}$ and the FFT order is more than 5 for forward FFT algorithms.  Because the precision of conventional floating-point representation is at $10^{-16}$, adding Gaussian noises with the deviation of $10^{-17}$ should have little effect on the input data.  For the same reason, the output tracking ratios are stable only when the input deviation is more than $10^{-14}$.  When the FFT order is $2$, a FFT calculation actually involves no arithmetic calculation between input data.  For the same reason, when the FFT order is less than 5, there is not enough arithmetic calculation for the result tracking ratios to reach equilibrium.}
\label{fig: Prec4_For_AvgSig_vs_FFTOrder_InDev}
\end{figure}

The estimated average output tracking ratio can then be compared with the measured ones to evaluate the predictability of the uncertainty-tracking mechanism.  One example of measured average output tracking ratios is shown in Figure \ref{fig: Prec4_For_AvgSig_vs_FFTOrder_InDev}, which shows that the average output tracking ratios using precision arithmetic are a constant despite that both average output uncertainty deviations and value errors increase linearly with the input deviation and exponentially with the FFT order.  Formula \eqref{eqn: fitting error} and Formula \eqref{eqn: fitting significand} are found empirically to be a good fit for any FFT algorithm with any input signal using any arithmetic.

The Reverse FFT algorithm is identical to the Forward FFT algorithm, except when:

\begin{itemize}
\item  The Reverse FFT algorithm uses constant (-i) instead of (+i) in Formula \eqref{eqn: Danielson-Lanczos}.

\item  The Reverse FFT algorithm divides the result further by $2^{L}$.  
\end{itemize}
Thus, the average output deviations and value errors of the reverse FFT algorithm are expected to obey Formula \eqref{eqn: fitting error} and Formula \eqref{eqn: fitting reverse coefficient} , in which $(\alpha_{for}$, $\beta_{for})$ are corresponding fitting parameters of Formula \eqref{eqn: fitting error} for the forward FFT, while the average output tracking ratios are expected to obey Formula \eqref{eqn: fitting significand coefficient} with the same $\alpha$ and $\beta$ as those of the forward FFT.  
\begin{equation}
\label{eqn: fitting reverse coefficient} 
\alpha = \alpha _{for}; \eqspace  \beta = \beta _{for}/2;
\end{equation}

The Round-trip FFT is the forward FFT followed by the reverse FFT, with the output of the forward FFT as input to the reverse FFT.  Thus, both its average output deviations and value errors are expected to fit Formula \eqref{eqn: fitting error} and Formula \eqref{eqn: fitting round-trip coefficient}, in which $(\alpha _{for}, \beta _{for})$ and $(\alpha _{rev}, \beta _{rev})$ are corresponding fitting parameters of Formula \eqref{eqn: fitting error} for the forward FFT and the reverse FFT, respectively.  Its tracking ratios are expected to fit Formula \eqref{eqn: fitting significand} and Formula \eqref{eqn: fitting round-trip coefficient}, in which $(\alpha _{for}, \beta _{for})$ and $(\alpha _{rev}, \beta _{rev})$ are corresponding fitting parameters of Formula \eqref{eqn: fitting significand} for the forward FFT and the reverse FFT, respectively.
\begin{equation}
\label{eqn: fitting round-trip coefficient} 
\alpha = \alpha _{for} \alpha _{rev}; \eqspace \beta = \beta _{for} \beta _{rev};
\end{equation}

\begin{figure}
\centering
\includegraphics[width=4.5in,height=2.75in]{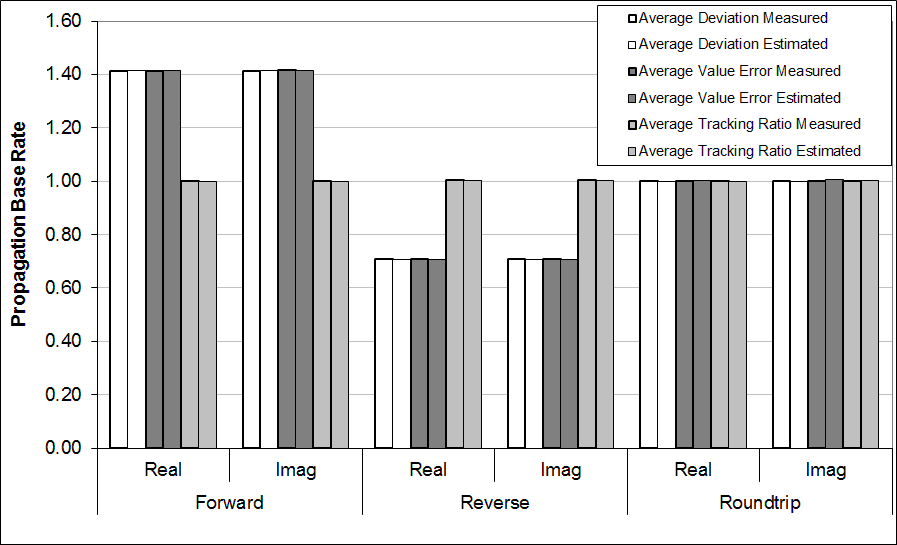} 
\captionof{figure}{ Empirical and theoretical $\beta$ for fitting average output deviations, value errors and tracking ratios for forward, reverse and roundtrip FFT using independence arithmetic on noisy sine signals.  In the chart, ``Real'' means real part, and ``Imag'' means imaginary part.}
\label{fig: Indp_PropBase_AvgSig}
\end{figure}

\begin{figure}
\centering
\includegraphics[width=4.5in,height=2.75in]{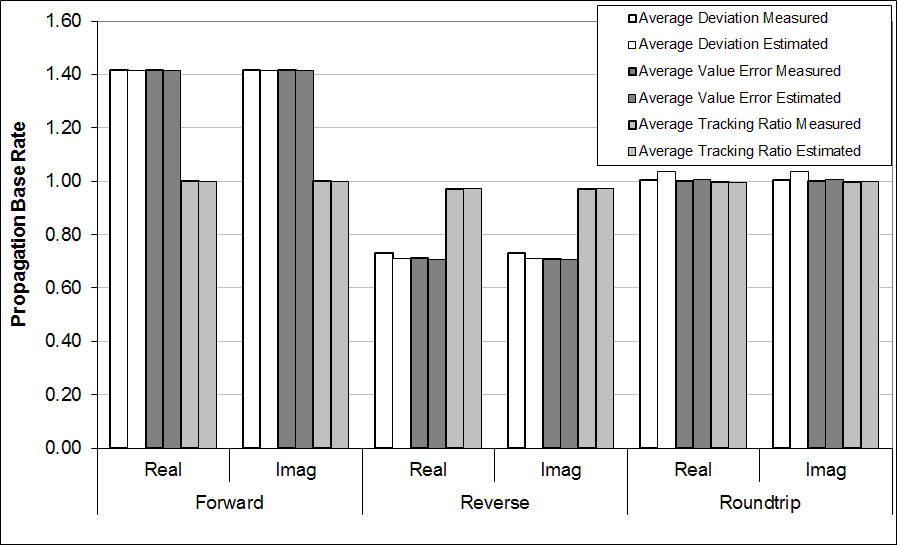} 
\captionof{figure}{ Empirical and theoretical $\beta$ for fitting average output deviations, value errors and tracking ratios for forward, reverse and roundtrip FFT using precision arithmetic on noisy sine signals.  In the chart, ``Real'' means real part, and ``Imag'' means imaginary part.}
\label{fig: Prec4_PropBase_AvgSig}
\end{figure}

\begin{figure}
\centering
\includegraphics[width=4.5in,height=2.75in]{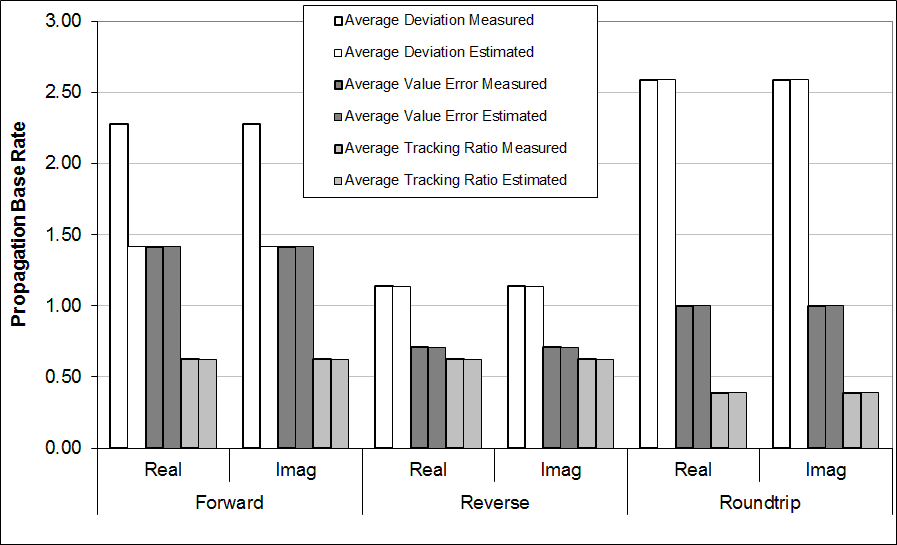} 
\captionof{figure}{ Empirical and theoretical $\beta$ for fitting average output deviations, value errors and tracking ratios for forward, reverse and roundtrip FFT using interval arithmetic on noisy sine signals.  In the chart, ``Real'' means real part, and ``Imag'' means imaginary part.}
\label{fig: Intv_PropBase_AvgSig}
\end{figure}

Figure \ref{fig: Indp_PropBase_AvgSig}, Figure \ref{fig: Prec4_PropBase_AvgSig} and Figure \ref{fig: Intv_PropBase_AvgSig} show the fitting of $\beta$ for independent, precision and interval arithmetic for all the three algorithms, respectively.  These three figures show that all measured $\beta$ make no distinction between input signals for any algorithms using any arithmetic, e.g., there is no difference between the real part and the imaginary part for a sine signal.  The estimated $\beta$ for average tracking ratios is obtained from Formula \eqref{eqn: fitting significand coefficient}.  The estimated $\beta$ for average uncertainty deviations and value errors for the reverse FFT and the roundtrip FFT are obtained from Formula \eqref{eqn: fitting reverse coefficient} and Formula \eqref{eqn: fitting round-trip coefficient}, respectively.  The estimated $\beta$ for average uncertainty deviations for the forward FFT is $\sqrt{2}$, which will be demonstrated later.  The measured $\beta$ and the estimated $\beta$ agree well with each other in all cases.  This confirms that uncertainty-tracking is a simple competition between the error propagation and uncertainty propagation:
\begin{itemize}
\item  Figure \ref{fig: Indp_PropBase_AvgSig} confirms that independence arithmetic is ideal for uncertainty-tracking for FFT algorithms: 1) $\beta$ for tracking ratios is a constant 1; and 2) $\beta$ for both the average output deviations and value errors is both 1 for the round-trip FFT because the result signal after the round-trip FFT should be restored as the original signal.  Thus, theoretical $\beta$ for the forward FFT and the reverse FFT are $\sqrt{2}$ and $1/\sqrt{2}$, respectively.

\item  Precision arithmetic has $\beta$ for average output deviations slightly larger than those of value errors, resulting in $\beta$ for average output tracking ratios to be a constant slightly less than 1.  Its $\beta$ for average output deviations is slightly larger than the corresponding $\beta$ of independence arithmetic, so its average output deviations propagate slightly faster with an increased FFT order than those of independent arithmetic.  Such slightly faster increase with the amount calculation is anticipated by the difference between Formula \eqref{eqn: uncertainty *} and Formula \eqref{eqn: stat *}  with $\gamma=0$.

\item  The $\beta$ for average output deviations using interval arithmetic is always much larger than $\beta$ for average output value errors, resulting in $\beta$ for average output tracking ratios of about $0.62$ for the forward and reverse FFT, and about $0.39  \cong {0.62}^{2}$ for the roundtrip FFT.  Consequently, using interval arithmetic, the average output deviations propagate much faster with the amount of calculations than the value error does.  Such fast propagation of uncertainty ranges is intrinsic to interval arithmetic due to its worst-case assumption.  
\end{itemize}

\begin{figure}
\centering
\includegraphics[width=4.5in,height=2.75in]{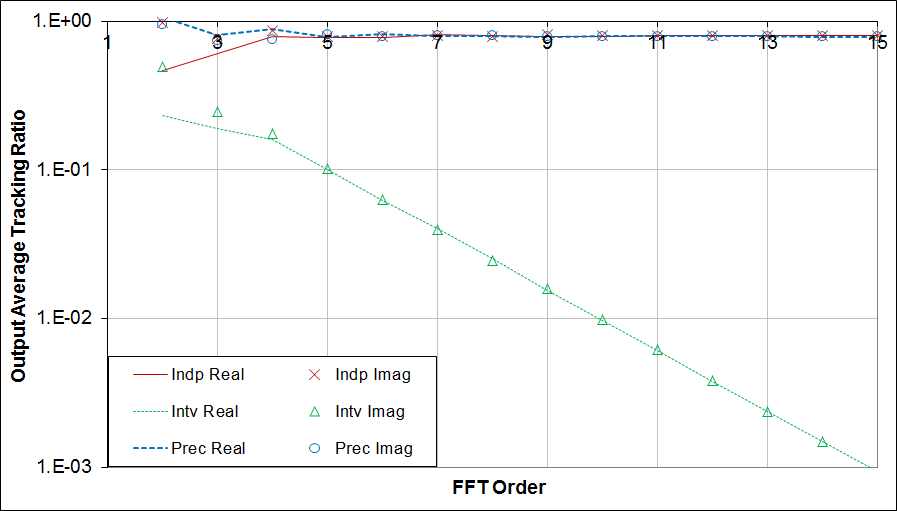} 
\captionof{figure}{ The empirical output average tracking ratios vs. the FFT order of the forward FFT for all three arithmetics when the input uncertainty deviation is $10^{-3}$.  In the legend, "Intv" means interval arithmetic, "Indp" means independence arithmetic, "Prec'' means precision arithmetic, ``Real'' means real part, and ``Imag'' means imaginary part.}
\label{fig: All_For_AvgSig_vs_FFTOrder}
\end{figure}

Figure \ref{fig: All_For_AvgSig_vs_FFTOrder} shows that for the forward FFT, the measured average output tracking ratios using either precision arithmetic or independence arithmetic are approximately constant of 0.8 in both cases, regardless of the FFT order.  In contrast, Figure \ref{fig: All_For_AvgSig_vs_FFTOrder} shows that using interval arithmetic the measured average output tracking ratios decrease exponentially with the FFT order L.  Such trends of average tracking ratios hold for all three FFT algorithms and all input signals.  Thus, in this case, the direct uncertainty tracking provided by precision arithmetic is better than the indirect uncertainty tracking provided by interval arithmetic.   

\begin{figure}
\centering
\includegraphics[height=3.00in]{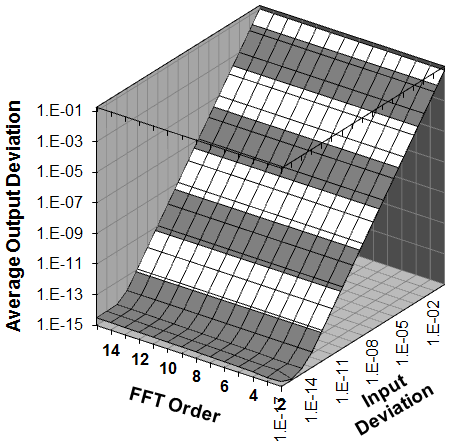} 
\captionof{figure}{The empirical average output deviations vs. the FFT order and input deviations using precision arithmetic for the round-trip FFT algorithm.}
\label{fig: Prec4_Rnd_AvgErr_vs_FFTOrder_InPrec}
\end{figure}

\begin{figure}
\centering
\includegraphics[height=3.00in]{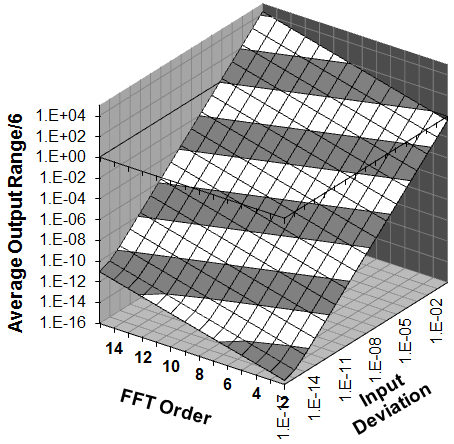} 
\captionof{figure}{ The empirical average output deviations vs. the FFT order and input deviations using interval arithmetic for the round-trip FFT algorithm.}
\label{fig: Intv_Rnd_AvgErr_vs_FFTOrder_InPrec}
\end{figure}

Figure \ref{fig: Prec4_Rnd_AvgErr_vs_FFTOrder_InPrec} shows that using precision arithmetic, each average output uncertainty deviation equals the corresponding input uncertainty deviation for all FFT orders after a round-trip operation.  Thus, after each round-trip operation, precision arithmetic restores the original signal and the corresponding uncertainty for FFT.  Such behavior seems ideal for a reversible algorithm.  In contrast, Figure \ref{fig: Intv_Rnd_AvgErr_vs_FFTOrder_InPrec} shows that using interval arithmetic, the average output uncertainty deviations increase exponentially with FFT orders, which means the undesirable broadening of uncertainty in the restored signal after a round-trip operation.

\subsection{Evaluating Uncertainty-Bounding }

\begin{figure}
\centering
\includegraphics[width=4.5in,height=2.75in]{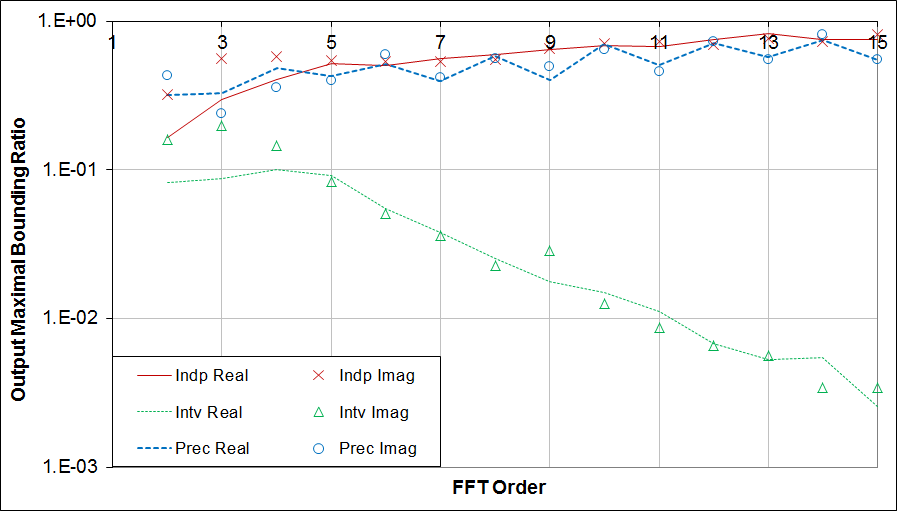} 
\captionof{figure}{ The empirical maximal output bounding ratios vs. the FFT order of the forward FFT for all three arithmetics.  In the legend, "Intv" means interval arithmetic, "Indp" means independence arithmetic, "Prec'' means precision arithmetic, ``Real'' means real part, and ``Imag'' means imaginary part.}
\label{fig: All_For_MaxBnd_vs_FFTOrder}
\end{figure}

\begin{figure}
\centering
\includegraphics[height=3.0in]{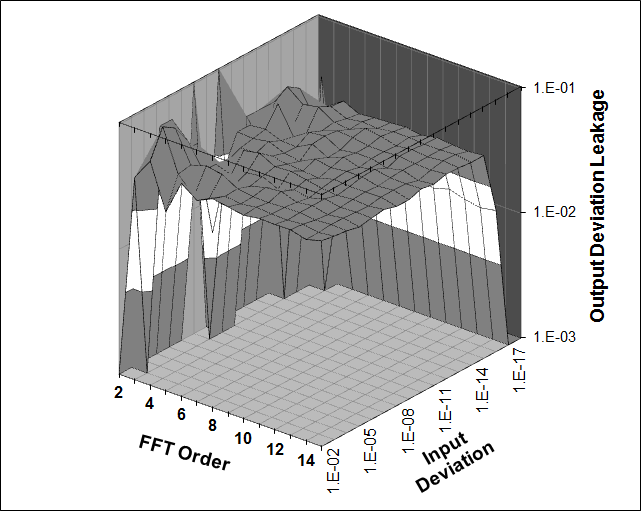} 
\captionof{figure}{ The empirical deviation leakages vs. the FFT order and input deviations using precision arithmetic for the forward FFT algorithm.}
\label{fig: Prec4_DevLeak_vs_FFTOrder_InPrec}
\end{figure}

While uncertainty tracking is the result of the propagation competition between average output deviations and average values errors with increased amount of calculations, uncertainty bounding is the result of the propagation competition between output bounding ranges and maximal value errors, both of which still fit Formula \eqref{eqn: fitting error} well using any arithmetic experimentally.  Formula \eqref{eqn: fitting significand} and Formula \eqref{eqn: fitting significand coefficient}  can be used to estimate the maximal bounding ratio as well.  For example, Figure \ref{fig: All_For_MaxBnd_vs_FFTOrder} shows that the maximal output bounding ratios using precision arithmetic fit Formula \eqref{eqn: fitting significand} well.  Unlike average output tracking ratios in Figure \ref{fig: All_For_AvgSig_vs_FFTOrder}, the maximal output bounding ratios increase slowly with the FFT order using either precision arithmetic or independent arithmetic.  In contrast, interval arithmetic has its maximal bounding ratios decreasing exponentially with the increased FFT order for all algorithms while keeping its bounding leakages at constant 0.  Detailed analysis shows that in interval arithmetic, $\beta$ for the maximal uncertainty bounding ranges exceeds $\beta$ for the maximal value error, suggesting the source of over-estimating uncertainty range with the increased amount of calculations.  Defining empirical \emph{deviation leakage} as the frequency of the value errors to be outside the range of mean $\pm$ deviation, Figure \ref{fig: Prec4_DevLeak_vs_FFTOrder_InPrec} shows that the deviation leakages is roughly a constant using precision arithmetic, suggesting the statistical nature of uncertainty bounding using precision arithmetic.  Whether precision arithmetic is better than interval arithmetic in uncertainty bounding depends on the statistical requirements for the uncertainty bounding:
\begin{itemize}
\item  In the situation when absolute bounding is required, interval arithmetic is the only choice.

\item  In the range estimation \cite{Statistical_Methods} involving low-resolution measurements whose sources of uncertainty are unclear, interval arithmetic is a better choice because the independence uncertainty assumption of precision arithmetic may not be satisfied. 

\item  Otherwise, precision arithmetic should be more suitable for normal usages.  
\end{itemize}

\clearpage
\section{Comparison Using Matrix Inversion}
\label{sec: matrix}

\subsection{Uncertainty Propagation in Matrix Determinant}

Let vector $[p_{1}, p_{2} \dots p_{n}]_{n}$ denote a permutation of the vector $(1,2\dots n)$ \cite{Linear_Algebra}.  Let $\$[p_{1}, p_{2} \dots p_{n}]_{n}$ denote the permutation sign of $[p_{1}, p_{2} \dots p_{n}]_{n}$ \cite{Linear_Algebra}.  For a $n$-by-$n$ square matrix M with the element $x_{i,j}, i,j=1,2\dots n$, let its determinant be defined as Formula \eqref{eqn: determinant} \cite{Numerical_Recipes} and let the sub-determinant at index $(i, j)$ be defined as Formula \eqref{eqn: sub-determinant} \cite{Linear_Algebra}:
\begin{align}
\label{eqn: determinant}
|M| \equiv 
\sum _{[p_{1}\dots p_{n}]_{n}} \$ [p_{1}\dots p_{n}]_{n} 
    \prod _{k} x_{k,p_{k}}; \\
\label{eqn: sub-determinant}
|M|_{i,j} \equiv 
\sum _{[p_{1}\dots p_{n}]_{n}}^{p_{i} = j} \$ [p_{1}\dots p_{n}]_{n} 
    \prod _{k}^{k \ne i} x_{k,p_{k}};
\end{align}
$(-1)^{i+j} |M_{(i,j)}|$ is the determinant of the $(n-1)$-by-$(n-1)$ matrix that results from deleting the row $i$ and column $j$ of $M$ \cite{Numerical_Recipes}.  Formula \eqref{eqn: determinant sum 1} holds for the arbitrary row index $i$ or the arbitrary column index $j$ \cite{Numerical_Recipes}:
\begin{equation}
\label{eqn: determinant sum 1}
|M| =\sum _{j=1}^{n} |M_{i,j}| x_{i,j} = \sum _{i=1}^{n} |M_{i,j}| x_{i,j};
\end{equation}

Assuming $p_{1}, p_{2} \in \{1,2...n\}$, let $[p_{1}, p_{2}]_{n}$ denote the length-2 unordered permutation which satisfies $p_{1} \neq p_{2}$, and let $<p_{1},p_{2}>_{n}$ denote the length-2 ordered permutation which satisfies $p_{1} < p_{2}$.  Letting ${<i_1,i_2>}_n$ be an arbitrary ordered permutation, Formula \eqref{eqn: determinant sum 1} can be applied to $M_{i,j}$, as:
\begin{equation}
|M_{<i_{1} ,i_{2}>_{n}[j_{1} ,j_{2}]_{n}}| \equiv \sum _{[p_{1} \dots p_{n}]_{n}}^{p_{i_{1}}=j_{1}, p_{i_{2}}=j_{2}} \$ [p_{1}\dots p_{n}]_{n} \prod _{k}^{k \ne i_{1}, k\ne i_{2}} x_{k,p_{k}};
\end{equation}
\begin{equation}
\label{eqn: determinant sum 2}
|M| = \sum _{j_{1}} x_{i_{1}, j_{1}} |M_{i_{1}, j_{1}}| = 
\sum _{j_{1}} \sum _{j_{2}}^{i_{2} \ne i_{1} , j_{2} \ne j_{1}} x_{i_{1}, j_{1}} x_{i_{2}, j_{2}} |M _{<i_{1}, i_{2}>_{n} [j_{1}, j_{2}]_{n}}|;
\end{equation}
Because $|M_{<i_{1} ,i_{2}>_{n}[j_{1} ,j_{2}]_{n}}|$ relates to the determinant of the $(n-2)$-by-$(n-2)$ matrix that results from deleting the row $i_{1}$ and $i_{2}$, and the column $j_{1}$ and $j_{2}$ of M.  This leads to Formula \eqref{eqn: sub-determinant equivalence}.
\begin{equation}
\label{eqn: sub-determinant equivalence}
||M_{{<i_{1},i_{2}>}_{n} {[j_{1},j_{2}]}_{n}}|| = ||M|_{{<i_{1},i_{2}>}_{n} {[j_{2},j_{1}]}_{n}}||  ;
\end{equation}
The definition of a sub-determinant can be extended to Formula \eqref{eqn: sub-determinant generic}, in which $m \in \{1,2...n\}$.  Formula \eqref{eqn: determinant sum 2} can be generalized as Formula \eqref{eqn: determinant sum}, in which $m \in \{1,2...n\}$ and $<i_{1} \dots i_{m}>_{n}$ is an arbitrary ordered permutation. Formula  \eqref{eqn: determinant sum} can be viewed as the extension for both Formula \eqref{eqn: determinant sum 1} and Formula \eqref{eqn: determinant}.
\begin{equation} 
\label{eqn: sub-determinant generic}
|M_{<i_{1} \dots i_{m}>_{n}[j_{1} \dots j_{m}]_{n}}| \equiv \sum _{[p_{1} \dots p_{n}]_{n}}^{p_{i_{k}}=j_{k}, k \in \{1 \dots m \}} \$ [p_{1}\dots p_{n}]_{n} \prod _{k \in \{1 \dots n \}}^{k \not\in \{i_{1} \dots i_{m}\}} x_{k,p_{k}};
\end{equation}
\begin{equation}\label{eqn: determinant sum}
|M| = \sum _{{[j_{1} \dots j_{m}]}_{n}}  
    |M _{<i_{1} \dots i_{m}>_{n} {[j_{1} \dots j_{m}]}_{n}}| 
    \prod _{k=1}^{m} x_{i_{k}, j_{k}};
\end{equation}

According to the basic assumption of precision arithmetic, the uncertainty of each element $x_{i,j}$ is independently and symmetrically distributed.  Let $ \widetilde{y}_{i,j}$ denote a random variable at the index $(i, j)$ symmetrically distributed with the deviation $\delta x_{i,j}$.  Let $\widetilde{|M|}$ denote the determinant of the matrix $\widetilde{M}$ whose element is $(x_{i,j} +  \widetilde{y}_{i,j})$.  Applying Taylor expansion to Formula \eqref{eqn: determinant sum} results in Formula \eqref{eqn: determinant Taylor expansion}, which results in Formula \eqref{eqn: determinant uncertainty expansion} after applying Formula \eqref{eqn: uncertainty variance}:
\begin{align}
\label{eqn: determinant Taylor expansion}
|\widetilde{M}| - |M| = &
\sum _{m=1}^{n} \sum _{<i_{1} \dots i_{m}>_{n}} \sum _{[j_{1} \dots j_{m}]_{n}}
  |M _{<i_{1} \dots i_{m}>_{n}{[j_{1} \dots j_{m}]}_{n}}| 
    \prod _{k=1}^{m}  \widetilde{y}_{i_{k}, j_{k}}; \\
\label{eqn: determinant uncertainty expansion}
{\delta |M|}^{2} = &
\sum _{m=1}^{n} \sum _{<i_{1} \dots i_{m}>_{n}} \sum _{[j_{1} \dots j_{m}]_{n}}
  |M _{<i_{1} \dots i_{m}>_{n}{[j_{1} \dots j_{m}]}_{n}}|^{2} 
    \prod _{k=1}^{m} \delta x_{i_{k}, j_{k}}^{2};
\end{align}
Defining $|M_{<>_{n} <>_{n}}| \equiv |M|$, Formula \eqref{eqn: determinant uncertainty recursion} is an recursive form of Formula \eqref{eqn: determinant uncertainty expansion}:
\begin{multline}
\label{eqn: determinant uncertainty recursion}
\delta |M_{<p_{1} \dots p_{k} >_{n} <q_{1} \dots q_{k} >_{n}}|^{2} = 
\sum _{p_{i}} \sum _{q_{j}} \delta x_{p_{i} ,q_{j}}^{2} \\ 
  (|M_{<p_{1} \dots p_{i} \dots p_{k}>_{n} <q_{1} \dots q_{j} \dots q_{k} >_{n}}|^{2} +  
   \delta |M_{<p_{1} \dots p_{i} \dots p_{k}>_{n} <q_{1} \dots q_{j} \dots q_{k} >_{n}}|^{2});
\end{multline}
When using Formula \eqref{eqn: determinant sum 1} to calculate determinant in conventional floating-point arithmetic:
\begin{itemize}
\item The input uncertainty can not be accounted for.
\item One path is chosen out of many possible paths, such as selecting a different sub-determinant to start with. 
\item Because of the rounding error, each path may result in a different result even if all elements of the determinant are precise, and the spread of all results is expected to be inversely proportional to the stability of the matrix \cite{Condition_Number}.  
\end{itemize}
In another word, using conventional floating-point arithmetic, the calculation of determinant is one leap of faith. Instead, Formula \eqref{eqn: determinant uncertainty recursion} shows that the result uncertainty is the aggregation of uncertainties from all possible path of Formula \eqref{eqn: determinant sum 1}.  To accounts for all such uncertainties, Formula \eqref{eqn: determinant uncertainty recursion} starts from all 1x1 sub-determinants, and constructs all sub-determinants whose size is 1 larger, until reaches the determinant itself.  Thus, uncertainty-bearing calculation should be order-of-magnitude more complex and time-consuming than the correspond calculation using conventional floating-point arithmetic.

The element $z_{i,j}$ at the index $(i,j)$ of the inverted matrix $M^{-1}$ is calculated as \cite{Linear_Algebra}:
\begin{equation}
\label{eqn: invert matrix}
z_{i,j} = \frac{|M_{j,i}|}{|M|};
\end{equation}
Formula \eqref{eqn: invert matrix} shows that the uncertainty of the matrix determinant $|M|$  propagates to every element of the inverted matrix $M^{-1}$.  Instead, the matrix which consists of the element $|M_{j,i}|$ at the index $(i, j)$ is defined as the adjugate matrix $M^{A}$ \cite{Linear_Algebra}, whose elements are not directly affected by $M^{-1}$.  $M^{A}$ is recommended to replace $M^{-1}$ whenever the application allows \cite{Numerical_Recipes}. 

\subsection{Matrix Testing Algorithm}

A matrix $\widehat{M}$ is constructed using random integers between [-16384, + 16384].  Its adjugate matrix $\widehat{M}^{A}$ and its determinant $|\widehat{M}|$ are calculated precisely using integer arithmetic.  $\widehat{M}$, $|\widehat{M}|$ and $\widehat{M}^{A}$ are all scaled proportionally as \textbf{$M$}, \textbf{$|M|$} and \textbf{$M^{A}$} so that the elements of \textbf{$M$} are 2's fractional numbers randomly distributed between [-1, +1].  The scaled matrix \textbf{$M$} is called a clean testing matrix.  \textbf{$M^{-1}$} is calculated from \textbf{$|M|$} and \textbf{$M^{A}$} using Formula \eqref{eqn: invert matrix}.  Floating-point arithmetic is used to calculate $M^{A}$ and $M^{-1}$ from M, and the results are compared with the corresponding precise results for value errors.  Gaussian noises corresponding to different deviations between $10^{-17}$ and $10^{-1}$ may be added to each clean testing matrix, to result in noisy testing matrix.  Each combination of matrix size and input deviation is tested by 32 different noisy matrices.

\subsection{Testing Matrix Stability}

Each matrix has a different stability \cite{Condition_Number}, which means how stable the inverted matrix is in regard to small value changes of the original matrix elements.  It is well known that more mutual cancellations in Formula \eqref{eqn: determinant} mean less stability of the matrix \cite{Arithmetic_Digital_Computers}\cite{Numerical_Recipes}, with the Hilbert matrix \cite{Hilbert_Matrix} being the most famous unstable matrix.  The condition number has been defined to quantify the stability of a matrix \cite{Condition_Number}.  Even though the definition of the condition number excludes the effects of rounding errors, in reality most calculations are done numerically using conventional floating-point arithmetic so that the combination effect of rounding errors and matrix instability cannot be avoided in practice.  When a matrix is unstable, the result is more error prone due to rounding errors of conventional floating-point arithmetic \cite{Arithmetic_Digital_Computers}.  Consequently, there are no general means to avoid the mysterious and nasty ``numerical instability'' in numerical applications due to rounding errors \cite{Arithmetic_Digital_Computers}.  For example, the numerical value of the calculated condition number of a matrix may have already been a victim of ``numerical instability'', and there is no sure way to judge this suspicion, so this value may not be very useful in judging the stability of the matrix in practice.  On the other hand, the rounding errors of conventional floating-point arithmetic can be used to test the stability of a matrix.  Rounding errors effectively change the item values of a matrix, so they produce a larger effect on a less stable matrix.  If the inverted matrix and the adjugate matrix are calculated using conventional floating-point arithmetic, larger value errors indicate that the matrix is less stable.

\begin{figure}
\centering
\includegraphics[width=4.5in,height=2.75in]{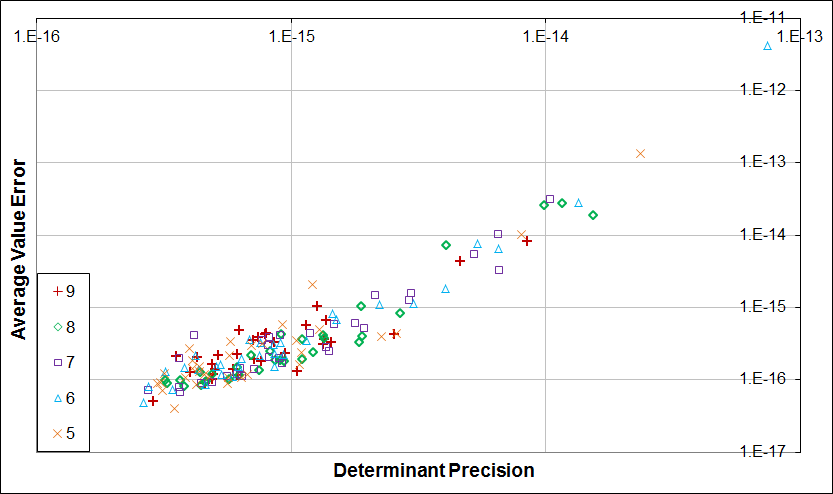} 
\captionof{figure}{ The empirical average value errors of the inverted matrix using conventional floating-point arithmetic vs. matrix determinant precision using precision arithmetic for clean matrices of different sizes (as shown in legend).}
\label{fig: Prec4_Inv_AvgErr_vs_DetSig_MatSize}
\end{figure}

\begin{figure}
\centering
\includegraphics[width=4.5in,height=2.75in]{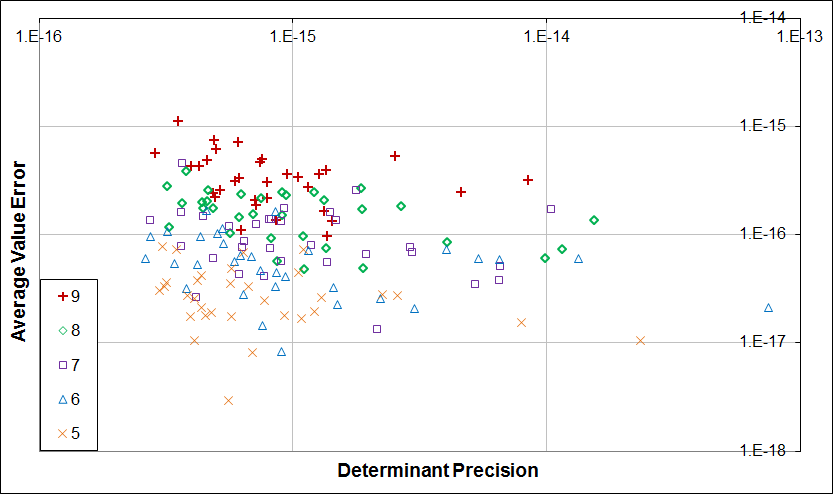} 
\captionof{figure}{ The empirical average value errors of the adjugate matrix using conventional floating-point arithmetic vs. matrix determinant precision using precision arithmetic for clean matrices of different sizes (as shown in legend).}
\label{fig: Prec4_Adj_AvgErr_vs_DetSig_MatSize}
\end{figure}

Precision arithmetic accounts for all rounding error with stable characterization of result uncertainties.  More mutual cancellations in Formula \eqref{eqn: determinant} will result in a smaller absolute value related to the uncertainty deviation of the determinant.  Thus, the precision of the determinant $|M|$ of a matrix $M$ calculated using precision arithmetic measures the amount of mutual cancellations, and it may measure the stability of a matrix.  Particularly, if $|M|$ is of coarser precision, then each element of $M^{-1}$ should tend to have a larger value error, according to Formula \eqref{eqn: invert matrix}.  This hypothesis is confirmed by Figure \ref{fig: Prec4_Inv_AvgErr_vs_DetSig_MatSize}, which shows a good linear relation between the precision of $|M|$ and the average value error of its inverted matrix $M^{-1}$, regardless of the matrix size.  The maximal output values errors are related to the precision of $|M|$ in the same fashion.  In contrast, Figure \ref{fig: Prec4_Adj_AvgErr_vs_DetSig_MatSize} shows that the value errors of the adjugate matrix $M^{A}$ do not depend noticeably on the precision of $|M|$.  Thus, the precision of the denominator in Formula \eqref{eqn: invert matrix} determines the overall stability in matrix inversion, confirming the validity of common advice to avoid matrix inversion operations in general \cite{Numerical_Recipes}.

\begin{figure}
\centering
\includegraphics[width=4.5in,height=2.75in]{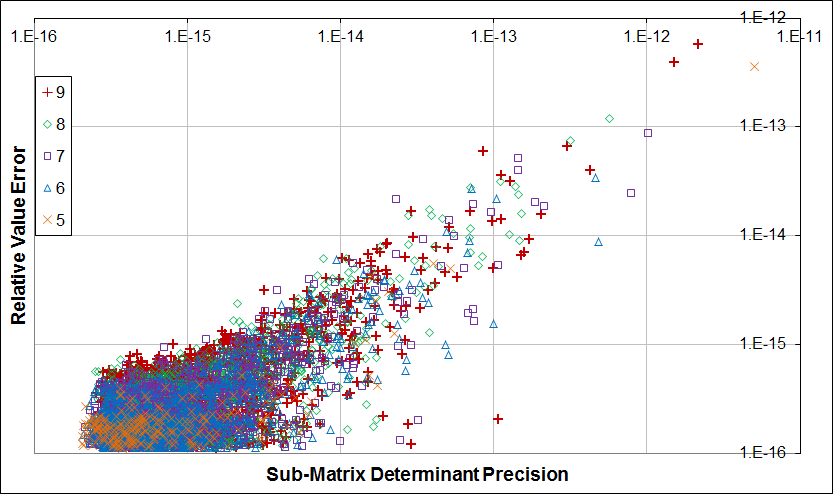} 
\captionof{figure}{ Empirical relative value errors of the adjugate matrix using conventional floating-point arithmetic vs. corresponding sub-matrix determinant precision using precision arithmetic for clean matrices of different sizes (as shown in legend).}
\label{fig: Prec4_Adj_RelErr_vs_DetSig_MatSize}
\end{figure}

Such a linear relation between the precision and the value error also extends to the calculation of the adjugate matrix.  Let the relative value error be defined as the ratio of the value error divided by the expected value.  The relative error is expected to correspond to the result precision linearly.  Figure \ref{fig: Prec4_Adj_RelErr_vs_DetSig_MatSize} compares each precision of the sub-matrix determinant $|M_{j,i}|$ with the corresponding relative error of the element at the index $(i, j)$ of the adjugate matrix $M^{A}$ of the clean matrix of different sizes.  It shows that larger relative errors of adjugate matrix elements indeed correspond to coarser precisions of the sub-matrix determinant.   

While each condition number \cite{Condition_Number} only gives the result sensitivity to one matrix element, Formula \eqref{eqn: determinant uncertainty expansion} contains the result sensitivity to any matrix element, any combination of matrix elements, as well as the aggregated result uncertainty deviation.  Therefore, Formula \eqref{eqn: determinant uncertainty expansion} and Formula \eqref{eqn: determinant uncertainty recursion} may be better than the condition numbers for describing matrix stability.

\subsection{Testing Uncertainty Propagation in Adjugate Matrix}

\begin{figure}
\centering
\includegraphics[height=3.0in]{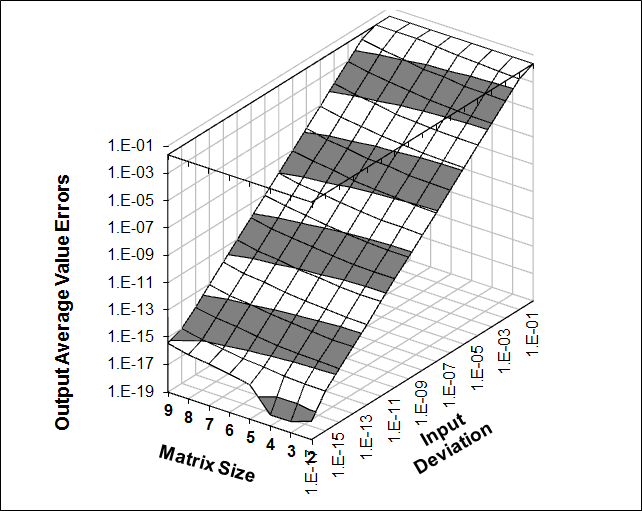} 
\captionof{figure}{Using precision arithmetic, the average output deviations of the adjugate
matrix vs. input precision and the matrix size.}
\label{fig: Prec4_AvgErr_vs_MatSize_InPrec}
\end{figure}

When the adjugate matrix is calculated using precision arithmetic, Figure \ref{fig: Prec4_AvgErr_vs_MatSize_InPrec} shows that the average output deviations for the adjugate matrix increase linearly with the input deviation, which is in good agreement with Formula \eqref{eqn: fitting error}.  Such relation is also true for maximal and average output values errors.  Formula \eqref{eqn: fitting error} is expected to describe the general value error propagation for linear algorithms in which $L$ is the amount of calculations \cite{Chaotic_Dynamics}.  The question is what value $L$ should be when calculating the adjugate matrix of a square matrix of size $N$.  Figure \ref{fig: Prec4_AvgErr_vs_MatSize_InPrec} suggests that $L$ increases with $N^{2}$ for the average output precision and average output error\footnote{The amount of calculation $L$ does not mean the calculation complexity using the Big O notation \cite{Big_O_Notation}.  It is just a measurement of how output uncertainty increases with a dimension of calculation according to \eqref{eqn: fitting error} \cite{Chaotic_Dynamics}. For example, any sorting algorithm will not change the uncertainty distribution, so that $L$ is always 0 regardless the calculation complexity for the sorting algorithm. The measured calculation time suggests calculation complexity of $O(2^N)$ for using Formula \eqref{eqn: determinant uncertainty recursion} to calculate the matrix determinant.}. 

\begin{figure}
\centering
\includegraphics[height=3.0in]{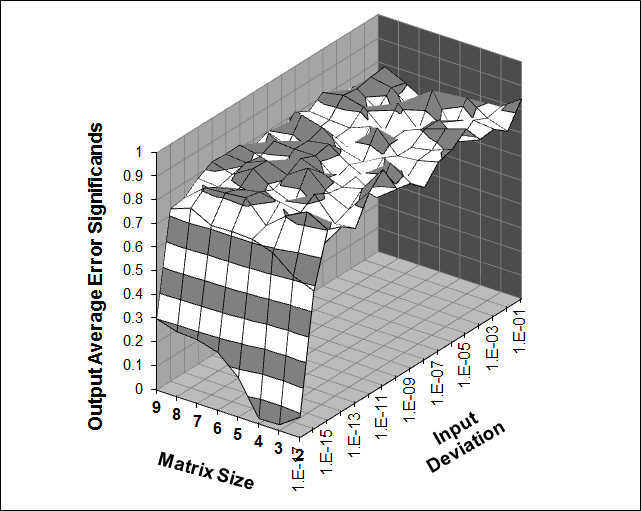} 
\captionof{figure}{ Using precision arithmetic, the average output tracking ratios of the adjugate matrix vs. input precision and the matrix size.}
\label{fig: Prec4_Inv_AvgSig_vs_MatSize_InPrec}
\end{figure}

Figure \ref{fig: Prec4_Inv_AvgSig_vs_MatSize_InPrec} shows that the average output tracking ratio of the adjugate matrix using precision arithmetic is approximately a constant of 0.8.  Figure \ref{fig: Prec4_Inv_AvgSig_vs_MatSize_InPrec} is very similar to Figure \ref{fig: Prec4_For_AvgSig_vs_FFTOrder_InDev}.  Similar to the maximal output bounding ratios of FFT algorithms, the maximal output bounding ratios for the adjugate matrix using precision also obey Formula \eqref{eqn: fitting significand} well, with $\beta$ of 1.005, meaning a slow increase with the matrix size.  Added to the similarity is the normalized uncertainty distribution shown in Figure \ref{fig: Prec4_Matrix9_NormDist}, which is very similar to Figure \ref{fig: Prec4_Sin_NormDist}.  Even though FFT and the calculating adjugate matrix are two very different sets of linear transformational algorithms, their uncertainty propagation characteristics are remarkably similar even in quantitative details.  This similarity indicates that precision arithmetic is a generic arithmetic for linear algorithms.

\subsection{Calibration}
\label{sec: calibration}

\begin{figure}
\centering
\includegraphics[width=4.5in,height=2.75in]{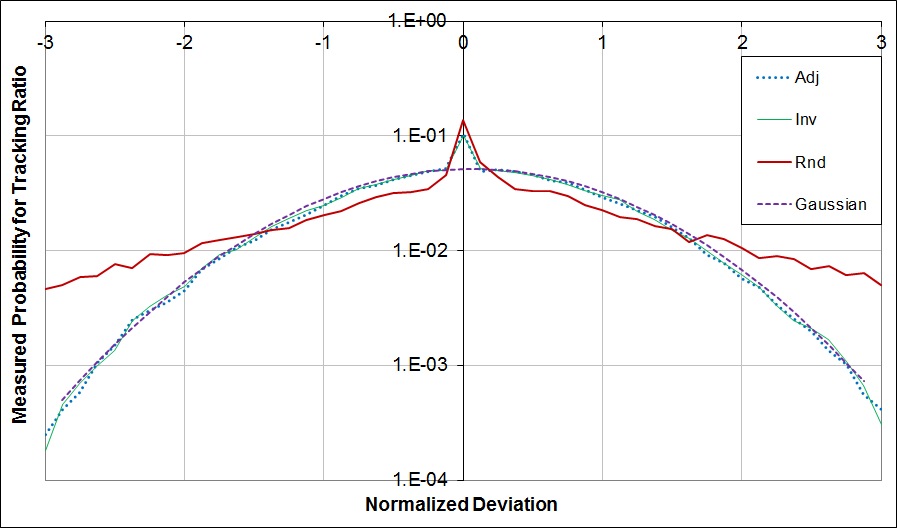} 
\captionof{figure}{ The measured tracking ratio distributions using precision arithmetic for matrix calculations of matrix size 9.  They are best fitted by a Gaussian distribution with the mean of 0.06 and deviation of 0.96.  In the legend, "Adj" means calculating adjugate $M^{A}$, "Inv" means calculating inverted $M^{-1}$, and "Rnd" means calculating double inverted ($M^{-1})^{-1}$. }
\label{fig: Prec4_Matrix9_NormDist}
\end{figure}

\begin{figure}
\centering
\includegraphics[height=3.0in]{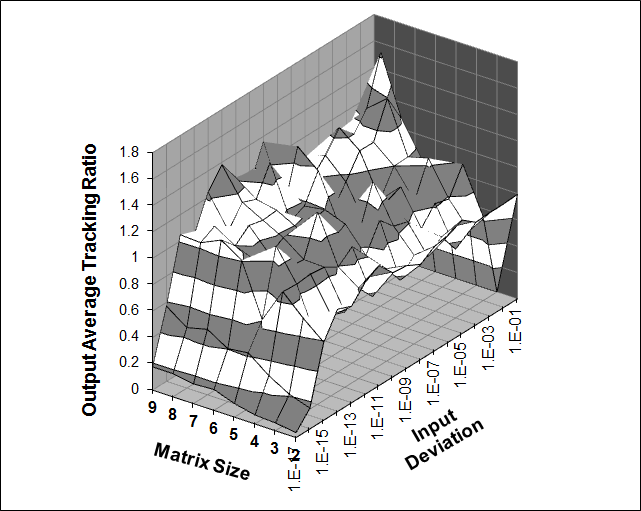} 
\captionof{figure}{ Using precision arithmetic, the average output tracking ratios of the double inverted matrix vs. input precision and the matrix size.}
\label{fig: Prec4_Rnd_AvgSig_vs_MatSize_InPrec}
\end{figure}

\begin{figure}
\centering
\includegraphics[width=4.5in,height=2.75in]{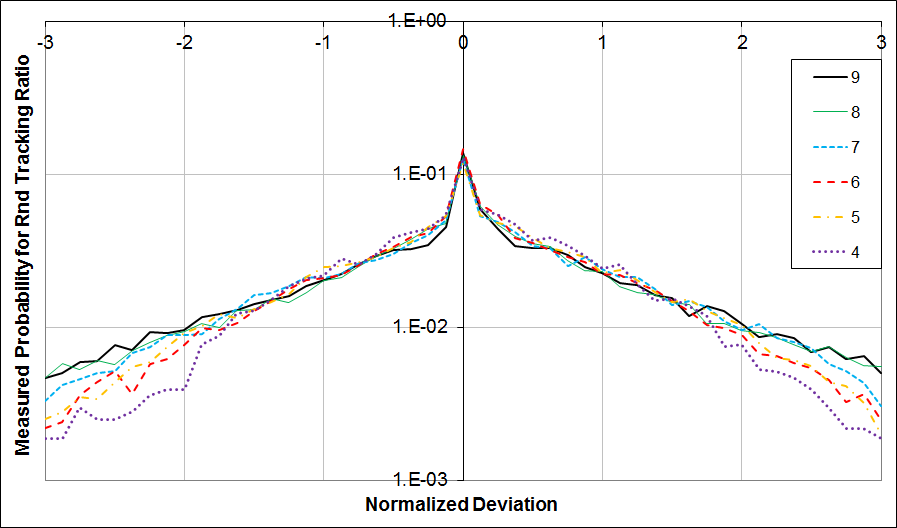} 
\captionof{figure}{ The measured tracking ratio distributions using precision arithmetic for matrix calculations of matrix size 9.  They are best fitted by a Gaussian distribution with the mean of 0.06 and deviation of 0.96.  In the legend, "Adj" means calculating adjugate $M^{A}$, "Inv" means calculating inverted $M^{-1}$, and "Rnd" means calculating double inverted ($M^{-1})^{-1}$. }
\label{fig: Prec4_Matrix_Rnd_NormDist}
\end{figure}

Because \textbf{$|M_{j,i}|$} and \textbf{$|M|$} are not independent of each other, \textbf{$M^{-1}$} calculated by Formula \eqref{eqn: invert matrix} contains the dependency problem. Figure \ref{fig: Prec4_Matrix9_NormDist} shows that the tracking ratios for the adjugate matrix and the inverted matrix are both standard distributed, while they are exponentially distributed when the inverted matrix is inverted again.  Because the inverted matrix has the same tracking ratio distribution as that of the adjugated matrix, which has no dependency problem, the inverted matrix contains hardly any dependency problem.  In contrast, Figure \ref{fig: Prec4_Matrix9_NormDist} shows that the double inverted matrix is severely affected by the dependency problem, such that its tracking ratio increases with matrix size as shown in Figure \ref{fig: Prec4_Rnd_AvgSig_vs_MatSize_InPrec}.  Figure \ref{fig: Prec4_Matrix_Rnd_NormDist} shows that average tracking ratios for different matrix sizes follows a same exponential distribution, but with different extend, e.g., the distribution for matrix size 4 has yet reaches stable distribution beyond 2.5, which causes the increase of the average tracking ratio with the matrix size as shown in Figure \ref{fig: Prec4_Rnd_AvgSig_vs_MatSize_InPrec}.

Applying the same algorithms twice results in so much differences, which shows that the dependency problem has been embedded in the data, and which shows the importance of calibration.

\clearpage
\section{Comparison Using Recursive Calculation of Sine Values}
\label{sec: recursion}

Starting from Formula \eqref{eqn: phase boundary}, Formula \eqref{eqn: phase sin} and Formula \eqref{eqn: phase cos} can be used recursively to calculate the phase array $\varphi[n]$ in Formula \eqref{eqn: FFT phase array}.  
\begin{align}
\label{eqn: phase boundary}
& \sin(0) = \cos(\frac{\pi}{2}) = 0; & \sin(\frac{\pi}{2}) = \cos(0) = 1; \\
\label{eqn: phase sin}
& \sin \left(\frac{\alpha + \beta}{2} \right) = \sqrt{\frac{1 - \cos \left(\alpha + \beta \right)}{2}} = & \sqrt{\frac{1 - \cos(\alpha) \cos \left(\beta) + \sin(\alpha \right) \sin(\beta)}{2}}; \\
\label{eqn: phase cos}
& \cos \left(\frac{\alpha + \beta}{2} \right) = \sqrt{\frac{1 + \cos \left(\alpha + \beta \right)}{2}} = & \sqrt{\frac{1 + \cos(\alpha) \cos(\beta) - \sin(\alpha) \sin(\beta)}{2}};
\end{align}

This algorithm is very different from both FFT and matrix inversion in nature because Formula \eqref{eqn: phase sin} and Formula \eqref{eqn: phase cos} are no longer linear, and the test presents a pure theoretical calculation without input uncertainty.  The recursion iteration count $L$ is a good measurement for the amount of calculations.  Each repeated use of Formula \eqref{eqn: phase sin} and Formula \eqref{eqn: phase cos} accumulates calculation errors to the next usage so that both value errors and uncertainty are expected to increase with $L$.  Each recursion iteration $L$ corresponds to $2^{L-2}$ outputs, which enables statistical analysis for large $L$.  

\begin{figure}
\centering
\includegraphics[width=4.5in,height=2.5in]{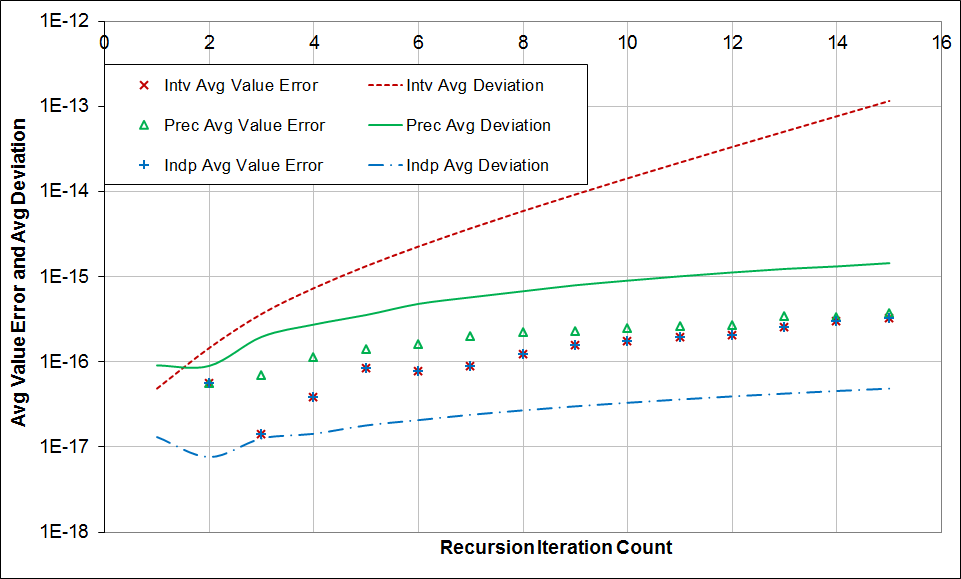} 
\captionof{figure}{ The empirical output average value errors and corresponding average output deviations vs. the recursion iteration count of the regressive calculation of sine values using interval arithmetic, precision arithmetic and independent arithmetic.  The x-axis indicates the recursion iteration count L, while the y-axis indicates either the average value errors or average uncertainty deviations.  In the legend, "Intv" means interval arithmetic, "Indp" means independence arithmetic, and "Prec'' means precision arithmetic.}
\label{fig: AvgValErr_vs_Regression}
\end{figure}

\begin{figure}
\centering
\includegraphics[width=4.5in,height=2.5in]{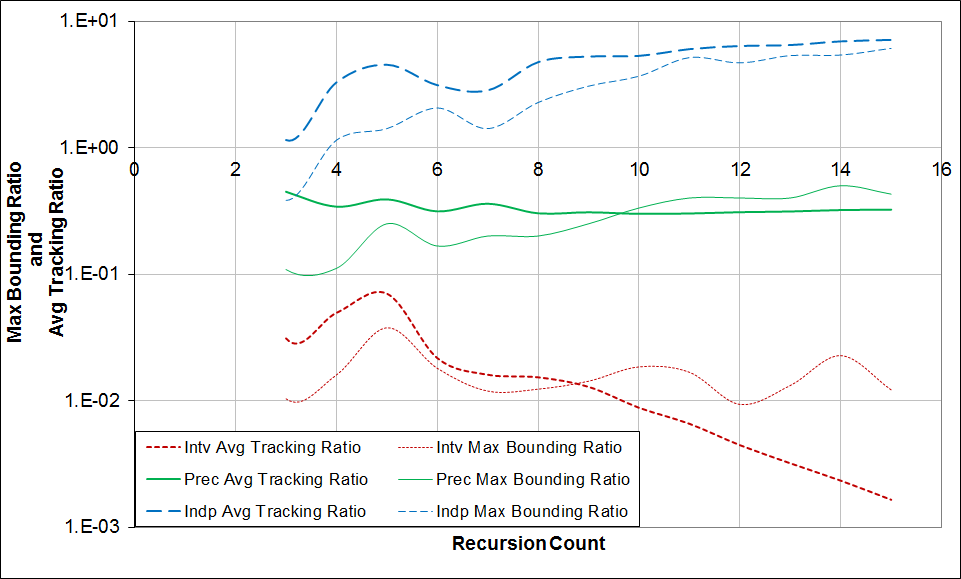} 
\captionof{figure}{ The empirical output maximal bounding ratios and average tracking ratios vs. the recursion iteration count of the regressive calculation of sine values using interval and precision arithmetic.  In the legend, "Intv" means interval arithmetic, "Indp" means independence arithmetic, and "Prec'' means precision arithmetic.}
\label{fig: AvgErrSig_MaxBndRat_vs_Regression}
\end{figure}

Figure \ref{fig: AvgValErr_vs_Regression} shows that both average output value errors and the corresponding average output deviation increase exponentially with the recursion count for all three arithmetics, and Figure \ref{fig: AvgErrSig_MaxBndRat_vs_Regression} shows that in response to the increased amount of calculations:
\begin{itemize}
\item The average tracking ratio for precision arithmetic is a constant about 0.25;

\item The maximal output bounding ratio for precision arithmetic increases slowly; 

\item The average tracking ratio for interval arithmetic decreases exponentially; and 

\item The maximal output bounding ratio for interval arithmetic remains roughly a constant.
\end{itemize}
Unlike FFT algorithms, the initial precise sine values participate in every stage of the recursion, which results in few small output deviations at each recursion.  Detailed inspection shows that the maximal output bounding ratios for interval arithmetic are all obtained from small output deviations, and bounding ratios using interval arithmetic in general decrease exponentially with the amount of calculations.  Thus, the result uncertainty propagation characteristics of the regressive calculation of sine values are very similar to those of both FFT and the calculating adjugate matrix; even though all these algorithms are quite different in nature.  This may indicate again that the stability of precision arithmetic is generic, regardless of the algorithms used.

\clearpage
\section{Validation Using Taylor Expansion}
\label{sec: taylor expansion}

When a Taylor expansion is implemented using conventional floating-point arithmetic, the rounding errors are ignored, so that the result of a higher order of expansion is assumed to be more precise, because the Cauchy estimator of the expansion, which gives an upper bound for the remainder of the expansion, decreases with the order of the expansion for analytic expressions.  A subjective upper limit is chosen for the Cauchy estimator, to stop the expansion at limited order \cite{Numerical_Recipes}.  However, such arbitrary upper limit may not be achievable with the amount of rounding errors accumulated during calculation, so that such upper limit may actually gives a false expansion precision.   

Using precision arithmetic, the rounding errors as well as the input uncertainties are all accounted for, so that the maximal expansion order when applying a Taylor expansion of Formula \eqref{eqn: Taylor 1d} or Formula \eqref{eqn: Taylor 2d}  is no longer subjective.  Formula \eqref{eqn: polynomial uncertainty} is decomposed into the contribution of each successive term for Tylor expansion, as Formula \eqref{eqn: polynomial uncertainty for Taylor expansion}:
\begin{multline}
\label{eqn: polynomial uncertainty for Taylor expansion}
(\delta \sum_{j=0}^{J+1} a_j x^j)^2 
= \int (\sum_{j=0}^{J} a_j (x + \widetilde{y})^j - \sum_{j=0}^{J} a_j x^j + a_{J+1} (x + \widetilde{y})^{J+1} - a_{J+1} x^{J+1})^2 \rho(\widetilde{y}) d \widetilde{y} \\
= \int (\sum_{j=0}^{J} a_j (x + \widetilde{y})^j - \sum_{j=0}^{J} a_j x^j)^2 \rho(\widetilde{y}) d \widetilde{y} +
    a_{J+1}^2 \int (\sum_{k=1}^{J+1} C_{J+1}^{k} \widetilde{y}^k x^{J+1-k})^2 \rho(\widetilde{y}) d \widetilde{y} \\
+2 \int (\sum_{j=0}^{J} a_j \sum_{k=1}^{j} 
   C_{j}^{k} \widetilde{y}^k x^{j-k})(a_{J+1} \sum_{k=1}^{J+1} C_{J+1}^{k} \widetilde{y}^k x^{J+1-k})
   \rho(\widetilde{y}) d \widetilde{y} \\
(\delta \sum_{j=0}^{J+1} a_j x^j)^2 - (\delta \sum_{j=0}^{J} a_j x^j)^2 = 
\sum_{k_1=1}^{J+1} \sum_{k_2=1}^{J+1} a_{J+1}^2 C_{J+1}^{k_1} C_{J+1}^{k_2} M(k_1+k_2) (\delta x)^{k_1+k_2} x^{2J+2-k_1-k_2} \\
+ 2 \sum_{k_1=1}^{J+1} \sum_{j=0}^{J} \sum_{k_2=1}^{j} a_j a_{J+1} C_{J+1}^{k_1} C_{j}^{k_2} 
   M(k_1+k_2) (\delta x)^{k_1+k_2} x^{J+1+j - k_1 -k_2};
\end{multline} 
Applying Formula \eqref{eqn: polynomial uncertainty for Taylor expansion} to Taylor expansion:
\begin{enumerate}
\item Formula \eqref{eqn: polynomial uncertainty for Taylor expansion} provides the deviation at $n$-th expansion order, which becomes stabilized when the \emph{delta deviation} at $n$-th expansion order (which is the contribution of the $n$-th expansion order to the deviation) is much less than the deviation at $n$-th expansion order. 

\item The \emph{resolution} of precision arithmetic is the deviation divided by $2^\chi$, in which $\chi$ is the constant bits calculated inside uncertainty.

\item The maximal expansion order of a Taylor expansion is reached when the Cauchy estimator is less than the resolution of precision arithmetic, after which the changes in Cauchy estimator is no longer detectable.  Ideally, the Taylor expansion reminder should also become zero when the expansion order is larger than the maximal expansion order.
\end{enumerate}

Formula \eqref{eqn: polynomial uncertainty for Taylor expansion} also shows that the deviation of Taylor expansion may decrease at certain expansion order.  For example, at $x = 1 \pm \delta x$, $1 - 2x + x^2$ is equivalent to $y^2$ at $y = 0 \pm \delta x$, thus it has smaller result variance than $1 - 2x$ at $x = 1 \pm \delta x$.  

Formula \eqref{eqn: Taylor expansion test} provides an example test in Taylor expansion, in which $n$ is a positive integer.  
\begin{equation}
\begin{split}
\label{eqn: Taylor expansion test}
f_n(x) = \sum _{j=0}^{n} (-x)^j; & \eqspace \lim _{n \to \infty} f_n(x) = 1/(1+x);
\end{split}
\end{equation}
In Formula \eqref{eqn: Taylor expansion test}, the absolute value of $(n+1)$th term in the expansion is the Cauchy remainder estimator of the $n$th order expansion.  Formula \eqref{eqn: Taylor expansion test} is analytic when $|x|$ is less than 1, and a smaller value $|x|$ means faster convergence to the correct value $1/(1 + x)$. 

\begin{figure}
\centering
\includegraphics[width=4.5in,height=2.5in]{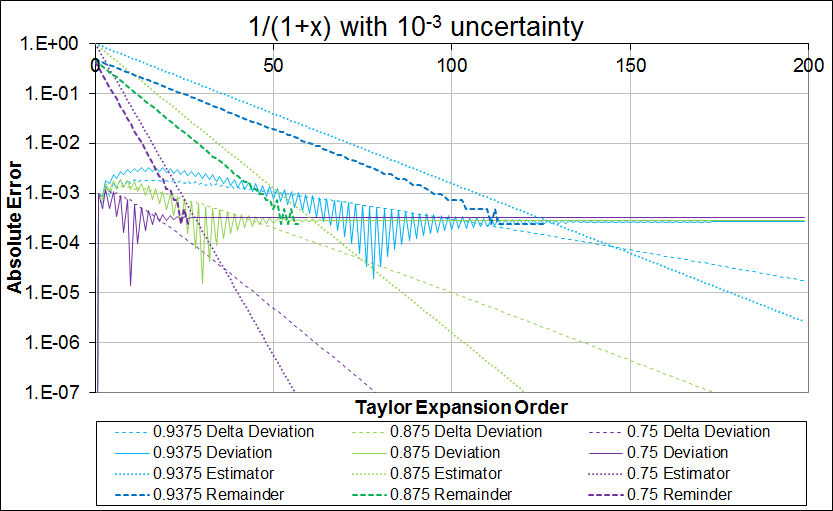} 
\captionof{figure}{ The delta deviation, deviations, Cauchy estimator and remainders of a Taylor expansion vs. the expansion orders for different input value with $10^{-3}$ input uncertainty using precision arithmetic with 0-bit calculated inside uncertainty. Different inputs are displayed using different color. }
\label{fig: Prec0_Taylor_1E-3}
\end{figure}

Using Formula \eqref{eqn: Taylor expansion test} as a test case, Figure \ref{fig: Prec0_Taylor_1E-3} confirms the above Taylor expansion process using precision arithmetic with 0-bit calculated inside uncertainty and with input uncertainty at $10^{-3}$.  For smaller $|x|$, in addition to faster decrease of both reminder and Cauchy estimator, delta deviation also decreases faster, thus deviation reaches its stable values faster.  Once the maximal expansion order is reached, the reminder also becomes to zero.  Figure \ref{fig: Prec0_Taylor_1E-3} repeats the above process with 4-bit calculated inside uncertainty, which only differs from Figure \ref{fig: Prec0_Taylor_1E-3} by having resolution smaller than deviation and larger maximal expansion order.

\begin{figure}
\centering
\includegraphics[width=4.5in,height=2.5in]{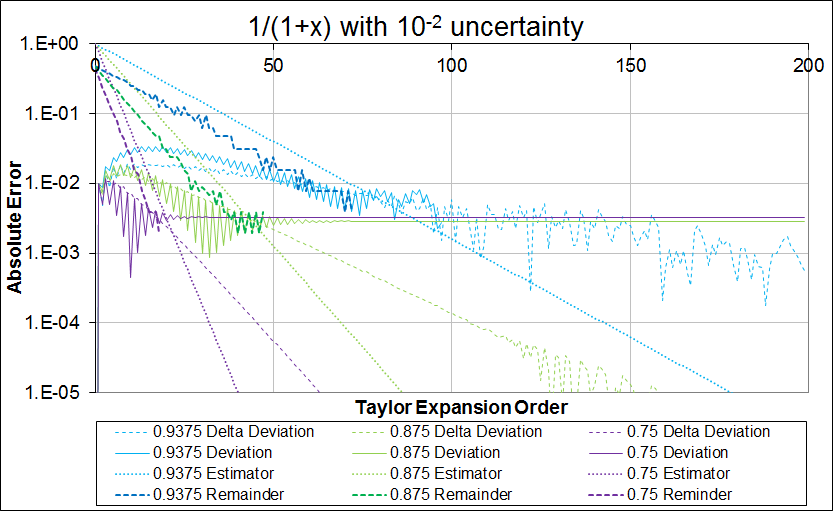} 
\captionof{figure}{ The delta deviation, deviations, Cauchy estimator and remainders of a Taylor expansion vs. the expansion orders for different input value with $10^{-2}$ input uncertainty using precision arithmetic with 0-bit calculated inside uncertainty. Different inputs are displayed using different color. }
\label{fig: Prec0_Taylor_1E-2}
\end{figure}

\begin{figure}
\centering
\includegraphics[width=4.5in,height=2.5in]{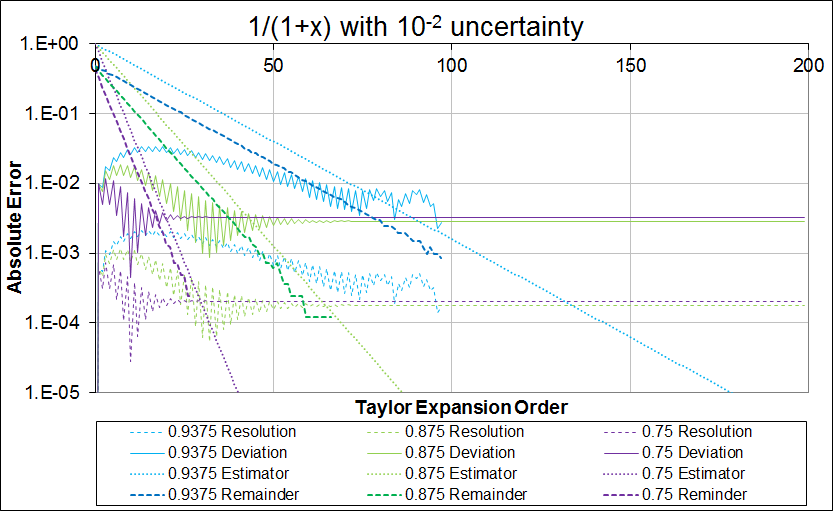} 
\captionof{figure}{ The deviations, resolutions, Cauchy estimator and remainders of a Taylor expansion vs. the expansion orders for different input value with $10^{-2}$ input uncertainty using precision arithmetic with 4-bit calculated inside uncertainty. Different inputs are displayed using different color. }
\label{fig: Prec4_Taylor_1E-2}
\end{figure}

When input has larger uncertainty, deviation reaches to its stable value much slower, which is show in Figure \ref{fig: Prec0_Taylor_1E-2} for 0-bit calculated inside uncertainty: 
\begin{itemize} 
\item When $x=0.75$, deviation barely reaches its stable value when the Cauchy estimator reaches resolution.
\item When $x=0.875$, deviation has not reaches its stable value when the Cauchy estimator reaches resolution, and reminder does not become zero at the maximal expansion order but a few orders beyond.
\item When $x=0.9375$, deviation has no stable value and becomes imaginative eventually. Nevertheless, reminder becomes zero beyond the maximal expansion order.
\end{itemize}
In contrast, with 4-bit calculated inside uncertainty as shown in Figure \ref{fig: Prec4_Taylor_1E-2}:
\begin{itemize} 
\item When $x=0.75$, the maximal expansion order is reached later when the resolution is stabilized.
\item When $x=0.875$, the maximal expansion order is reached later when the resolution is stabilized, however reminder still does not become zero at the maximal expansion order but a few orders beyond.
\item When $x=0.9375$, resolution has no stable value and becomes negative eventually, after which the precision representation becomes undefined.  Because Cauchy estimator never reaches resolution, the maximal expansion order is not defined either.
\end{itemize}
Judged from the above simple cases of Taylor expansion, calculating inside uncertainty brings no clear-cut benefit.

\clearpage
\section{Validation of Precision Arithmetic Using Numerical Integration}
\label{sec: integration}

In numerical integration over the variable $x$ using conventional floating-point arithmetic, a finer sampling of the function to be integrated $f(x)$ is associated with a better result \cite{Numerical_Recipes}, and it is assumed that $f(x)$ can be sampled at infinitive fine intervals of $x$.  In reality, floating-point arithmetic has limited significant bits, so that rounding errors will increase with finer sampling of $f(x)$.  However, such limitation of numerical integration due to rounding errors is seldom studied seriously. In this paper:
\begin{enumerate}
\item The function to be integrated is treated as a black-box function.
\item The numerical integration is carried out using the rectangular rule \cite{Numerical_Recipes}.
\item The residual error is estimated locally as the difference between using the rectangular rule and using the trapezoidal rule \cite{Numerical_Recipes}.
\item The sampling is localized using simplest depth-first binary-tree search algorithm.
\item The sampling stops when the residual error is no longer significant.
\end{enumerate}

Specifically, for each integration interval $[x_{start}, x_{end}]$, define:
\begin{align}
& x_{mid} \equiv (x_{start} + x_{end})/2; \\
& f_{err} \equiv (f(x_{start}) + f(x_{end}))/2 - f(x_{mid}); \\
\label{eqn: integration delta}
& f_{\Delta} \equiv f(x_{mid}) (x_{end} - x_{start});
\end{align}
If $f_{err}$ becomes insignificant, the interval $[x_{start}, x_{end}]$ is considered to be fine enough, and $f_{\Delta}$ is added to the total integration.  Otherwise, the search continues on the intervals $[x_{start}, x_{mid}]$ and $[x_{mid}, x_{end}]$, which is the next depth for searching.  This searching algorithm is very adaptive, with the local search depth depending only on how $f(x)$ changes locally.  However, such adaptation to the local change of $f(x)$ brings one weakness to this searching algorithm: when $f(0)=f'(0)=0$, the algorithm spends the majority of the execution time around $x=0$, searching in tiny intervals of great depth, and adding tiny significant values to the result each time.  This weakness is called zero trap here.  It cannot be removed by simply offsetting $f(x)$ by a constant because doing so will change the precision of each sampling of $f(x)$, and increase the output uncertainty deviation.  For a proof-of-principle demonstration, zero trap is avoided in this paper.

Formula \eqref{eqn: integration of power} provides an example test for the above simple algorithm, in which $n$ is a positive integer.  
\begin{equation}
\label{eqn: integration of power}
\frac{4^{n+1} - 10^{-6(n+1)}}{n+1} = \int _{10^{-6}}^{4} x^{n} dx;
\end{equation}
Table \ref{tab: numerical integration} shows that the result of numerical integration is very comparable to the expected value.  It shows that the above integration algorithm introduces no broadening of result uncertainty, so the above algorithm always selects optimal integration intervals when calculating the best possible result for a numerical integration.  Tests of integration using different polynomials with different integration ranges all confirm the above result.

\begin{table}[h]
\centering
\begin{tabular}{|c|c|c|c|} 
\hline 
Power n & Search Depth & $\delta \left( \int _{10^{-6} }^{4}x^{n} dx \right) $ & $\int _{10^{-6} }^{4}x^{n} dx -\frac{4^{n+1} -10^{-6(n+1)} }{n+1} $ \\ 
\hline 
2 & [25, 47] & 1.32x10${}^{-14}$ & -0.705x10${}^{-14}$ \\ 
\hline 
3 & [25, 47] & 2.52x10${}^{-14}$  & -1.42x10${}^{-14}$ \\ 
\hline 
4 & [26, 47] & 1.16x10${}^{-13}$ & -1.13x10${}^{-13}$ \\ 
\hline 
5 & [26, 48] & 5.08x10${}^{-13}$ & -6.82x10${}^{-13}$ \\ 
\hline 
6 & [26, 48] & 1.92x10${}^{-12}$ & -2.72x10${}^{-12}$ \\ 
\hline 
\end{tabular}
\captionof{table}{Uncertainty deviation and value error of numerical integration vs. expected results using precision arithmetic for different power function.  The search range is deepest near $10^{-6}$.}
\label{tab: numerical integration}
\end{table}

One thing worth noticing in Table \ref{tab: numerical integration} is that even though Formula \eqref{eqn: integration delta} consistently underestimates integration for each integration interval $[x_{start}, x_{end}]$, the final underestimation is quite small and comparable to the uncertainty deviation.  This example shows that the bias inside the uncertainty range has insignificant contribution to the final result using precision arithmetic.

\clearpage
\section{Comparison Using Progressive Moving-Window Linear Regression}
\label{sec: Comparison Using Progressive Moving-Window Linear Regression}

\subsection{Progressive Moving-Window Linear Regression Algorithm}

Formula \eqref{eqn: linear regression} gives the result of the least-square line-fit of $Y = \alpha + \beta X$ between two set of data ${Y_j}$ and ${X_j}$, in which $j$ is an integer index to identify $(X, Y)$ pairs in the sets \cite{Numerical_Recipes}.

\begin{equation}
\begin{split}
\label{eqn: linear regression}
& \alpha = \frac{\sum _{j} Y_{j} }{\sum _{j} 1}; \\
& \beta = \frac{\sum _{j} X_{j} Y_{j} \; \sum _{j} 1 - \sum _{j} X_{j} \; \sum _{j} Y_{j}}
    {\sum _{j} X_{j} X_{j} \; \sum _{j} 1 - \sum _{j} X_{j} \; \sum _{j} X_{j} };
\end{split}
\end{equation}

In many applications data set ${Y_j}$ is an input data stream collected with fixed rate in time, such as a data stream collected by an ADC (Analogue-to-Digital Converter) \cite{Electronics}.  ${Y_j}$ is called a time-series input, in which $j$ indicates time.  A moving window algorithm \cite{Numerical_Recipes} is performed in a small time-window around each $j$.  For each window of calculation, ${X_j}$ can be chosen to be integers in the range of $[-H, +H]$ in which $H$ is an integer constant specifying window’s half width so that $\sum _{j} X_{j} = 0$, to reduce \eqref{eqn: linear regression} into \eqref{eqn: time-series linear regression}:

\begin{equation}
\begin{split}
\label{eqn: time-series linear regression}
& \alpha _{j} = \alpha \; 2 H = \sum _{X=-H+1}^{H} Y_{j-H+X}; \\
& \beta _{j} = \beta \; \frac{H (H+1)(2H+1)}{3} = \sum _{X=-H}^{H} X Y_{j-H+X}; \\
\end{split}
\end{equation}

\begin{figure}
\centering
\includegraphics[height=2.5in]{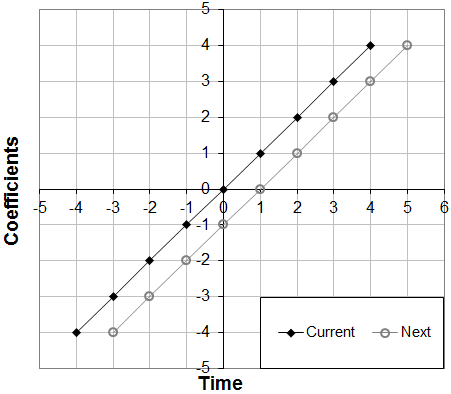} 
\captionof{figure}{ Coefficients of X in \eqref{eqn: time-series linear regression} at current and next position in a time series of the least square linear regression. Except the two end points at $X=-H$ and $X=H+1$, respectively, the coefficient difference between the current and then next position in a time series are all by 1 in the overlapping region from $X=-H+1$ to $X=H$, which results in \eqref{eqn: moving-window linear regression}. }
\label{fig: Moving_Window_Linear_Fit}
\end{figure}

According to Figure \ref{fig: Moving_Window_Linear_Fit}, in which $H$ takes an example value of 4, the calculation of $(\alpha _{j}, \beta _{j})$ can be obtained from the previous values of $(\alpha _{j-1}, \beta _{j-1})$, to reduce the calculation of \eqref{eqn: time-series linear regression} into a progressive moving-window calculation of \eqref{eqn: moving-window linear regression}:

\begin{equation}
\begin{split}
\label{eqn: moving-window linear regression}
& \beta _{j} = \beta _{j-1} - \alpha _{j-1} + H(Y_{j-2H-1} + Y_{j}); \\
& \alpha _{j} = \alpha _{j-1} - Y_{j-2H-1} + Y_{j};
\end{split}
\end{equation}

\subsection{Dependency Problem in a Progressive Algorithm} 

\eqref{eqn: moving-window linear regression} uses each input multiple times, so it will have dependency problem for all the three uncertainty-bearing arithmetic.  The question is how the overestimation of uncertainty evolves with time. 

\begin{figure}
\centering
\includegraphics[height=2.5in]{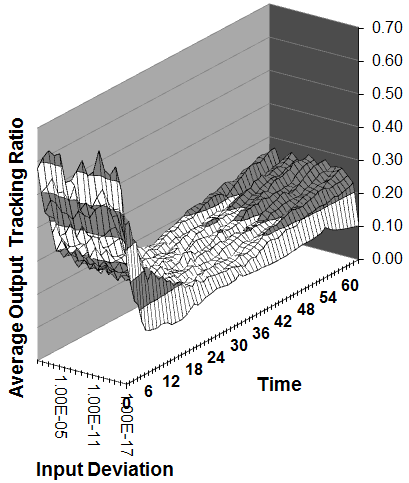} 
\captionof{figure}{ The average tracking ratio vs. time and the input uncertainty deviations for the progressive moving-window linear regression of a straight line using precision arithmetic with 4-bit calculated inside uncertainty.  }
\label{fig: Simple_Prec4_AvgErrSig_vs_InDev_Time}
\end{figure}

\begin{figure}
\centering
\includegraphics[height=2.5in]{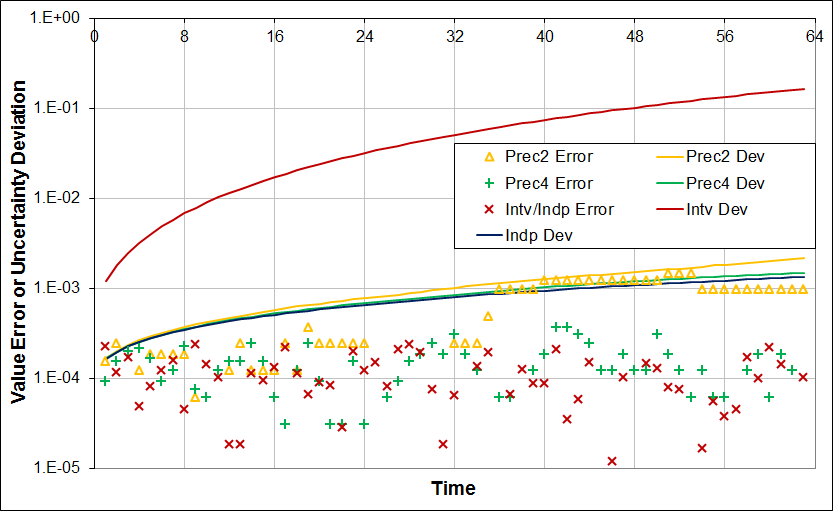} 
\captionof{figure}{ The output uncertainty deviations and the value errors vs. time for the progressive moving-window linear regression of a straight line.  In the legend,  "Indp" means independent arithmetic, "Intv" means interval arithmetic, "Prec4" and "Prec2" means the precision arithmetic with 4-bit and 2-bit calculated inside uncertainty, respectively. }
\label{fig: Simple_Err_Dev_vs_Time}
\end{figure}

\begin{figure}
\centering
\includegraphics[height=2.5in]{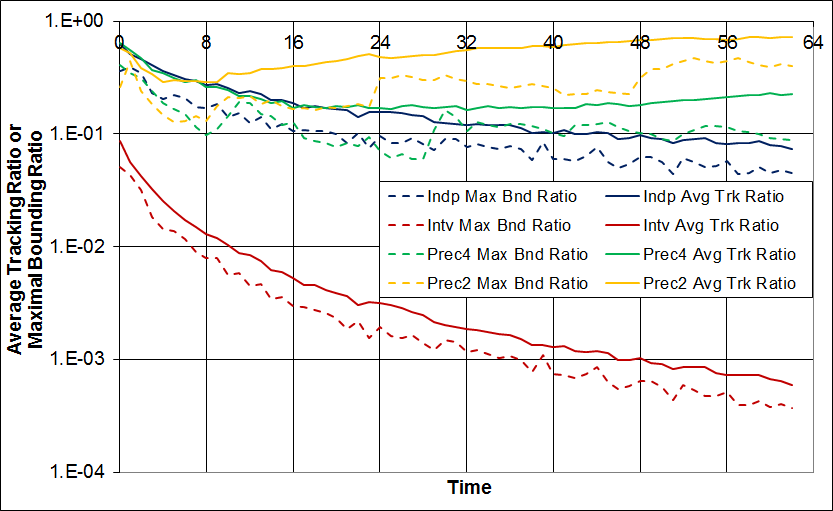} 
\captionof{figure}{ The average tracking ratios and the max bounding ratios vs. time for the progressive moving-window linear regression of a straight line.  In the legend, "Indp" means independence arithmetic, "Intv" means interval arithmetic, "Prec4" and "Prec2" means the precision arithmetic with 4-bit and 2-bit calculated inside uncertainty, respectively.  "Max Bnd Ratio" is the abbreviation for the maximal bounding ratio, and "Avg Trk Ratio" is the abbreviation for the average tracking ratios. }
\label{fig: Simple_AvgErrSig_MaxBndRat_vs_Time}
\end{figure}

The moving-window linear regression is done on a straight line with a constant slope of exactly $1/1024$ for each advance of time, with a full window width of 9 data points, or $H=4$.  Both average output value errors and deviations of all three arithmetic increases linearly with input deviations, and increase monotonically with time.  Thus both the average output tracking ratio and the maximal output bounding ratio are largely independent of input precisions, e.g., Figure \ref{fig: Simple_Prec4_AvgErrSig_vs_InDev_Time} shows such trend for the average output tracking ratio using precision arithmetic.  Such independence to input precision is expected for linear algorithms in general \cite{Numerical_Recipes}.  Therefore, only results with the input deviation of $10^{-3}$ are shown for the remaining discussions unless otherwise specified.  Figure \ref{fig: Simple_Err_Dev_vs_Time} shows the output deviation and the value errors vs. time while Figure \ref{fig: Simple_AvgErrSig_MaxBndRat_vs_Time} shows the output average tracking ratios and the maximal bounding ratios vs. time for all three arithmetics.  

For interval arithmetic and independence arithmetic, the output value errors remain on a constant level, while the output deviations increase with time, so that both output average tracking ratios and maximal bounding ratios decrease with time.  The stable linear increase of output deviation with time using either interval arithmetic or independence arithmetic in Figure \ref{fig: Simple_Err_Dev_vs_Time} suggests that the progressive linear regression calculation has accumulated every input uncertainty, which results in the monotonic decrease of both the maximal bounding ratios and the average output tracking ratios with time using both arithmetics in Figure \ref{fig: Simple_AvgErrSig_MaxBndRat_vs_Time}.

\begin{figure}
\centering
\includegraphics[height=2.5in]{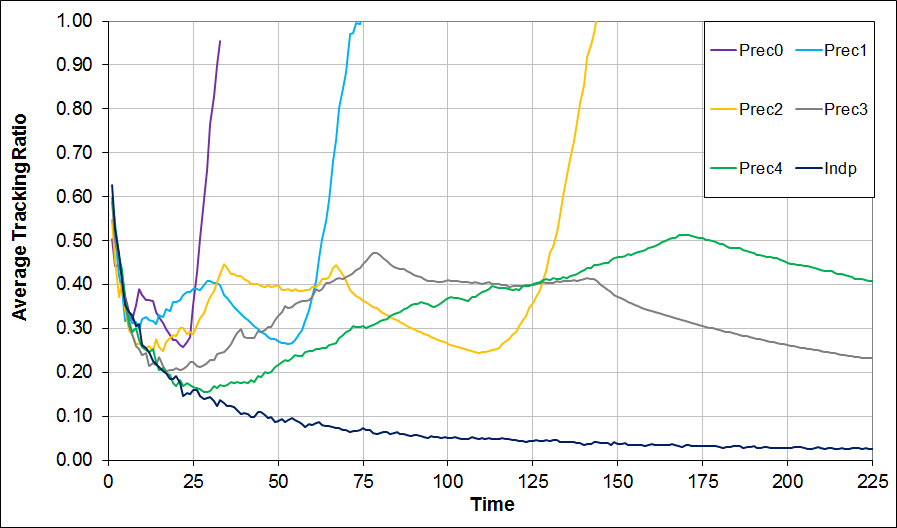} 
\captionof{figure}{ The average tracking ratios vs. time and the bits calculated inside uncertainty using precision arithmetic for the progressive moving-window linear regression of a straight line.  In the legend, "Indp" means independence arithmetic, "PrecX" means the precision arithmetic with X-bit calculated inside uncertainty. }
\label{fig: Simple_ErrSig_vs_Time_Prec}
\end{figure}

\begin{figure}
\centering
\includegraphics[height=2.5in]{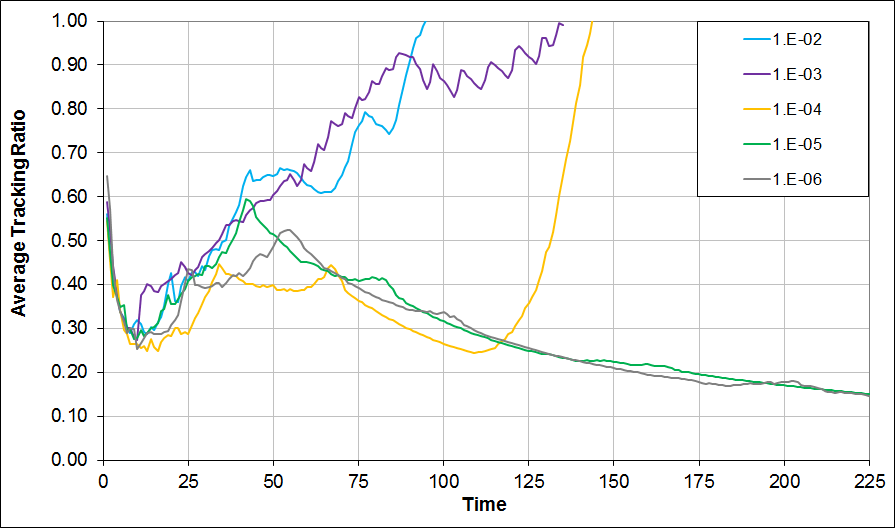} 
\captionof{figure}{ The average tracking ratios vs. time and the input precision using precision arithmetic with 2-bit calculated inside uncertainty for the progressive moving-window linear regression of a straight line for different input uncertainty deviations.  }
\label{fig: Simple_ErrSig_vs_Time_Prec2}
\end{figure}

In contrast, while precision arithmetic has slightly larger output deviations than those of independence arithmetic, its output value errors follows its output deviations, so that both its tracking ratios and bounding ratios remain between 0.1 and 0.9.  The reason for such increase of output value errors with time is due to the fact that precision arithmetic calculates only limited bits inside uncertainty, and uses larger granularity of values in calculation for larger uncertainty deviation. Such granularity of calculation is evident when comparing 2-bit or 4-bit calculation inside uncertainty using precision arithmetic in Figure \ref{fig: Simple_Err_Dev_vs_Time}.  This mechanism of error tracking in precision arithmetic is also demonstrated in Figure \ref{fig: Simple_ErrSig_vs_Time_Prec} and Figure \ref{fig: Simple_ErrSig_vs_Time_Prec2}.  Figure \ref{fig: Simple_ErrSig_vs_Time_Prec} shows that for fewer bits calculated inside uncertainty, the output value errors follow the output deviation closer in time, but such usage of larger granularity of values in calculation causes the result to become insignificant sooner, while for more bits calculated inside uncertainty, the average tracking ratios initially follow the result using independence arithmetic longer, and then follow the output deviation for longer duration.  The similarity in patterns of the average tracking ratios for different bits calculated inside uncertainty using precision arithmetic in Figure \ref{fig: Simple_ErrSig_vs_Time_Prec} suggests that they are all driven by a same mechanism but on different time scale, which is expected when smaller granularity of error needs more time to accumulate to a same level.  From the definition of tracking ratio, the granularity of error is actually measured in term of granularity of precision, e.g., Figure \ref{fig: Simple_ErrSig_vs_Time_Prec2} shows that for same bits calculated inside uncertainty, smaller input uncertainty deviations results in longer tracking of the output value errors to the output deviations.  The similar pattern of average tracking ratios is repeated on slower time scale for smaller input uncertainty deviations in Figure \ref{fig: Simple_ErrSig_vs_Time_Prec2}, revealing similar underline error-tracking mechanism in both cases.  Figure \ref{fig: Simple_ErrSig_vs_Time_Prec2} also shows that for the same bits calculated inside uncertainty, the average tracking ratios deviate from independence at exactly the same time.  Figure \ref{fig: Simple_ErrSig_vs_Time_Prec} and Figure \ref{fig: Simple_ErrSig_vs_Time_Prec2} thus demonstrate a uniform and granular error tracking mechanism of the precision arithmetic for different bits calculated inside uncertainty.  

\begin{figure}
\centering
\includegraphics[height=2.5in]{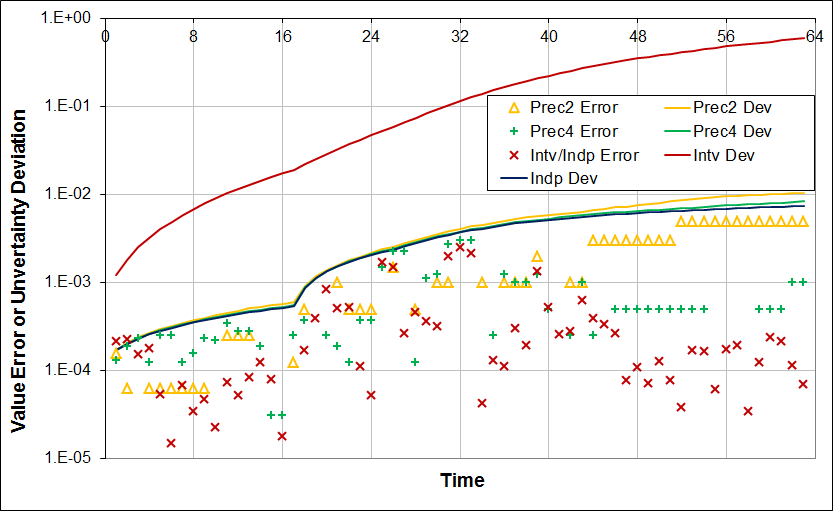} 
\captionof{figure}{ The output deviations and the value errors vs. time for the progressive moving-window linear regression of a straight line with 10-fold increase of both input noise and input uncertainty in the middle. }
\label{fig: Changed_Err_Dev_vs_Time}
\end{figure}

\begin{figure}
\centering
\includegraphics[height=2.5in]{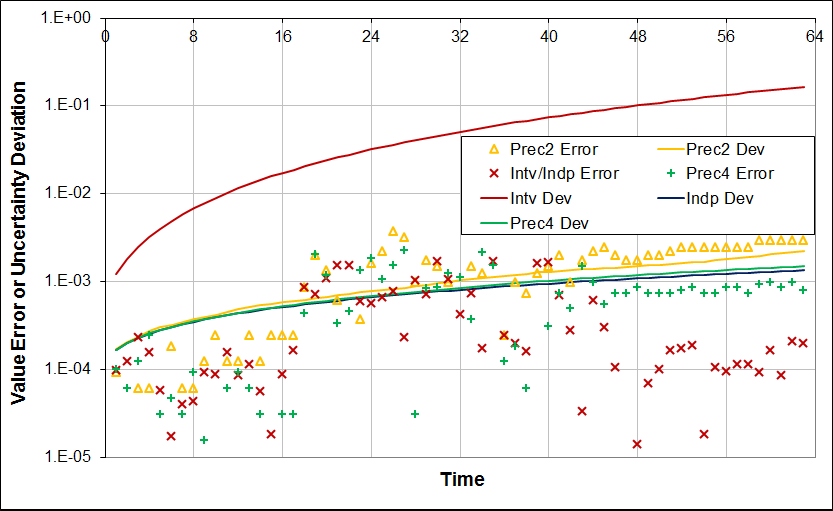} 
\captionof{figure}{ The output deviations and the value errors vs. time for the progressive moving-window linear regression of a straight line with only 10-fold increase of both input noise in the middle. }
\label{fig: Noiser_Err_Dev_vs_Time}
\end{figure}

\begin{figure}
\centering
\includegraphics[height=2.5in]{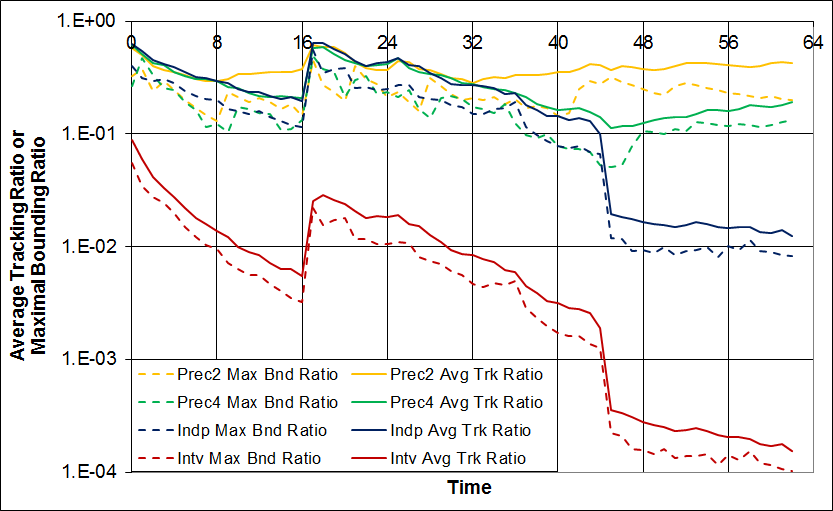} 
\captionof{figure}{ The average tracking ratios and the max bounding ratio vs. time for the progressive moving-window linear regression of a straight line with 10-fold larger input noise and deviation in the middle, to simulate larger noise following the straight line.  }
\label{fig: Changed_AvgErrSig_MaxBndRat_vs_Time}
\end{figure}

\begin{figure}
\centering
\includegraphics[height=2.5in]{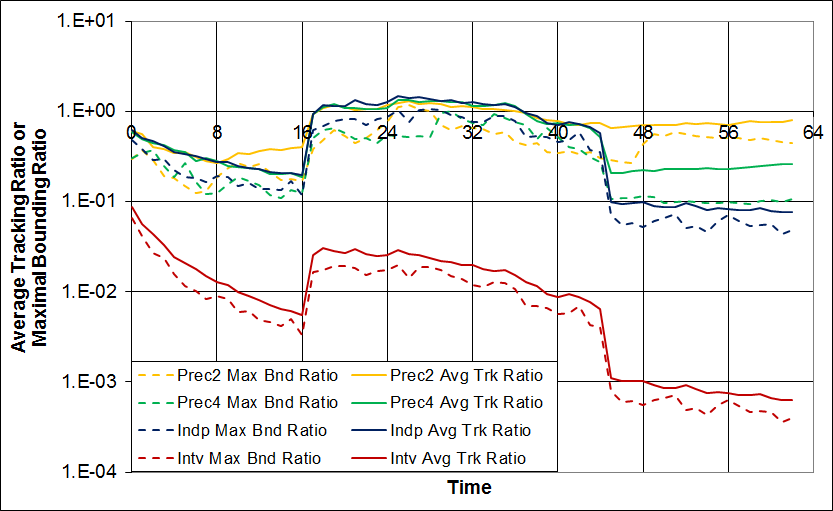} 
\captionof{figure}{ The average tracking ratios and the max bounding ratio vs. time for the progressive moving-window linear regression of a straight line with 10-fold larger input noise but same input deviation in middle, to simulate defects in obtaining the corresponding uncertainty deviations. }
\label{fig: Noiser_AvgErrSig_MaxBndRat_vs_Time}
\end{figure}

Is such increase of the value errors with the increase of uncertainty deviation using precision arithmetic desired?  First, in real calculations the correct answer is not known, and the reliability of a result depends statistically on the uncertainty of the result, so that there is no reason to assume that calculating more bits inside uncertainty is any better.  Conceptually, when the uncertainty of a calculation increases, the value error of the calculation is also expected to increase, which agrees with the trend shown by precision arithmetic.  Second, the stability of the average output tracking ratios and the maximal bounding ratios of precision arithmetic is quite valuable in interpretation results.  For example, even the output deviation may have unexpectedly changed, as in this case if dependency problem were not known and expected, such stability still gives a good estimation of the value errors in the result using precision arithmetic.  Third, such stability ensures that the result of algorithm at each window does not depend strongly on the usage history of the algorithm, which makes precision arithmetic the only practically usable uncertainty-bearing arithmetic for this progressive algorithm.  To test the effect of usage history on each uncertainty-bearing arithmetic, noise is increased by 10-fold at the middle 1/3 duration of the straight line, to result in additional two test cases:
\begin{itemize}
\item \emph{Changed}: In Figure \ref{fig: Changed_Err_Dev_vs_Time} and \ref{fig: Changed_AvgErrSig_MaxBndRat_vs_Time}, the input deviation is also increased by 10-fold to simulate an increase in measured uncertainty.
\item \emph{Defective}: In Figure \ref{fig: Noiser_Err_Dev_vs_Time} and \ref{fig: Noiser_AvgErrSig_MaxBndRat_vs_Time}, the input deviation remains the same to simulate the defect in obtaining the uncertainty deviations.  
\end{itemize}  
Accordingly, the original case of linear regression on a line with fixed slope is named as \emph{Simple}.

The question is how each uncertainty-bearing arithmetic responses to this change of data in the last 1/3 duration of calculation.  Using either independence or interval arithmetic, both the average output tracking ratios and the maximal output bounding ratios are decrease by about 10-fold in Figure \ref{fig: Changed_AvgErrSig_MaxBndRat_vs_Time} while they are not affected at all in Figure \ref{fig: Noiser_AvgErrSig_MaxBndRat_vs_Time}.  They show extreme sensitivity to the usage history.  Because the real input data are neither controllable nor predictable, the result uncertainty for this progressive algorithm using either interval arithmetic or independence arithmetic may no longer be interpretable.  In contrast, using precision arithmetic, both the average output tracking ratios and the maximal output bounding ratios are relatively stable, while the output deviations and value errors are sensitivity to usage history, so that the result using precision arithmetic is still interpretable.

\subsection{ Choosing a Better Algorithm for Imprecise Inputs }

\begin{figure}
\centering
\includegraphics[height=2.5in]{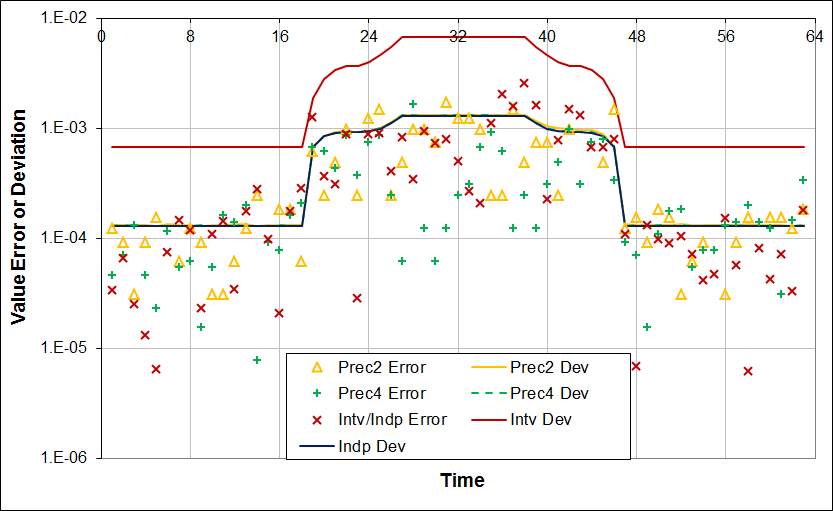} 
\captionof{figure}{ The output deviations and the value errors vs. time for the expressive moving-window linear regression of a straight line with 10-fold increase of both input noise and input uncertainty in the middle using precision arithmetic with 4-bit calculated inside uncertainty. }
\label{fig: ChangedDirect_Err_Dev_vs_Time}
\end{figure}

Formula \eqref{eqn: moving-window linear regression} has much less calculations than Formula \eqref{eqn: time-series linear regression}, and it seems a highly efficient and optimized algorithm according to conventional criterion \cite{Numerical_Recipes}.  However, from the perspective of uncertainty-bearing arithmetic, Formula \eqref{eqn: moving-window linear regression} is progressive while Formula \eqref{eqn: time-series linear regression} is expressive, so that Formula \eqref{eqn: time-series linear regression} should be better.  Figure \ref{fig: Changed_Err_Dev_vs_Time} and Figure \ref{fig: ChangedDirect_Err_Dev_vs_Time} respectively show the output deviations and the value errors vs. time for using either Formula \eqref{eqn: moving-window linear regression} or Formula \eqref{eqn: time-series linear regression} of a straight line with 10-fold increase of input uncertainty in the middle 1/3 duration.  They show that while the progressive algorithm carries all the historical calculation uncertainty into future, the expressive algorithm is clean from any previous results.  For example, at the last 1/3 duration when the moving window is already out of the area for the larger input uncertainty, the progressive algorithm still gives large result uncertainty, while the expressive algorithm gives output result only relevant to the input uncertainty within the moving window.  So instead of Formula \eqref{eqn: moving-window linear regression}, Formula \eqref{eqn: time-series linear regression} is confirmed to be a better solution for this linear regression problem.  

\subsection{ Modelling Dependency Problem }

\begin{figure}
\centering
\includegraphics[height=2.5in]{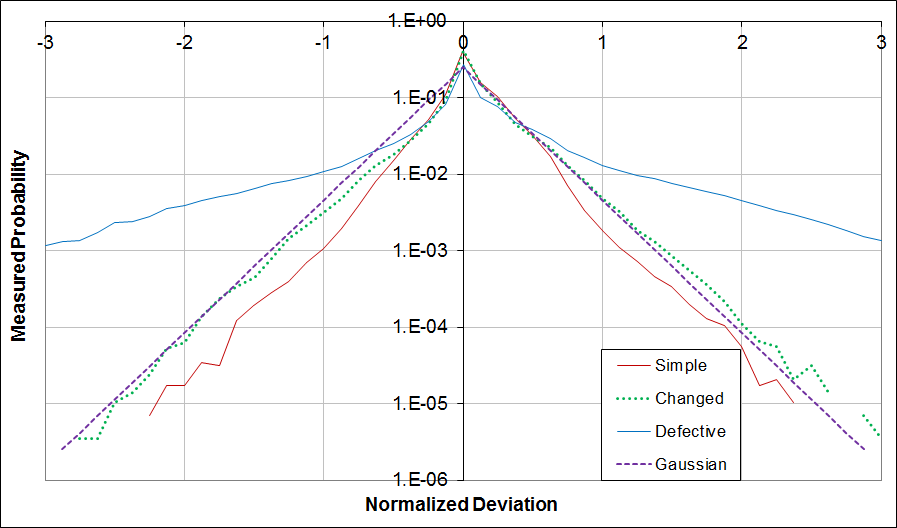} 
\captionof{figure}{ The measured tracking ratio distributions of the progressive moving-window linear regression for different cases (as shown in legend) using precision arithmetic with 4-bit calculated inside uncertainty.  The case of "Changed" is best fitted by a exponential distribution with the mean of 0 and deviation of 0.25. }
\label{fig: Prec4_LineFit_NormDist}
\end{figure}

\begin{figure}
\centering
\includegraphics[height=2.5in]{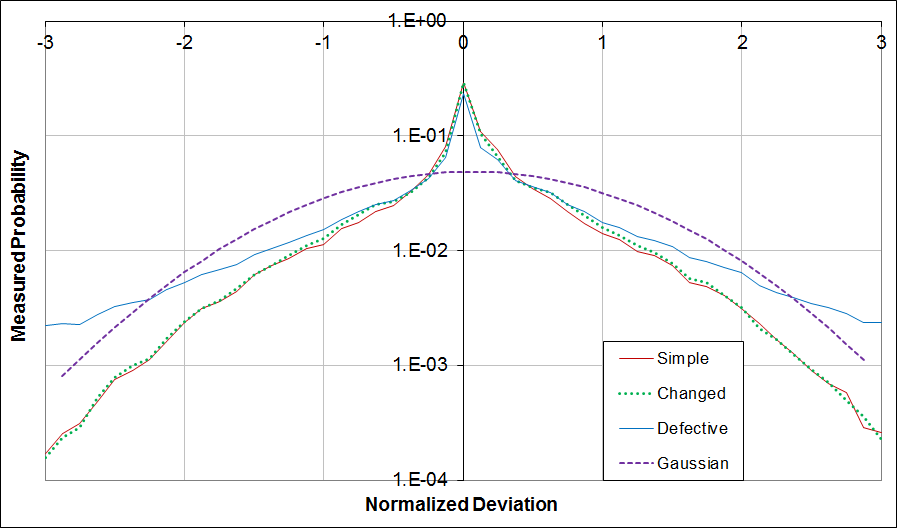} 
\captionof{figure}{ The measured tracking ratio distributions of the expressive moving-window linear regression using Formula \eqref{eqn: moving-window linear regression} for different cases (as shown in legend) using precision arithmetic with 4-bit calculated inside uncertainty.  The cases of "Simple" and "Changed" are best fitted by a Gaussian distribution with the mean of 0.06 and deviation of 0.97.  }
\label{fig: Prec4_LineFitDirect_NormDist}
\end{figure}

However, the majority algorithms used today are progressive.  Most practical problems are not even mathematical and analytical in nature, so that they may have no expressive solutions.  Expressive algorithms are simply just not always avoidable in practice.  With known expressive counterpart, the progressive moving-window linear regression algorithm can serve as a model for studying progressive algorithms.  For example:
\begin{itemize}
\item The progressive moving-window linear regression shows that the dependency problem of independence and interval arithmetic can manifest as dependency on the usage history of an algorithm.  Because of its stability, precision arithmetic should be used generally in progressive algorithms.  
\item Figure \ref{fig: Prec4_LineFit_NormDist} shows that the result tracking ratios of the progressive linear regression is exponentially distributed, while Figure \ref{fig: Prec4_LineFitDirect_NormDist} shows that the result tracking ratios of the expressive linear regression is Gaussian distributed only when the uncertainty deviation is characterized correctly, e.g., the result is Gaussian distributed for the "Changed" case but not for the "noisier" case.  Thus, the exponentially distributed tracking ratios does not necessarily imply dependency problem.
\end{itemize}

\clearpage
\section{Conclusion and Discussion}
\label{sec: conclusion and discussion}

\subsection{Summary}

The starting point of precision arithmetic is the uncorrelated uncertainty assumption, which requires input data to have decent precision for each or small overall correlation among them, as shown in Figure \ref{fig: Independent_Uncertainty_Assumption}, which quantifies the statistical requirements for input data to precision arithmetic.  In addition, it requires that the systematic errors is not the major source of uncertainty, and all of its input data do not have confused identities.    

Due to the uncorrelated uncertainty assumption and central limit theorem, the rounding errors of precision arithmetic are shown to be bounded by a Gaussian distribution with a truncated range.  The rounding error distribution is extended to describe the uncertainty distribution in general, with the uncertainty deviation of a single precision value given by Formula \eqref{eqn: uncertainty distribution}, and the result uncertainty deviation of a function given by Formula \eqref{eqn: uncertainty 1d} and its multi-dimension extensions such as Formula \eqref{eqn: uncertainty 2d}.  

Formula \eqref{eqn: fitting error} is shown to describe the general uncertainty deviation propagation in precision arithmetic.  The average tracking ratios and the maximal bounding ratio using precision arithmetic are shown to be independent of input precision, and stable for the amount of calculations for a few very different applications.  In contrast, both average tracking ratios and the maximal bounding ratio using interval arithmetic are shown to decrease exponentially with the amount of calculations in all tests.  Such stability is the major reason why precision arithmetic is better than interval arithmetic in all tests done so far.  

The statistical nature of precision arithmetic provides not only quantitative explanation for the dependency problem, but also solutions to the dependency problem, which is in form of either Taylor expansion or calibration.  The treatment of dependency problem is another major advantage of precision arithmetic over interval arithmetic.

Statistical precision has a central role in precision arithmetic:
\begin{itemize}
\item Precision is regarded as information content of a uncertainty-bearing value, which is in par with information entropy in information theory.  Because of this, precision needs to be preserved when the uncertainty-bearing value is multiplied or divided by a constant, which results in the scaling principle.   
\item Precision arithmetic itself can be deduced from the scaling principle and the uncorrelated uncertainty assumption.
\item The convergence property of the result deviation using Taylor expansion method is determined by input precisions, such as for inversions and square roots.
\end{itemize}

\subsection{Efficiency of Precision Arithmetic}

Precision arithmetic tries to solve a different mathematical problem from conventional floating-point arithmetic. For example, to calculate the determinant of a matrix:
\begin{itemize}
\item Conventional floating-point arithmetic may use a Laplace method \cite{Numerical_Recipes}, namely, to randomly choose a row or a column, and then to sum up the products of each element within the chosen row or the column with the corresponding sub-determinant of the element. Each sub-determinant is calculated in the same fashion.  Depending on the choices of the row or the column in each stage, there are many paths to calculate the determinant of a matrix.  Because conventional floating-point arithmetic has unbounded rounding errors, each path may give a different result, and the spread of all the results depends on the stability of the matrix and each sub-matrix \cite{Condition_Number}.  In this perspective, by taking a random path and assuming to get the only correct result, conventional floating-point arithmetic can be viewed as a leap-of-faith approach.

\item In contrast, precision arithmetic also needs to calculate the spread of the result due to rounding error or input uncertainties, so it effectively has to cover all paths of the calculation.  For example, using Formula \eqref{eqn: determinant uncertainty recursion}, precision arithmetic starts from each elements of the matrix, and treat it as a 1x1 sub-determinant, then grow it to all possible 2x2 sub-determinants containing it, etc, until reach the determinant of the matrix.  Thus, precision arithmetic takes order-of-magnitude more time than a single leap-of-faith calculation.  
\end{itemize}   

However, it is wrong to conclude that precision arithmetic is less efficient than conventional floating-point arithmetic, because in most cases rounding errors and input uncertainty can not be ignored.  Because conventional floating-point arithmetic can not contain uncertainty in its value, it has to use another value to specify uncertainty, such as an interval of $[min, max]$ or a common statistical pair $value \pm deviation$, which may brings the following drawbacks:
\begin{itemize}
\item The most common way to calculate result spread using conventional floating-point arithmetic is sampling \cite{Stochastic_Arithmetic} \cite{Numerical_Recipes}.  Assuming the matrix size is $N \times N$, and a minimal 3-point sampling is performed on each matrix element, then the spread calculation of matrix determinant requires $N^6$ leap-of-faith calculations, which is still a lot.  In contrast, using Formula \eqref{eqn: determinant uncertainty recursion}, precision arithmetic only need one calculation.  Thus, conventional floating-point arithmetic may be less efficient than precision arithmetic in this context. 

\item During to unbounded rounding errors, a conventional floating-point value losses its precision gradually and silently, so that a interval or a statistical pair itself can become unknowingly invalid.  At least, it is not clear at what precision the interval or the statistical pair specifies.
\end{itemize}

\subsection{Choose a better algorithm}

Because precision arithmetic tries to solve a different problem than conventional floating-point arithmetic, it has completely different criterion when choosing algorithms or implementations of algorithms.  For example, for matrix inversion, because conventional floating-point arithmetic has unbounded rounding errors, it will choose certain flavour of LU-decomposition over Gaussian elimination and determinant division \cite{Numerical_Recipes}. The result difference of LU-decomposition, Gaussian elimination and determinant division shows that conventional floating-point arithmetic has strong dependency problem, which has been a way of life when using conventional floating-point arithmetic, e.g., different algorithms or different implementation of the same algorithm are expected to give different results, of which a best algorithm or implementation is always chosen for each usage context \cite{Numerical_Recipes}, even though they may be mathematically equivalent. In contrast, rounding errors are bounded in both precision arithmetic and interval arithmetic \cite{Worst_Case_Error_Bounds}, so they are no longer needed to be considered.  When interval arithmetic reformat a numerical question as "Given each input to be an interval, what the output would be?", it effectively states that the results for most numerical questions to be solved should be \emph{unique} to be either one or a few intervals that tightest bounds the results, \emph{regardless} of the algorithm to be used, \emph{unless} dependency problem is introduced in the implementation of an algorithm.  Same concept is true for precision arithmetic, which converges all input uncertainty distribution to \emph{ubiquitously Gaussian} at the outputs, and which further \emph{quantifies} the source of the dependency problem.  Using precision arithmetic instead of conventional floating-point arithmetic, the focus has shifted from minimizing rounding errors to minimizing dependency problem.  Of the three algorithms for matrix inversion, both LU-decomposition and Gaussian elimination are progressive, which means that each input may appear multiple times in different branch at different time, whose dependency problem is difficult to quantified.  On the other hand, a determinant of a $N \times N$ matrix can be treated as a $N$-order polynomial with $N^2$ variables, to be readily for the Taylor expansion, which results in Formula \eqref{eqn: determinant uncertainty recursion}, so that the determinant division method is chosen in this paper for matrix inversion.  For the same reason, in the moving-window linear regression, the worse method in conventional floating-point arithmetic, Formula \eqref{eqn: time-series linear regression}, becomes the better method in precision arithmetic, and vice versa.  

Due to the requirement of minimizing dependence problem, precision arithmetic has much less operational freedom than conventional arithmetic and may require extensive symbolic calculations, following practices in affine arithmetic \cite{Symbolic_Affine_Arithmetic}.  Also, the comparison relation in conventional arithmetic needs to be re-evaluated in precision arithmetic, which brings about another reason for different algorithm selection.

\subsection{Improving Precision Arithmetic}

Figure \ref{fig: Independent_Uncertainty_Assumption} uses a cut-off for the test of the uncorrelated uncertainty assumption among two uncertainty-bearing values.  A better approach is to associate the amount of the dependence problem with the amount of correlation between the uncertainties of the two values.  

There are actually three different ways to round up $(2S+1)?4R\; 2^E$:
\begin{enumerate}
\item always round up $(2S+1)?4R\; 2^E$ to $(S+1)-R\; 2^{E+1}$;
\item always round up $(2S+1)?4R\; 2^E$ to $S+R\; 2^{E+1}$;
\item randomly round up $(2S+1)?4R\; 2^E$ to either $(S+1)-R\; 2^{E+1}$ or $S+R\; 2^{E+1}$.
\end{enumerate}
The first method results in slightly slower loss of significand than the second method, while the third method changes precision arithmetic from deterministic to stochastic.  Because no empirical difference has been detected among these three different rounding up methods, the first method is chosen in this paper. Further study is required to distinguish the different rounding up methods.

The objectives of precision arithmetic need to be studied further.  For example, Formula \eqref{eqn: uncertainty variance} has rejected the effect of uncertainty on the expected value by incorporating the value shift due to uncertainty as increase of variance, such as in the case of calculating $f(x) = x^{2}$.  The effect of such asymmetrical broadening is unclear.  

The number of bits to be calculated inside uncertainty also needs to be studied further.  For example, when limited bits are calculated inside uncertainty, adding insignificant higher order term of a Taylor expansion may decrease the value error while increasing the uncertainty deviation, which may call for an optimal bits to be calculated inside uncertainty for the truncation rule.  

Because precision arithmetic is based on generic concepts, it is targeted to be a generic arithmetic for both uncertainty-tracking and uncertainty-bounding.  However, it seems a worthwhile alternative to interval arithmetic and the de facto independence arithmetic.  Before applying it generally, precision arithmetic still needs more groundwork and testing.  It should be tested further in other problems such as improper integrations, solutions to linear equations, and solutions to differential equations.

\subsection{Acknowledgements}

As an independent researcher, the author of this paper feels indebted to encouragements and valuable discussions with Dr. Zhong Zhong from Brookhaven National Laboratory, Prof. Hui Cao from Yale University, Dr. Anthony Begley from \emph{Physics Reviews B}, the organizers of \emph{AMCS 2005}, with Prof. Hamid R. Arabnia from University of Georgia in particular, and the organizers of \emph{NKS Mathematica Forum 2007}, with Dr. Stephen Wolfram in particular. Finally, the author of this paper is very grateful for the editors and reviewers of \emph{Reliable Computing} for their tremendous help in shaping this unusual paper from unusual source, with managing editor, Prof. Rolph Baker Kearfott in particular.

\bibliographystyle{unsrt}
\bibliography{PrecisionArithmetic}

\section{Figures}

\clearpage

\clearpage

\begin{figure}
\centering
\includegraphics[width=4.5in,height=2.5in]{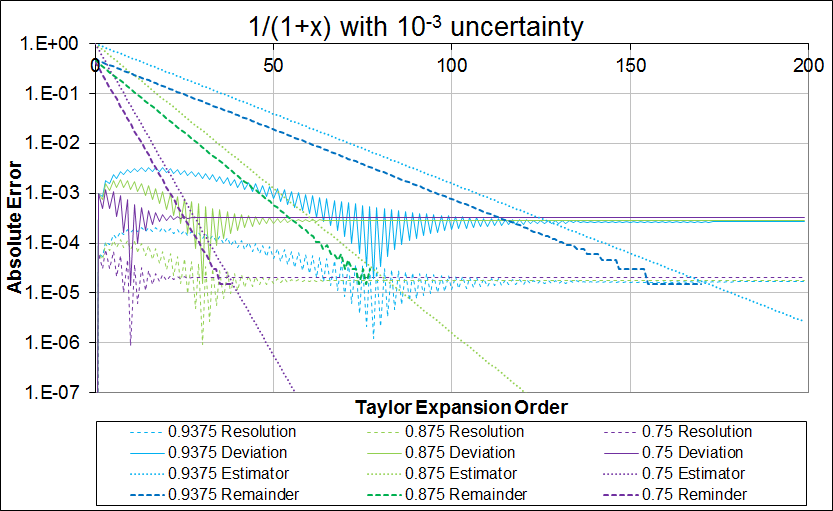} 
\captionof{figure}{ The deviations, resolutions, Cauchy estimator and remainders of a Taylor expansion vs. the expansion orders for different input value with $10^{-3}$ input uncertainty using precision arithmetic with 4-bit calculated inside uncertainty. Different inputs are displayed using different color. }
\label{fig: Prec4_Taylor_1E-3}
\end{figure}

\clearpage

\begin{figure}
\centering
\includegraphics[height=2.5in]{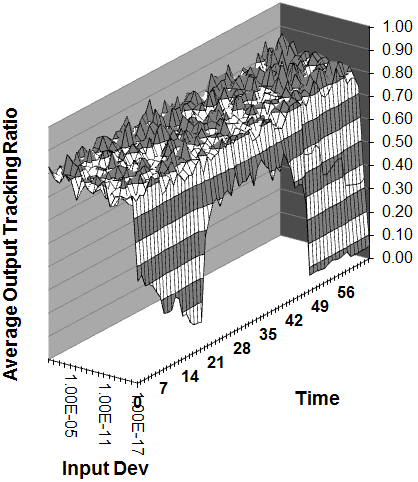} 
\captionof{figure}{ The average tracking ratio vs. time and the input uncertainty deviations for the expressive moving-window linear regression of a straight line with 10-fold increase of both input noise and input uncertainty in the middle using precision arithmetic with 4-bit calculated inside uncertainty.  }
\label{fig: ChangeDirect_Prec4_AvgErrSig_vs_InDev_Time}
\end{figure}

\end{document}